%% file: paper_IREmodel.tex
\pdfoutput=1
\documentclass[manuscript]{aastex}
\usepackage{amsmath}
\usepackage{comment}
\usepackage{grffile}
\usepackage{lscape}
\usepackage{arrayjobx}
\usepackage{multirow}

\usepackage{booktabs}
\usepackage{longtable}
\usepackage{threeparttablex}
\usepackage{tabularx}
\usepackage{arydshln}

\usepackage{lineno}

\newcommand{\bff}{\bf}
\renewcommand{\bff}{}

\usepackage[version=3]{mhchem}

\shorttitle{IRE model}
\shortauthors{Oya et al.}

\title{
FERIA: Flat Envelope Model with Rotation and Infall under Angular Momentum Conservation
}

\author{
Yoko Oya\altaffilmark{1, 2}, 
Hirofumi Kibukawa\altaffilmark{1, 3}, 
Shota Miyake\altaffilmark{4}, 
and Satoshi Yamamoto\altaffilmark{1, 2}
}

\email{oya@taurus.phys.s.u-tokyo.ac.jp}
\altaffiltext{1}{Department of Physics, The University of Tokyo, 7-3-1, Hongo, Bunkyo-ku, Tokyo 113-0033, Japan}
\altaffiltext{2}{Research Center for the Early Universe, The University of Tokyo, 7-3-1, Hongo, Bunkyo-ku, Tokyo 113-0033, Japan} 
\altaffiltext{3}{Institute for Cosmic Ray Research, University of Tokyo, 5-1-5, Kashiwanoha, Kashiwa, Chiba, 277-8582, Japan}
\altaffiltext{4}{Graduate School of Mathematical Sciences, The University of Tokyo, 3-8-1, Komaba, Meguro-ku, Tokyo 153-8941, Japan}

\begin{abstract}
Radio observations of low-mass star formation in molecular spectral lines have rapidly progressed 
since the advent of Atacama Large Millimeter/submillimeter Array (ALMA). 
A gas distribution and its kinematics within a few 100s au scale around a Class 0-I protostar are spatially resolved, 
and the region where {\bff a} protostellar disk is being formed is now revealed in detail. 
In such studies, 
it is essential to characterize the complex physical structure around a protostar 
consisting of an infalling envelope, a rotationally-supported disk, and an outflow. 
For this purpose, 
we have developed a general-purpose computer code `{\tt FERIA}' (Flat Envelope model with Rotation and Infall under Angular momentum conservation) 
generating the image cube data based on the infalling-rotating envelope model and the Keplerian disk model, 
both of which are often used in observational studies. 
In this paper, 
we present the description and the usage manual of {\tt FERIA} and summarize caveats in actual applications. 
This program outputs cube {\tt FITS} files, 
which can be used for direct comparison with observations. 
It can also be used to generate mock data for the machine/deep learnings. 
Examples of these applications are described and discussed to demonstrate 
how the model analyses work with actual observational data. 
\end{abstract}


\begin{document}

\input{part_preamble.tex}

\section{Introduction} \label{sec:intro}

Thanks to increasing capabilities of millimeter/submillimeter interferometers 
including Atacama Large Millimeter/submillimeter Array (ALMA), 
observational studies on a Solar-type (low-mass) protostellar sources 
in molecular spectral lines at a high angular resolution 
have substantially progressed during the last decade. 
A complex structure within a few 100s au around a protostar, 
consisting of a protostellar disk, an infalling envelope, and an outflow, 
has been revealed even for infant sources. 
Indeed, rotationally supported disks (Keplerian disks) have been detected 
not only in Class I sources but also in Class 0 sources 
\citep[e.g.][]{Tobin2012, Yen2013, Yen_diskALMA, Murillo2013, Ohashi2014, Aso2015, Oya_16293, Oya_16293Cyc4, Alves2017_BHB07-11, Lee2017_HH212, Tokuda2017_L1521F, Okoda_15398, Yang2020}. 
Infalling-rotating envelopes have also been characterized for many Class 0 and I sources 
\citep[e.g.][]{Sakai_1527nature, Sakai_1527apjl, Sakai_TMC1A, Ohashi2014, Aso2015, Sakai_TMC1A, Oya_16293, Oya_16293B, Oya_16293Cyc4, Hsieh2019, Imai_B335rotation, Jacobsen2019, Gaudel2020, Yang2020}, 
while the outflow structures near their launching zone have been studied in relation to the \desys\ 
\citep[e.g.][]{Oya_15398, Oya_1527, Oya_483outflow, Oya_16293Aoutflow, Tabone2017, Lee2018_HH212, Zhang2018}. 
These studies are tackling with a key issue when and how the disk structure is formed from an \ire. 
This is of great importance in astronomy and planetary science, 
because a protostellar disk is evolved into a protoplanetary disk, 
which is a birthplace of a planetary system. 
Since the above three components are mutually related, 
their individual characterizations are essential for comprehensive understanding of the disk formation. 

In addition to observational studies, 
theoretical magnetohydrodynamics (MHD) simulation has extensively been conducted 
to elucidate the disk formation process 
\citep[e.g.][]{Machida2011_magneticbreaking, MachidaMatsumoto2011, MachidaHosokawa2013, Tomida2015, Tsukamoto2017, Zhao2016, Zhao2018, Bate2018, MachidaBasu2019, Xu2021}. 
These studies reveal a series of formation processes of a low-mass protostar 
and a disk structure around it, 
where the outflow from the protostar and/or the disk is also reproduced. 
Such simulations are very useful to clarify the physics occurring in the vicinity of the protostar 
and have indeed played an important role in interpretation of observational results. 
On the other hand, 
they involve a number of parameters and assumptions, 
and hence, direct comparison with observed intensity distributions 
and/or kinematic structures is not always easy. 
For this reason, 
simple phenomenological models are often employed 
to interpret 
the observed data. 

As for the \desys, 
the Keplerian disk model and the ballistic model of the \ire\ \citep{Oya_15398} 
are successfully used to characterize the disk and envelope structures 
{\bff \citep[][]{Sakai_1527apjl, Sakai_TMC1A, Oya_16293, Oya_483, Lee2017_HH212, Okoda_15398, Imai_B335rotation}.} 
By use of these models, 
central mass, 
\sam\ of the gas, and the \ia\ of the \desys\ are evaluated. 
For the outflow structure, 
the parabolic model has often been employed 
\citep[e.g.][]{Lee_outflow, Hirano2010, Arce2013, Oya_15398, Oya_1527}. 
The model including the outflow rotation is also invoked, 
because the rotation motion of the outflow/jet has been detected {\bff in some sources} 
\citep[e.g.][]{Hirota2017, Zhang2018, Oya_483outflow}. 
With the outflow models, 
the \ia\ of the outflow axis with respect to the line of sight in the vicinity of the protostar can be derived, 
which helps the analysis of the \desys\ \citep[e.g.][]{Oya_15398, Oya_483, Oya_483outflow}. 
Moreover, a possible rotation of outflows can be discussed in relation to the \sam\ of the gas 
derived from the analysis of the \ire\ \citep{Oya_483outflow, Oya_16293Aoutflow}. 
Information derived with the simple models will be useful as initial constraints for detailed MHD simulations. 

In the ALMA era, 
the above models of the \desys\ and the outflow system 
are becoming more and more important for characterization of observed results. 
To make such analyses easier, 
general simulation tools for these systems are useful. 
With this in mind, 
we developed a computer program `\model' (\modellong)
for calculating data cubes of the \ire\ and Keplerian-disk models by mock observations. 
This program is public in GitHub\footnote{{\url{\githubmodel}}}. 
The model is not a full simulation 
considering hydrodynamics, magnetic field, radiation transfer, and various other factors specific to each source. 
However, 
its simplicity rather helps us to extract the essence of observed kinematic structures. 
Since this program outputs a cube \fits\ file, 
it is easy to compare with the observational data. 
This program can also be used to generate mock data for the analyses using supervised machine learnings. 

In this paper, 
we describe the model in Section~\ref{sec:formulae}. 
We introduce how to use the codes in Section~\ref{sec:howto}, 
and summarize some examples of the model results in Section~\ref{sec:examples}. 
Moreover, 
we {\bff present} some applications of \model\ to actual observational data in Section~\ref{sec:obs}. 
We discuss the applicability of this model in Section~\ref{sec:disc} 
with some caveats for a practical use. 
Finally, Section~\ref{sec:summary} summarizes the paper.

\section{Basic Formulae and Key Assumptions of the Envelope Model} \label{sec:formulae}

\subsection{Overview} \label{sec:formulae_overview}

{\bff 
\model\ assumes the axisymmetric structure of the \desys\ of a finite size 
with/without flared structure (Figure~\ref{fig:scheme}). 
Gas motion is approximated by a ballistic motion and/or the Keplerian rotation of the central mass, 
where the self-gravity of the gas is not considered (Figure~\ref{fig:scheme}e). 
Distributions of the gas density and molecular emissivity are given a priori in the power-law form. 
\model\ calculates the data cube 
of a mock observation for a \desys\ with a given \ia, taking a finite spatial and spectral resolutions into account. 

Specifically, 
\model\ defines the mesh (Figure~\ref{fig:scheme}f) and calculates the velocity vector of the gas element in each mesh. 
Then, \model\ calculates the line intensity for each position on the 2-dimensional (position-position) plane of the sky 
by using the assumed distributions of gas density and molecular emissivity. 
Each position has a 1-dimensional (velocity) spectrum, 
which is obtained by projecting the contribution from the gas elements along the \los\ onto the \pos. 
Finally, 
convolutions for the limited beam size and spectral resolution are applied for the position-position plane and the velocity axis, respectively, 
to obtain the output cube in the position-position-velocity manner. 
With this cube data, comparison with the observation data can be done in various ways. 
\model\ can provides position-velocity (PV) diagrams, 
which are often used for the comparison. 

Table \ref{tb:params} summarizes the physical parameters used in the disk/envelope model, 
as well as additional parameters necessary for model calculations and header information. 
Their details are described in Section~\ref{sec:howto}. 
}

\subsection{Geometric and Kinematic Structures} \label{sec:formulae_vel}


{\bff 
Figure~\ref{fig:scheme} shows the geometric structure assumed in \model. 
The thickness of the \desys\ is assumed 
to be constant or flared as the distance from the protostar (Figure~\ref{fig:scheme}d). 
Mock observation is performed by incorporating the effect 
of the \ia\ and the position angle (P.A.) of the major axis in the \pos. 

\model\ employs the envelope model with infall and rotation motion of a gas element 
under the conservation of the energy and the angular momentum, i.e. the ballistic motion. 
As well, \model\ outputs 
a mock observation of a Keplerian disk model 
and that of a combination of a ballistic model and a Keplerian disk model. 

{\bff The theoretical model for an envelope with the ballistic motion was 
first developed by \citet{CassenMoosman1981}. 
\citet{Ohashi1997} and \citet{Momose1998} presented a simplified model} 
in order to interpret the rotation structure of low-mass protostellar cores, 
where the infall motion and the rotation motion are assumed to be independent. 
\citet{Sakai_1527nature} reported the observational results of the \cycCH\ line showing the kinematic structure 
which can be interpreted as the ballistic motion described below. 
A three-dimensional model of an \ire\ was constructed 
to simulate its kinematic structure by \citet{Oya_15398}, 
which in fact reproduced 
the kinematic structure observed for the protostellar source L1527 \citep{Sakai_1527apjl}. 
The kinematic structure employed in \model\ is 
essentially the same as the `\ire\ model' reported by \citet{Oya_15398}. 

In this model, 
the ballistic motion {\bff to the central source} is assumed; 
i.e., the gas is simply assumed to be falling and rotating under the gravity of a central protostar. 
Thus, the motion of the gas is approximated by the particle motion, 
ignoring effects of gas pressure, magnetic field, and self-gravity. 
Because of the energy and angular momentum conservation, 
the gas cannot fall inward of a certain radius, or the periastron. 
This position is called as the `\cb' in this model {\bff (Figure~\ref{fig:scheme}e).} 
The radius of the \cb\ (\rcb) is represented as; 
\begin{align}
	r_{\rm CB} 	&= \frac{j^2}{2 G M}, \label{eq:rcb}
\end{align}
where $G$ is the gravitational constant, 
$M$ is the central (protostellar) mass, 
and $j$ is the \sam\ of the gas. 
It is the radius at which all the kinetic energy is converted to the rotational energy. 
It is a half of the \centr\ (\rcr), 
where the gravitational force and the centrifugal force balances with each other; 
\begin{align}
	r_{\rm CR} 	&= \frac{j^2}{G M} \\ 
				&= 2 r_{\rm CB}. \label{eq:rcr}
\end{align}
A basic concept of the \cb\ is introduced to explain 
the kinematic structure of the \ire\ of a low-mass protostellar source L1527 \citep{Sakai_1527nature}. 

Rotation and infall velocities (\vrot\ and \vfall) of the gas 
at the distance of $r$ to the protostar are represented as follows; 
\begin{align}
	v_{\rm rot}		&= \frac{j}{r} \\
				&= \frac{1}{r} \sqrt{2 G M r_{\rm CB}}, \label{eq:vrot} \\
	v_{\rm fall}	&= \sqrt{\frac{2 G M}{r} - v_{\rm rot}^2} \\
				&= \frac{1}{r} \sqrt{2 G M \left( r - r_{\rm CB} \right)}. \label{eq:vfall}
\end{align}
Thus, the velocity field is determined by $M$ and \rcb. 
The \ia\ ($i$) of the \desys\ also affects the apparent velocity along the \los. 
At the \cb, \vfall\ equals to 0, and \vrot\ takes its maximum value. 
On the other hand, 
\vfall\ takes its maximum value at the \centr. 
This situation is shown in Figure~\ref{fig:velocityprofile}.

\subsection{Calculation of the Line Intensity} \label{sec:formulae_int}


For intensity simulations, 
the gas distribution has to be provided. 
In \model, 
a power-law radial distribution is adopted, 
where no gas is assumed inside {\bff a specific radius (see Section \ref{sec:howto_inp_cube}).} 
The power-law of $r^{-1.5}$ corresponds to the density profile of an infalling cloud 
\citep[e.g.][]{Shu1977, Ohashi1997, Harvey2003}. 
An optically thin condition is also assumed in this model, 
where the intensity of the line emission is proportional to the column density along the line of sight. 
Namely, excitation effects and radiative transfer effects are not considered. 
These assumptions are rather arbitrary. 
However, 
these effects can effectively be incorporated by changing the power-law index of the gas distribution. 
If the main purpose of the simulation is to reproduce the velocity field of the gas, 
these assumptions do not affect the results seriously. 
Nevertheless, ones may want to consider the above effects. 
For such a purpose, 
the scripts of \model\ is open to modify by themselves. 

A spectral line is assumed to have an intrinsic Gaussian profile with a certain line width, 
and an intensity distribution is spatially convolved with a Gaussian beam 
with a certain full width at half maximum (FWHM). 

\section{Practical Information in Using \model} \label{sec:howto}

\model\ is distributed in GitHub\footnote{{\url{\githubmodel}}}, 
which consists of seven {\tt C++} files and one header file. 
{\tt Makefile} to build this program and a template file to input 
are also available in GitHub. 
{\bff As well, a python script to run \model\ recursively with various combinations of the parameter values is provided.} 

\model\ outputs a \fits\ file 
for easy comparison between model results and observational data. 
For this purpose, this program requires \cfitsio\ library installed in advance. 
\cfitsio\ is a C and Fortran subroutines for reading and writing \fits\ files \citep{Pence_CFITSIO}, 
which is distributed by NASA\footnote{{\url{https://heasarc.gsfc.nasa.gov/fitsio/}}}. 
Users of \model\ may need to modify the {\tt Makefile} 
according to the configuration of \cfitsio\ library in their environments. 

\subsection{Mesh in \model} \label{sec:formulae_mesh}

{\bff 

\model\ considers a three-dimensional (position-position-position; PPP) gas structure, 
and it outputs a mock observational result as a three-dimensional (position-position-velocity; PPV) \fits\ file, i.e. a data cube. 
Its geometrical structure is shown in Figures~\ref{fig:scheme}(a$-$d), 
while its kinematic structure is in Figure~\ref{fig:scheme}(e) (see Section~\ref{sec:formulae_vel}). 
They are divided into small elements as shown in Figure~\ref{fig:scheme}(f). 
Here, nested gridding is not employed, 
which would cause artifact features (see Section~\ref{sec:formulae_caveats}). 
This is because the major purpose of \model\ is to output a pile of data cubes with its agility 
rather than to perform an attentive simulation (see Section~\ref{sec:obs}). 

In the convolution processes mentioned in Section~\ref{sec:formulae_int}, 
the intrinsic line width and the beam size are used, 
which depend on each observation. 
Thus, mesh sizes for simulations need to be adjusted for each observation project 
(see Section~\ref{sec:howto_inp_grid}).  

}

\subsection{Input Parameters} \label{sec:howto_inp}

Table~\ref{tb:params} summarizes the parameters for the model. 
These parameters are specified in the input file (see \fileinput), 
except for two parameters (\logNpix\ and \logNvel) specified in the header file (\filehead). 
Some of them are key free parameters, 
while the others need to be given 
as the header information for the output \fits\ file.

\subsubsection{Header Information for the Output FITS File} \label{sec:howto_inp_head}

The output \fits\ file is named after the input values of the physical parameters by default. 
Alternatively, the output file name can be given voluntarily 
as long as its length does not exceed the limit given in the header file ({\tt lfilename} in \filehead; 256 as the default). 
When a \fits\ file with the output file name already exists, 
it will be replaced by the new model result with the `overwrite' parameter of `y' (yes). 
With the `overwrite' parameter of `n' (no), 
the file is replaced {\bff only} if its numbers of the mesh are different from the current setting (see Section~\ref{sec:howto_inp_grid}). 

For easy comparison between model results and observational data of an actual target source, 
the output \fits\ file of the model has some header information. 
The field center of the \fits\ file is given by its frame (e.g. ICRS, J2000) 
and coordinates (the right ascension and the declination). 
The central source is assumed to be at {\bff these} coordinates. 
The line-of-sight velocity is obtained as the combination of the gas motion and the systemic velocity of the source. 
Thus, the output \fits\ file can directly be overlaid on the observational data 
in usual softwares for viewing and analyzing data cubes. 

In addition, the model has parameters as optional notes; 
the name of the source, the molecular line transition, and its rest frequency. 
These parameters do not affect the model result.

\subsubsection{Input Parameters for the Geometric/Kinematic Structures} \label{sec:howto_inp_cube}

As seen in the {\bff Eqs.} (\ref{eq:rcb})$-$(\ref{eq:vfall}), 
the kinematic structure can be obtained for given $M$ and \rcb. 
The distance to the object ($d$) and the \ia\ of the \desys\ ($i$; Figure~\ref{fig:scheme}b) also affect the modeled data cube. 
The model takes into account the position angle (P.A.) of the \desys\ for easy comparison with observational data; 
the P.A. of the \desys\ is defined as the P.A. of the line along which the mid-plane of the \desys\ extends (Figure~\ref{fig:scheme}c). 
The direction of the rotation motion is specified by an integer ($1$ or $-1$); 
`1' and `-1' stand for the counterclockwise and clockwise rotation, respectively, 
with the inclination angle of 0\degr\ {\bff (Figure~\ref{fig:scheme}e).} 
These are the main parameters for calculating the kinematic structure.

The physical structure is defined by the following parameters; 
the outer and inner radii of the envelope (\rout, \rin) in au, 
the thickness of the gas structure (\henv) in au, 
and the flare angle of the scale height (\fenv) in degree. 
{\bff Figure~\ref{fig:scheme}(d)} schematically shows what these parameters denote. 
No molecular emission is assumed if the distance from the protostar ($r$) 
is larger than the outer radius 
or smaller than the inner radius. 
The gas structure is assumed to be a thin disk with the constant thickness of \henv, 
or a flared one with a flare angle of \fenv. 
If both thickness and flare angle are set to be non-zero values, 
the model incorporates both of them as shown in {\bff Figure~\ref{fig:scheme}(d);} 
the thickness of the gas structure increases with the input flare angle (\fenv), 
where its extrapolated value at the protostar equals to the input thickness (\henv).

\model\ can model {\bff a Keplerian disk} as well as {\bff an} \ire. 
{\bff To obtain a} Keplerian disk model, {\bff \rcb\ is set to be} larger than \rout. 
The following physical parameters are for the Keplerian disk component 
corresponding to those for the \ire\ component: 
\hdisk\ and \fdisk. 

Moreover, 
{\bff a combination of the \ire\ and the Keplerian disk inside it can be modeled by \model\ (Figure~\ref{fig:scheme}e).} 
{\bff To obtain such a model, 
\rcb\ is set to be smaller than \rout\ and larger than \rin.} 
In this case, the transition zone from the envelope to the disk is at the \cb.

\subsubsection{Input Parameters for the Line Emissivity} \label{sec:howto_inp_emit}

{\bff 
The molecular density is assumed to have a power-law radial distribution. 
Its value at the \cb\ is denoted as \nmolCB. 
\dprofenv\ denotes the power-law index of the molecular density in the envelope component. 
As well, the gas temperature at the \cb\ and its power-law index are denoted as \tCB\ and \tprofenv, respectively. 
\dprofdisk\ and \tprofdisk\ are the corresponding parameters for the Keplerian disk component. 
}

In \model, the emissivity is simply assumed to be proportional 
to {\bff the product of} the column density of the molecule {\bff and the gas temperature}. 
The absolute value of the calculated intensity is not scaled to meet the actual line intensity. 
Thus, the absolute value of the calculated intensity in each pixel does not have physical meaning, 
while the relative intensity between the pixels has some meaning. 
The line intensity projected onto the \pos\ can be normalized by its maximum value 
by setting the `Normalize' parameter to be `y'. 
Users of this model may want to take into account the excitation effect and/or radiation transfer for calculating the molecular intensity. 
They can modify the \filesky\ file to meet their needs. 
As well, any density and temperature distributions can be incorporated by editing the \fileenv\ file.

\subsubsection{Input Parameters for Convolution} \label{sec:howto_inp_conv}

The line intensity is convolved with the intrinsic line width of the gas and the Gaussian beam. 
A spectral line at each three-dimensional position is assumed to have an intrinsic Gaussian profile 
whose FWHM is given by the parameter `Linewidth' in the input file. 
After the projection onto the \pos, 
the line intensity is convolved with the Gaussian beam. 
The Gaussian beam is defined by the FWHM of major and minor axes and the P.A. of the major axis.

\subsubsection{Input Parameters for the Mesh} \label{sec:howto_inp_grid}

The numbers of the mesh for the three position-axes and the velocity axis 
are specified in the header file (\filehead) to input to the program. 
The number of the mesh is set to be 
$2^{\rm \logNpix}$ for the position axes 
and $2^{\rm \logNvel}$ for the velocity axis. 
See Section~\ref{sec:howto_resource} for the limitation for these values. 

The mesh size of the position and velocity axes 
are specified in the input file to the program (\fileinput). 
The mesh size of the position axis (`pixel size') is given in arcsecond, 
while that of the velocity axis (`velocity resolution') is given in \kmps. 
It should be noted that the mesh size of the velocity axis does not always assure the resolution for the velocity, 
if the image suffers from coarse sampling along the position {\bff axes,} 
as described in Section~\ref{sec:formulae_caveats}. 

The field size of the map is obtained as (pixel size $\times 2^{\rm \logNpix}$), 
while the velocity range is from $-$(velocity resolution $\times 2^{{\rm \logNvel} - 1})$ to $+$(velocity resolution $\times 2^{{\rm \logNvel} - 1})$ 
centered at the systemic velocity of the source.

\subsubsection{Input Parameters for Position-Velocity Diagrams} \label{sec:howto_inp_PV}

\model\ outputs a PV 
diagram as well as a data cube.  
The position axis of the PV diagram is defined 
with its P.A. and central position. 
The central position is given as the offsets from the protostar (i.e. the field center) in au. 
When the cube \fits\ file with the output file name already exists and is not overwritten (Section~\ref{sec:howto_inp_head}), 
only a new \fits\ file of a PV diagram is generated by importing the existing cube \fits\ file. 
{\bff If a \fits\ file of a PV diagram with the same file name already exists, it is overwritten.}

\subsection{Notes on Computer Resources} \label{sec:howto_resource}

The calculation time and required memory 
mostly depend on the numbers of the mesh defined by `\logNpix' and `\logNvel' (Section~\ref{sec:howto_inp_grid}). 
With a computer with {\tt macOS Mojave} and the CPU of 3.7 GHz (Intel Xeon E5), 
the time for generating one new cube \fits\ file is about 90 seconds  
with `\logNpix' and `\logNvel' of 8. 
The available values for these parameters are restricted by the size of the random access memory (RAM) of users' computers. 
For instance, 
the maximum value for these parameters is 7 for computers with RAM of 8, 16, and 32 GB, 
and is 8 for computers with RAM of 64 and 128 GB. 
This program does not employ the parallel computing. 

\subsection{Caveats for Artificial Features} \label{sec:formulae_caveats}

\model\ employs the discrete Fourier translation 
to convolve the intensity with the Gaussian beam and the Gaussian line profile. 
This can make fringes for coarse meshes. 

In addition, 
the mesh sizes in this model are taken to be uniform for the whole cube, 
so that the velocity resolution can be effectively worse than that specified 
if the velocity vector of the gas steeply changes within a small scale. 
This situation happens, for instance, {\bff near the protostar.} 
For this reason, a clumpy artifact feature often appears 
in the model 
{\bff around} the protostar, 
if the {\bff inner radius} 
is small.

\section{Examples of the Model Results} \label{sec:examples}

Figure~\ref{fig:result_example} shows an example of the \ire\ model result. 
Figure~\ref{fig:result_example}(a) shows the integrated intensity map of the model, 
while Figure~\ref{fig:result_example}(b) shows the PV diagram along the blue arrow shown in Figure~\ref{fig:result_example}(a). 
A protostar with the central mass 
($M$) of \Mex\ is located 
at the central position in Figure~\ref{fig:result_example}(a). 
The distance to the protostar from the Sun ($d$) is set to be \Dex. 
The envelope has an outer radius (\rout) of \Routex, 
and the radius of the \cb\ is \CBex. 
Note that there is no molecular emission outside \rout\ and 
inside \rin\ (\rin\ $=$ \rcb) 
according to the specification of \model\ (Section~\ref{sec:howto_inp_cube}). 
The envelope is assumed to have an edge-on configuration (\ia\ of \Iex) 
extending along the east-west axis. 
The integrated intensity relative to its peak value in the panel is shown in a gray scale. 
The integrated intensity is the highest around the \cb. 

In the modeled PV diagram (Figure~\ref{fig:result_example}b), 
the angular offset of 0\arcsec\ corresponds to the protostellar position. 
The vertical axis represents the line-of-sight velocity of the molecules 
involving the systemic velocity of the source. 
In Figure~\ref{fig:result_example}(b), 
a spin-up feature can be seen along the east-west axis; 
the rotation velocity increases as approaching to the \cb. 
The velocity takes its maximum and minimum value around the \cb, 
where the velocity is red-shifted and blue-shifted in the eastern {\bff and western} sides of the protostellar position, respectively. 
Since no gas resides inside the \cb\ in this model, 
the spin-up feature disappears at the \cb. 
This is in contrast to the Keplerian disk case described later (see Section~\ref{sec:examples_Kep}). 
In addition, 
the \cvelcomp\ due to the infall motion can be recognized. 
Toward the protostellar position, 
only a velocity shift due to the infall motion can be seen, 
since the rotation motion is perpendicular to the \los.

\subsection{Envelope Models with Various Physical Parameters} \label{sec:examples_params}

Figures~\ref{fig:varparams_mom0}$-$\ref{fig:varparams_PV} show the model results with various parameter values. 
Figure~\ref{fig:varparams_mom0} shows the integrated intensity (moment 0) maps of the models with various \ia s ($i$) and the radii of the \cb\ (\rcb), 
while Figure~\ref{fig:varparams_mom1} shows their velocity field (moment 1) maps. 
Figure~\ref{fig:varparams_PV} shows the PV 
diagrams corresponding to Figure~\ref{fig:varparams_mom1}. 
The position axis in the PV diagrams is along the mid-plane of the \desys\ (P.A. 270\degr), 
which is shown by the blue arrow in Figure~\ref{fig:result_example}(a). 
The other parameters are common for these models. 
Some parameters do not affect the calculation, 
as described in the caption of Figures~\ref{fig:varparams_mom0}$-$\ref{fig:varparams_PV}, 
and thus they are set arbitrarily. 
The employed parameter values are summarized in the captions of these figures. 
The images are obtained by convolving the emission 
with the intrinsic Gaussian profile with the FWHM of 0.2 \kmps\ 
and the Gaussian beam of (\bmaj $\times$ \bmin) $=$ (0\farcs2 $\times$ 0\farcs2) (P.A. 0\degr).

In the following subsections, 
physical implications of some characteristic features seen in the calculated images are described. 
This is useful not only to understand the model but also to interpret the observed images.

\subsubsection{Integrated Intensity Maps and Velocity Field Maps} \label{sec:examples_params_mom01}

In Figure~\ref{fig:varparams_mom0}, 
the maps for the edge-on configuration (a highly inclined configuration; $i$ = 90\degr) 
show double-peaked intensity distributions. 
At $i$ of 30\degr\ and 60\degr, 
an elliptic hole of the intensity distribution is seen. 
With the face-on configuration ($i$ = 0\degr), 
the intensity distributions show ring-like structures. 
These holes appear because there is no molecular emission inside the radius of the \cb\ in this model (Section~\ref{sec:howto_inp_cube}). 
For each case, the intensity takes the maximum value near the \cb. 

The velocity of the gas in the model depends on the central mass ($M$) as well as $i$ and \rcb. 
The central mass only scales the value of velocity up and down, 
and it does not essentially affect the overall appearance of the velocity field map. 
Thus, we show the results for various $i$ and \rcb\ with the fixed central mass in Figure~\ref{fig:varparams_mom1}. 

In the edge-on configuration ($i$ = 90\degr) case in Figure~\ref{fig:varparams_mom1}, 
the averaged velocity in the eastern and western sides of the protostar is red- and blue-shifted, respectively. 
These velocity shifts represent the rotation motion around the protostar. 
The maximum velocity shifts are seen around the \cb, 
which is consistent with the velocity profile of the \irm\ (Figure~\ref{fig:velocityprofile}). 
The velocity does not show any gradient along the north-south axis. 

On the other hand, a velocity gradient along the north-south axis is seen 
in the panels for $i$ of 30\degr\ and 60\degr. 
These velocity gradients are due to the infall motion. 
As shown in Figure~\ref{fig:scheme}, 
the northern edge of the envelope is close to us 
in the cases with positive $i$ and a P.A. of 90\degr. 
Thus, the line emission is red-shifted in the northern side of the protostar. 
In these panels, 
the velocity fields show skewed features; 
the most red- and blue-shifted components are seen in the northeastern and southwestern sides of the protostar, respectively. 
This is because that the velocity components of the rotation and infall motion 
projected along the \los\ have the same direction {\bff (see Figure~\ref{fig:scheme}e)} at these positions. 
Such a skewed feature was first revealed in the observation of L1551 IRS 5 by \citet{Momose1998}. 
The results for the face-on configuration ($i$ = 0\degr) are trivial; 
the velocity shift is completely symmetric to the mid-plane of the envelope, 
and thus, the averaged velocity is infinitesimal everywhere. 

\subsubsection{Position-Velocity Diagrams} \label{sec:examples_params_PV}

Figure~\ref{fig:varparams_PV} shows the PV diagrams along the {\bff major axis of the} \desys\ (P.A. 270\degr). 
For the edge-on configuration ($i$ = 90\degr), 
the spin-up feature toward the \cb\ is seen regardless of \rcb. 
The maximum velocity-shift seen at the \cb\ is larger for a smaller \rcb, 
if the other parameters are fixed (see {\bff Eq.} (\ref{eq:vrot})). 
The infall motion can be confirmed as the counter-velocity components. 
The velocity shifts at the angular offset of 0\arcsec\ (i.e. the position of the protostar) 
also reflect the infall motion. 
For $i$ of 30\degr\ and 60\degr, 
the counter-velocity components are not seen in most of panels, 
while they can slightly be seen in the case for \rcb\ of 30 au at $i$ of {\bff 60\degr.} 
The gas having considerable infall motion in the models for \rcb\ of 100 and 300 au 
are distant from the protostellar position on the plane of the sky. 
Therefore, these components are almost outside the beam, 
and do not make effective contributions to these PV diagrams. 
Thus, investigating the velocity gradient perpendicular to the mid-plane of the \desys\ is essential for inclined cases, 
as demonstrated below and presented by \citet{Oya_16293} for instance. 
For the face-on configuration ($i$ = 0\degr), 
it is natural that a velocity gradient due to the infall and rotation motion cannot be seen regardless of \rcb. 

Figure~\ref{fig:varparams_PV-PA} shows the model results of the PV diagrams 
prepared along six P.A.s for $i$ of 0\degr, 30\degr, 60\degr, and 90\degr. 
The \rcb\ value is fixed to be 100 au. 
The P.A.s of the position axes are taken for every 30\degr, 
as shown by arrows in Figure~\ref{fig:varparams_mom1}. 
The P.A.s of `270\degr' and `0\degr' represent the direction along the mid-plane of the envelope 
and that perpendicular to it, respectively. 
For the edge-on configuration ($i$ = 90\degr), 
the distributions look concentrated around the protostar in all the PV diagrams except for the P.A. of 270\degr. 
A slight velocity gradient can be seen for these P.A.s, 
except for the P.A. of 0\degr\ (the direction perpendicular to the mid-plane). 
For the diagram with the P.A. of 0\degr, 
only the velocity shift due to the infall motion near the `\centr' is visible, 
which is a half of the peak velocity-shift of the rotation motion at the \cb\ (Figure~\ref{fig:velocityprofile}). 

At $i$ of 30\degr\ and 60\degr, 
the kinematic structure systematically changes from P.A. to P.A., 
which is caused by the complex combination of the rotation and infall motions. 
As shown in {\bff Figures~\ref{fig:scheme}(e)} and~\ref{fig:varparams_mom1}, 
the rotation and infall motions cancel each other in the southeastern and northwestern sides of the protostar, 
while they strengthen each other in the northeastern and southwestern sides. 
For this reason, the absolute values of the velocity shift tend to be higher 
in the diagrams with the P.A. of 30\degr\ and 60\degr\ (southwest-northeast) 
than those with the P.A. of 300\degr\ and 330\degr\ (southeast-northwest). 
The velocity shifts due to the infall motion is seen for the P.A. of 0\degr. 
Since the infall motion has its maximum velocity at the \centr, 
the maximum velocity shift appears at a position with a slight offset from the protostar. 
For the face-on configuration ($i$ = 0\degr), 
no velocity gradient is seen for any P.A.

\subsection{Comparison with the Keplerian Disk Model} \label{sec:examples_Kep}

In analyses of rotating structures around young stellar objects, 
a thin-disk model with the Keplerian rotation is frequently employed. 
\model\ can also generate a \fits\ file of a data cube for the Keplerian disk (Section~\ref{sec:howto_inp_cube}). 
Figure~\ref{fig:result_kepler} shows the model results of the Keplerian disk. 
Although the results are trivial and well known, 
we present these figures to compare them with the results for the \irm\ described in Section~\ref{sec:examples_params}. 

In the Keplerian disk model, 
the velocity of the gas at the radius $r$ from the protostar is represented as; 
\begin{align}
	v_{\rm Kep} &= \sqrt{\frac{GM}{r}}, \quad v_{\rm fall} = 0. \label{eq:kep}
\end{align}
As shown in Figure~\ref{fig:velocityprofile}, 
the Keplerian velocity takes a smaller value than \vrot\ with the \irm\ by a factor of ${\displaystyle \sqrt{2}}$ 
at the \cb\ ($r$ = \rcb) for the same central mass, 
while it equals to \vrot\ and \vfall\ of the \irm\ at the \centr\ ($r$ = 2\rcb).

Figure~\ref{fig:result_kepler} shows 
the integrated intensity maps, the velocity field maps, and the PV diagrams 
of the Keplerian disk models for various \ia s ($i$). 
In these models, 
the central mass ($M$) and the outer radius (\rout) are fixed to 0.3 \Msun\ and 500 au, respectively. 
The constant thickness (\hdisk) of 50 au is assumed. 
The other parameters are summarized in the caption. 

In the integrated intensity maps (Figure~\ref{fig:result_kepler}a), 
the distributions look compact and are concentrated around the protostar. 
Since the density of the gas is assumed to be proportional to $r^{-1.5}$ in this case, 
the contribution from the vicinity of the protostar is dominant. 
This appearance is in contrast to the \ire\ model case described in Section~\ref{sec:examples_params_mom01}, 
where the peak intensities are seen near the \cb. 
In the velocity field maps (Figure~\ref{fig:result_kepler}b), 
the rotation motion is clearly visible. 
It is well-known 
that the velocity field of the Keplerian disk has reflection symmetry with respect to the disk mid-plane 
regardless of the \ia. 
This is again in contrast to the \ire\ model case, 
which shows a skewed feature (Figure~\ref{fig:varparams_mom1}). 

In the PV diagrams along the disk mid-plane (P.A. 270\degr), 
the spin-up feature is seen, 
except for the model with a face-on configuration ($i$ = 0\degr). 
When the kinematic structure is well resolved by a beam, 
no counter-velocity component is seen in contrast to the \ire\ model case (Figure~\ref{fig:varparams_PV}) 
because of the absence of the infall motion. 
In the PV diagrams along the direction perpendicular to the disk mid-plane (P.A. 0\degr), 
no velocity gradient is seen regardless of $i$, 
while the \ire\ model shows a velocity gradient along this direction due to the infall motion. 
These different behaviors between the Keplerian disk model and the \ire\ model 
correspond to the difference of their velocity field maps described above (Figures~\ref{fig:varparams_mom1} and~\ref{fig:result_kepler}b). 
On the basis of these features, 
we can discriminate between the Keplerian motion and the \ire\ in principle. 
However, such discrimination is not always obvious for observed data in practice, 
as shown later.

\section{Comparison with the Actual Observations} \label{sec:obs}

\subsection{General Aim of the Model} \label{sec:obs_aim}

The \ire\ model described in Section~\ref{sec:formulae} is a quite simplified one as mentioned in Section~\ref{sec:intro}; 
the model does not consider any excitation effects, 
radiative transfer effects, 
abundance variations of molecules, 
hydrodynamics effects, 
the effect of the self-gravity of the gas, 
and temporal variation of the angular momentum. 
In reality, 
the emission may be optically thick in some parts, 
and the distribution itself may be asymmetric around the protostar. 
Moreover, 
the \ire\ model is not appropriate in some cases (see Section~\ref{sec:disc_caveats}). 
Thus, it is not very fruitful to make a fine tuning of the model 
so as to better match with the observed intensity. 
Thus, these models have been used just for extracting the fundamental characteristics 
of the observed kinematic structure 
\citep[e.g.][]{Sakai_1527apjl, Oya_16293}. 

The advantage of \model\ is its agility and easiness to use in comparison with 
full hydrodynamics simulations including the above {\bff detailed} effects. 
Thus, the analysis with using \model\ helps observers 
to find reasonable physical parameter values for their target sources instantly, 
which can be used as the initial parameters for further analyses.

\subsection{Obtaining the Best Fit Parameters} \label{sec:obs_fit}

%
%

In this section, 
we present some examples of the model analysis by using \model\ for actual observational data taken with ALMA. 
So far, 
model simulations with a wide range of physical parameters and eye-based fitting 
have usually been employed. 
We here perform unbiased evaluation of the parameter values 
{\bff by using the cube data.} 


In model analyses, 
chi-squared (\chisq) tests are often employed to derive 
the best model parameters. 
For instance, 
\citet{Oya_16293Cyc4} performed the reduced \chisq\ tests for the PV diagrams. 
They prepared the PV diagrams of the observed line emissions and the models along the major axis of the disk mid-plane. 
We can perform similar tests for cube data by using \model. 
We here show trials of a \chisq\ test for the observational data 
toward three young low-mass protostellar sources: {\bff L1527 IRS (hereafter L1527), B335, and Elias 29.} 
{\bff The fitting for L1527 is performed to confirm the applicability of \model. 
B335 is employed as an example for an infall-dominated system with a slight rotation motion. 
To the contrast, Elias 29 has been reported to have a Keplerian disk.} 
Details of the observation and the analysis for L1527 are described in Appendix~\ref{sec:app_obs}, 
while those for {\bff B335 and Elias 29}  
were already reported by {\bff \citet{Imai_B335rotation} and \citet{Oya_Elias29},} respectively. 
For each target source, 
we prepare a data cube of a molecular line observed with ALMA concentrated in the vicinity of the protostar. 

Model simulations are performed by using \model\ for various sets of 
the central mass ($M$) and the \ia\ ($i$) as the free parameters. 
As well, 
the radius of the \cb\ (\rcb) or the inner radius (\rin) of the emitting region is employed as a free parameter 
for the \ire\ model or the Keplerian disk model, respectively. 
We calculate the reduced \chisq\ values between the observed and modeled data cubes\footnote{The intensities in each model cube are scaled to reduce the \chisq\ value, preserving the relative intensity among pixels.}, 
and find the best-fit parameter values. 
Contribution to the reduced \chisq\ values from imperfection of the model would overwhelm those of the statistical noise. 
Under such a constraint, 
it is difficult to quantitatively discuss the likelihood for the parameter ranges. 
Therefore, 
we just demonstrate unbiased screenings for the parameter sets in this study, 
while further interpretation of the results of the model analyses are left for 
future studies 
on individual sources. 

\input{part_chisq.tex}

\subsection{Combination with Machine/Deep Learnings} \label{sec:obs_fit_ml}

In the last decade, 
a plenty of observational data of ALMA have been archived for open use. 
Some of them have a high spatial and velocity resolutions enough 
to trace the kinematic structure in the vicinity of a protostar. 
With such a flood of observational data, 
it gets more and more important to analyze them effectively and automatically. 
In this context, 
it would be effective to adopt machine/deep learnings for analyzing the observational data 
\citep[see][]{Baron2019}. 

In fact, for instance, 
the principal component analysis (PCA) 
is often employed 
to analyze observational data in an unbiased way 
\citep[e.g.][]{Ungerechts1997, Meier2005, Neufeld2007, Watanabe2016, Spezzano2017, Okoda_15398PCA, Okoda_483PCA}. 
It is an unsupervised machine learning algorithm for dimensionality reduction \citep{Jolliffe1986}. 
Moreover, 
deep learning algorithms such as the convolutional neutral networks (CNN) 
and the conditional generative adversarial networks (cGAN) \citep[e.g.][]{Gillet2019, Shirasaki2019, Moriwaki2020} 
have recently been employed. 
These previous works using machine learnings and deep learnings were 
mostly performed for intensity distributions or line spectra, 
i.e. mainly for the studies of the morphology of objects or the chemical composition. 
On the other hand, 
there are still no successful attempt 
for the gas kinematics in protostellar sources 
as far as we know.

\model\ can generate one cube \fits\ file in a 10s seconds. 
It allows us to readily prepare a heap of mock data cubes, 
which can be used as the training datasets for supervised learnings. 
A possible application 
is the classification of the observed gas kinematic structure 
into 
the \ire\ motion 
and the Keplerian motion 
in the three-dimensional coordinate. 
Support vector classification (SVC) and three-dimensional convolutional neural network (3DCNN) 
are candidate supervised learning algorithms for this purpose. 
Since the actual gas motion is often not 
pure ballistic nor pure Keplerian, 
the classification may not be perfect. 
Nevertheless, 
such automatic discrimination between them, if possible, 
will be useful to deal with a large observational data as an initial classification for further detailed inspection.

We here present such an analysis with SVC 
to classify the \irm\ and the Keplerian motion. 
Hereafter, 
we employ the stochastic gradient descent (SGD) algorithm to train the classifier 
by using the library provided by SCIKIT-LEARN\footnote{\url{\scikitsgd}}. 
We apply this analysis method to the ALMA observations toward 49 Ceti and \iras\ Source A. 

\subsubsection{Analysis with Support Vector Classification: 49 Ceti Case} \label{sec:obs_fit_ml_49Ceti}
First, we apply the above analysis method to the ALMA observation toward the debris disk of 49 Ceti. 
Since the CO and [CI] emissions clearly trace the Keplerian motion in this source \citep{Higuchi_49Ceti-CI}, 
they can be used as 
a good test case to assess the usability of this analysis method. 

By using \model, 
we first prepare 9180 \ire\ models and 9180 Keplerian disk models as the mock data set, 
which are simulated by using various combination of 
the central mass 
($M$), the \ia\ ($i$), 
and the outer radius (\rout), 
as summarized in Table~\ref{tb:ML_params}. 
In addition, 
the radius of the \cb\ (\rcb) or the inner radius (\rin) is added in the combinations for an \ire\ or a Keplerian disk, respectively. 
Note that values of \rcb\ and \rin\ are employed only if they are smaller than \rout. 
The other parameters are fixed according to the characteristics of 49 Ceti 
in order to reduce the parameter space; 
for instance, 
we employ $-72\degr$ as the P.A. of its disk mid-plane \citep{Higuchi_49Ceti-CI}. 
80\%\ among the mock data are randomly picked up 
and are used as the training and validation data sets.  
The quality of the trained classifier is assessed by testing with the remaining 20\%\ mock data as the evaluation data set. 
The confusion matrix of this evaluation is shown in Table~\ref{tb:ML_cfmat-49Ceti}. 
The trained classifier marks the accuracy as high as 99.9\%, 
where only 4 mock data are wrongly classified. 
Two of the wrongly classified mock data have the completely face-on configuration ($i$ of 0\degr\ or 180\degr), 
and the other two have the smallest central mass 
($M$ of 0.50 \Msun) in the parameter range (Table~\ref{tb:ML_params}). 
Because they have small velocity shifts, the kinematic structures are not well resolved in the modeled data cubes. 
Thus, these 4 mock data are recognized as outliers among the mock data, 
which are difficult to classify. 
Therefore, the trained classifier is enough accurate to classify the two kinematic structures for ordinary cases. 

Then, we apply this classifier to the actual data cubes of the CO and [CI] emissions observed with ALMA \citep{Higuchi_49Ceti-CI}. 
The classifier successfully classifies the kinematic structure of these two molecular emissions as the Keplerian motion as expected. 
Therefore, 
this approach with a machine learning seems to work on the actual observation data.

\subsubsection{Analysis with Support Vector Classification: \iras\ Case} \label{sec:obs_fit_ml_16293}

Second, we apply this analysis method to the ALMA observation toward the Class 0 protostellar source \iras\ Source A, 
which is a binary or multiple system. 
According to \citet{Oya_16293Cyc4}, 
the \CO\ and \TFA\ emissions trace its circummultiple structure and circumstellar disk, respectively. 

We first prepared 
14400 \ire\ models and 14400 Keplerian disk models 
as the mock data set, 
whose parameter ranges are summarized in Table~\ref{tb:ML_params}. 
Values of \rcb\ and \rin\ are employed only if they are smaller than \rout. 
We employ $50\degr$ as the P.A. of the disk/envelope mid-plane of \iras\ Source A \citep{Oya_16293Cyc4}. 
Although \citet{Oya_16293Cyc4} reported that 
the gravitational centers for the circummultiple and circumstellar structures are slightly offset from each other, 
we employ the centeral position for the circummultiple structure as that for all the mock data in this study. 
Again, 80\%\ of the mock data set are used to train and tune the classifier, 
and the other 20\%\ are to evaluate it. 
The confusion matrix is shown in Table~\ref{tb:ML_cfmat-16293}. 
The classification accuracy is as high as 98.9\%, 
and thus this classifier works well. 
The classification is failed for 62 mock data out of 5760 mock data; 
they seem to be outliers among the mock data as we found in the analysis for the 49 Ceti. 
37 of the 62 mock data have the completely face-on configuration, 
and 10 of the others have the smallest central mass 
($M$ of 0.1 \Msun; Table~\ref{tb:ML_params}). 
In the other 15 mock data, 
the radius of the \cb\ (\rcb) and 
the outer {\bff radius} (\rout) are close to each other; 
for instance, 
\rcb\ of 140 au and \rout\ of 150 au. 
These mock data have only small emitting regions, 
and the classification may be difficult by the present method. 

Then, we apply it to the actual data cubes of the molecular emissions observed with ALMA, 
where the details of the observation were reported by \citet{Oya_16293Cyc4} and \citet{Oya_16293Aoutflow}. 
The results are summarized in Table~\ref{tb:ML_results-16293}. 
The \CO\ emission is successfully classified to the \ire\ as previously reported \citep{Oya_16293Cyc4}. 
The \TFA\ emission, which comes from both the circummultiple structure with the \irm\ and the circumstellar disk with the Keplerian motion, 
is also classified to the \ire. 
As well, 
all the other molecular lines except for the \ACE\ emission are classified to the \ire. 
A reason for this result would be that 
the contribution of the extended circummultiple structure is overwhelming for most molecular lines 
in comparison with the compact circumstellar disk. 
In contrast, 
the \ACE\ emission is expected to be suitable 
to investigate the circumstellar disk without the contamination from the circummultiple structure, 
and thus, its further investigation is awaited. 
Unbiased analyses with machine/deep learnings and model calculations 
are potentially useful to find out such unique cases. 
In addition, 
a survey for molecular species classified as the Keplerian disk 
is essential to study the disk chemistry in this source.

In this section, 
we have presented just a tentative demonstration of 
the analysis method using a machine learning SVC 
by using the mock data produced by \model. 
Utility of analyses introducing machine/deep learnings will rapidly be enhanced in the future 
with expanding archival observation data. 
We would like to emphasize that 
the agility and simpleness of \model\ are quite advantageous 
in preparing a huge pile of model dataset for such analyses.

\section{Discussion} \label{sec:disc}

\subsection{Why Is the Model Applicable?} \label{sec:disc_just}

While the \irm\ has long been considered \citep[e.g.][]{Ohashi1997}, 
the concept of its \cb\ was first reported in the ALMA observation by \citet{Sakai_1527nature}, 
and was formulated and modeled by \citet{Oya_15398}. 
An existence of the \cb\ is naturally expected 
from simple assumptions of the energy and the angular momentum conservation. 
In fact, its identification or hint was confirmed in observations of other young low-mass protostellar sources 
{\bff \citep[e.g.][]{Oya_15398, Oya_16293, Oya_483, Oya_16293B, Sakai_TMC1A, Lee2017_HH212, Alves2017_BHB07-11, Imai_B335rotation, Imai_FAUSTCB68}.} 
Moreover, 
disk structures seem to be formed inside the \cb s even at the earliest evolutionary stages (Classes 0-I). 
The \cb\ can be regarded as a boundary 
between the \ire\ outside it and the disk component inside it. 
In reality, 
the transition zone from an envelope to a disk would extend {\bff over} a considerable size near the \cb, 
and the envelope and the disk may intricately be contaminated with each other there. 
{\bff Numerical MHD calculation indeed show a complex structure of the transition zone in the disk forming region, 
which depends on assumed physical conditions.} 
Nevertheless, 
{\bff the \ire\ model well represents the kinematic structure of such regions at least in some sources mentioned above, where} 
the \cb\ stands for an approximate position of the transition zone. 

Generally, 
the size of a rotationally-supported disk, or a Keplerian disk, 
is thought to correspond to the `\centr' \citep{Hartmann_accretionbook}. 
It is the radius where the centrifugal force of the gas and the gravity under the central mass 
are balanced out, 
and is twice the radius of the \cb\ (Section~\ref{sec:formulae_vel}). 
Thus, the gas can stably keep rotation around the protostar at the \centr. 
The angular momentum of the gas infalling at a later time tends to be larger, 
which makes the \centr\ larger. 
This situation results in the smooth growth of the disk. 

However, 
it seems to be a rather quasi-static picture. 
In an \ire, the rotation speed equals to the infall speed at the \centr. 
Since the rotation speed is the same as the Keplerian speed there, 
the rotation motion can be continuous from the \ire\ to the Keplerian disk at the \centr. 
On the other hand, 
there remains the infall motion as far as the effect of the gas pressure and the magnetic field are not significant (see below), 
and the infalling gas tends to go further inward of the \centr. 
{\bff In this case, the} 
gas components between the \rcr\ and \rcb\ has a rotating speed higher than the Keplerian speed. 
A hint of such a `super-Kepler' like component is seen, 
for instance, 
in the numerical simulations by \citet{Machida2011_magneticbreaking} (their Figure 4) 
and \citet{Zhao2016} (their Figures 11 and 15). 
{\bff Although Figure 4 of \citet{Machida2011_magneticbreaking} reveals 
that the motion is mostly represented by the Keplerian motion, 
it also shows some cases where the rotation velocity slightly exceeds the Keplerian velocity 
in the transition zone from the envelope to the disk. 
This implies an intrinsic complexity of the transition zone.} 
According to \citet{Zhao2016}, 
a weak magnetic coupling due to removal of small dust grains is responsible for {\bff such a} feature. 

Some observational results {\bff indeed} indicate 
that the gas apparently keeps falling beyond the \centr\ toward the \cb\ \citep[e.g.][]{Sakai_1527apjl, Oya_16293}, 
and they have been reasonably reproduced by the \ire\ model. 
Furthermore, a jump of the rotation speed down to the Keplerian speed 
around the \cb\ is also suggested in the \FA\ observation 
toward BHB 07$-$11 by \citet{Alves2017_BHB07-11}. 
Thus, phenomenologically speaking, 
the \ire\ model captures an essence of the kinematic structure of the envelope and the transition zone 
{\bff at least partly.} 

{\bff Although the \ire\ model reproduces a major observed feature of the gas kinematics, 
we should note that this is a very simplified model.} 
In more realistic cases, 
the infall motion of the gas will be suppressed due to the gas pressure. 
This effect is, 
however, not always effective near the \centr. 
The static (\Pstat) and dynamic (\Pdyn) pressures of the gas are represented as follows 
\citep[e.g.][]{Chandrasekhar1953b, Hartmann_accretionbook}: 
\begin{align}
	P_{\rm stat} &= \rho c_s^2, \label{eq:Pstat} \\
	P_{\rm dyn} &= \frac{1}{2} \rho v_{\rm fall}^2 \label{eq:Pdyn}
\end{align}
where $\rho$, $c_s$, and \vfall\ denote 
the mass density of the gas, the sound speed, and the radial velocity of the gas motion, respectively. 
Thus, if $\rho$ can be assumed to be locally constant, 
the dynamic pressure is larger than the static pressure if \vfall\ is larger than $\sqrt{2} c_s$. 
Therefore, 
the gas is hardly supported by the static pressure for the case of a large \vfall, 
and keeps falling beyond the \centr\ toward the \cb\ to some extent. 

As for the L1527 case \citep[Section~\ref{sec:obs_fit_chi2_L1527}; ][]{Sakai_1527nature, Sakai_1527apjl}, 
such a condition is expected to be fulfilled until the gas reaches near the \cb, 
as shown in Figure~\ref{fig:pressure}. 
In fact, 
the static and dynamic pressures are roughly evaluated to be 
$0.5 \times 10^{-6}$ and $1.6 \times 10^{-6}$ dyn \cmsquare\ at the \centr, respectively, 
where the \vfall\ is 0.9 \kmps. 
Here, 
we employed $M$ of 0.18 \Msun\ and \rcb\ of 100 au reported by \citet{Sakai_1527nature} 
instead of the rough evaluation 
in the trial of the \chisq\ test described in Section~\ref{sec:obs_fit_chi2_L1527}. 
The gas number density is assumed to be $10^8$ \cmcubic, 
and the averaged particle mass to be $3.83 \times 10^{-24}$ g. 
$c_s$ is proportional to ${\displaystyle \sqrt{T}}$, 
where $T$ is the gas temperature. 
It is calculated to be 0.35 \kmps\ for the gas temperature of 30 K 
in the \ire\ of L1527 \citep{Sakai_1527nature}. 

On the other hand, 
the magnetic pressure is represented as: 
\begin{align}
	P_M &= \frac{B^2}{2 \mu_0}, \label{eq:PM}
\end{align}
where $B$ is the magnetic field and $\mu_0$ the magnetic permeability \citep{Chandrasekhar1953a}. 
Then, we obtain the following practical expression: 
\begin{align}
	P_M &= 3.98 \times 10^{-14} B \left( \mu G \right)^2 {\rm dyn\ cm}^{-2}. \label{eq:PM_practical}
\end{align}
Hence, 
the magnetic pressure can be comparable to the dynamic pressure, 
only when the magnetic field is of the order of 10 mG. 
In turn, 
the \ire\ model does not work under 
such a high magnetic field condition. 
The magnetic field strength in the transition zone has not been well understood observationally, 
although the effect of the magnetic breaking has often been invoked to account the infall motion 
\citep[e.g.][]{Machida2014a, Aso2015, Yen_diskALMA}. 
It should be noted that 
the ratio of the magnetic pressure to the static gas pressure is assumed to be unity or less 
in the MHD calculations \citep[e.g.][]{Machida2011_magneticbreaking}, 
otherwise the disk/envelope sturucture would be unstable \citep{Shibata_lowBeta, Machida2000_lowBeta}. 
Therefore, 
it is not clear how the magnetic pressure contributes to the force balance around the \centr. 

Since $\rho$ and $c_s$ will be increased around the \cb\ due to the stagnation of the gas 
and the relatively high temperature at the \cb\ \citep[60 K in L1527, $>$300 K in \iras;][]{Sakai_1527apjl, Oya_16293} 
in comparison with the temperature in the \ire\ (30 K in L1527, $<$200 K in \iras), 
the static pressure will be enhanced near the \cb. 
Thus, 
the gas may stop falling before it reaches at the \cb. 
This means that the edge of the rotationally supported disk will be somewhere between the \centr\ and the \cb.

{\bff 
Within the \rcb, \model\ assumes a disk component. 
This is justified by the following considerations. 
Because of the gradual increase of the \sam\ of the infall gas described above, 
the gas accreted before would have a smaller \rcb\ than 
the \rcb\ for the currently accreting gas 
(see Eq.~(\ref{eq:rcb})). 
This allows the existence of a disk component within the current \rcb. 
Moreover, the \am\ of the gas would be 
{\bff removed from} 
the stagnated gas around the \cb, 
which would help the transition of the infall gas to the rotationally-supported gas. 
This process is also potentially related to the outflow launching (see Section \ref{sec:disc_outflow}), 
which has been studied observationally by \citet{Oya_1527, Oya_483outflow, Oya_16293Aoutflow}. 
The \am\ of the accreting gas is transferred to the rotation of the outflowing gas 
\citep[e.g.][]{Blandford1982, Tomisaka2000, Anderson2003, Pudritz2007, MachidaHosokawa2013} (see Section \ref{sec:disc_outflow}). 
The mechanism of this kinematic transition occurring from \rcr/\rcb\ toward the disk 
is the remaining important issue 
for full understandings of the disk formation. 
}

\subsection{Caveats for Employing the Model} \label{sec:disc_caveats}

Although the \ire\ model has successfully been employed to investigate the observed kinematic structures, 
we expect some cases for which the \ire\ model could not be applied. 

The effects of the gas pressure and the magnetic pressure described above 
may not be negligible in some sources. 
The magnetic breaking effect could be overwhelming near the protostar, 
and the infalling gas would be stagnated before it reaches the \cb. 
As well, very young protostellar sources are not appropriate for \model. 
In such sources, 
the envelope mass is not negligible in comparison with the central mass. 
Thus, 
the self-gravity of the envelope gas needs to be considered, 
which is not taken into account in \model. 

In addition, sources with a very small or high central mass 
are not appropriate for the \ire\ model. 
With a small central mass, 
the infall velocity is so small that the dynamic pressure of the gas is small. 
This will result in the situation that the static pressure around the \centr\ is higher than the dynamic pressure. 
For instance, 
the Class 0 protostellar source \irass\ is reported to have a central mass 
as small as 0.007~\Msun\ \citep{Okoda_15398}, 
resulting in the infall velocity of 0.3~\kmps\ at its \centr\ (80 au). 
In this case, the dynamic pressure of ${\displaystyle \left( 0.2 \times 10^{-6} \right)}$~dyn~\cmsquare\ does not overwhelm 
the static pressure of ${\displaystyle \left( 0.5 \times 10^{-6} \right)}$~dyn~\cmsquare\ at the \centr, 
assuming the number density of $10^8$~\cmcubic\ and $c_s$ of 0.35~\kmps. 
Meanwhile, 
with a high central mass, 
the gas temperature around the \centr\ will be high due to the high luminosity of the protostar. 
Thus, 
the static pressure, which is proportional to $T$, 
could be higher than the dynamic pressure. 
In these situations, 
the infall gas may be supported by the static pressure and cannot fall toward the \cb. 
In addition, 
as for the more evolved sources, 
the infalling envelope gas may be exhausted or dissipated, 
and thus, the dynamic pressure may not be high enough to push the stagnated gas 
near 
the \cb\ any more. 
Therefore, 
the edge of the disk would eventually be extended to the \centr\ as expected 
\citep[e.g.][]{Hartmann_accretionbook}.

In contrast, 
the gas can never fall inward of the \cb\ unless it loses the angular momentum and the energy, 
and hence, 
it will be stagnated outside 
the \cb\ by colliding with the gas infalling afterwards. 
Such gas stagnation has been reported 
in the observation of L1527 with ALMA \citep{Sakai_1527_highres}. 
The above mechanism will cause a weak accretion shock around the \cb, 
which is indeed indicated by the emission of complex organic molecules 
and the high-excitation lines of \TFA\ in \iras\ Source A 
\citep{Oya_16293, Miura2017, Oya_16293Cyc4}. 
Thus, 
the disk formation, 
that is the transition from an \ire\ to a disk component, 
is not a straightforward process, 
but involves discontinuous physical processes. 
Conversely, 
volatile species highlighting an accretion shock 
helps the observers to find where the transition zone is 
\citep[e.g.][]{Oya_16293, Oya_483, Oya_16293B}. 
{\bff In any case, 
the structure of the transition zone is not as simple as that assumed in the models. 
Therefore, the models should be used with enough understandings of such complexity.}

\subsection{Relation to Outflow Studies} \label{sec:disc_outflow}

Outflow structures are one of the candidate mechanisms 
responsible for the angular momentum loss of the infalling gas. 
In fact, 
rotation motions of the outflowing gas have recently been reported for several sources 
\citep[e.g.][and literatures therein]{Hirota2017, Oya_483outflow, Oya_16293Aoutflow, Zhang2018, Tabone2020, Lee2021}, 
as predicted by theoretical models 
{\bff \citep[e.g.][]{Blandford1982, Tomisaka2000, Anderson2003, Pudritz2007, MachidaHosokawa2013}.} 
In the Class 0 protostellar binary \iras\ Source A, 
the \sam\ of the rotating outflow was evaluated and compared with 
those of the circummultiple structure and the circumstellar disk \citep{Oya_16293Aoutflow}, 
which were evaluated {\bff with} the aid of \model\ \citep{Oya_16293Cyc4}. 
They suggested that the outflow of this source has a larger \sam\ than the circumstellar disk, 
and that the outflow is indeed possible to extract the angular momentum from the \desys. 
\model\ allows us to quantitatively characterize the kinematic structure of the \desys, 
which can contribute to tackling with this important problem during the star-formation process.

\section{Summary and Future Prospects} \label{sec:summary}

We have developed the computer code `\model', 
which outputs the \fits\ files of the data cubes and PV diagrams 
of the \ire\ model and Keplerian disk model. 
The source codes of \model\ are open to the community via GitHub\footnote{{\url{\githubmodel}}}. 
We also distribute 
the source codes for the chi-squared test of \fits\ cubes, 
which will be helpful to compare the modeled results with \model\ and the observed data\footnote{{\url{\githubchi}}}.

We have described the basic formulae of the models (Section~\ref{sec:formulae}) 
and how to use this program (Section~\ref{sec:howto}). 
We have shown some examples of the model results (Section~\ref{sec:examples}), 
and have presented the application of this model to the actual observational data (Section~\ref{sec:obs}). 

\model\ is not only for young protostellar sources, 
but for other systems with accretion, 
because the kinematic structure employed in this model is quite basic. 
For instance, 
\citet{Aalto2020} have recently reported 
that the active galactic nucleus NGC1377 shows a non-circular motions.  
This structure may correspond 
to the \cvelcomp\ in the \ire\ described in Section~\ref{sec:examples_params_PV}.

The \ire\ model can reasonably explain the basic kinematic structure 
observed for Class 0-I protostellar sources. 
However, 
we would like to stress that 
\model\ is based on simple assumptions, 
and thus, there are several important caveats for employing it 
(see Sections~\ref{sec:formulae_caveats}, \ref{sec:obs_aim}, and~\ref{sec:disc_caveats}). 
In reality, these effects will affect 
{\it where the infall motion of the gas actually stops.} 
Recent progress in the observational study of young protostellar sources has clarified 
that the next important step to the star-formation process is 
to understand {\it what is occurring between the \centr\ and the \cb.} 
For such studies, 
the program \model\ introduced in this paper will be a powerful tool. 

\acknowledgements

{\bff The authors are grateful to an anonymous reviewer 
for useful comments and constructive suggestions.} 
The authors thank 
Dr. Yoshiyuki Kabashima and Dr. Takashi Takahashi 
for their invaluable discussion on the machine learning. 
The authors also thank Dr. Aya Higuchi 
for providing the observational data for 49 Ceti. 
This study used the ALMA data set 
ADS/JAO.ALMA\#2013.1.01102.S, 
ADS/JAO.ALMA\#2016.1.00457.S, 
ADS/JAO.ALMA\#2016.1.01203.S, 
ADS/JAO.ALMA\#2016.1.01376.S, 
and \\ ADS/JAO.ALMA\#2017.0.00467.S. 
ALMA is a partnership of the European Southern Observatory, 
the National Science Foundation (USA), 
the National Institutes of Natural Sciences (Japan), 
the National Research Council (Canada), 
and the NSC and ASIAA (Taiwan), 
in cooperation with the Republic of Chile. 
The Joint ALMA Observatory is operated by the ESO, the AUI/NRAO, and the NAOJ. 
The authors are grateful to the ALMA staff for their excellent support. 
This study is supported by a Grant-in-Aid from the Ministry of Education, 
Culture, Sports, Science, and Technologies of Japan 
(grant Nos. 18H05222, 19H05069, 19K14753, and 21K13954). 

\appendix

\section{Observation: L1527} \label{sec:app_obs}
The observations toward L1527 (ADS/JAO.ALMA\#2016.1.01203.S) 
were carrier out on November 19th, 2016 and September 3rd, 2017 with ALMA. 
The molecular lines of CS, $^{13}$CS, \FA, \cycCH, and SO were observed at the frequency from 137 to 151 GHz. 
The CS (\csl) line is used in this study. 

The field center of the observations was set to 
$\left( \alpha_{\rm ICRS}, \delta_{\rm ICRS} = 04^{\rm h} 39^{\rm m} 53.\!\!^{\rm s}9, 26^\circ 03^\prime 09.\!\!^{\prime\prime}6 \right)$ 
near the protostellar position based on our previous observations \citep{Sakai_1527nature, Sakai_1527apjl, Oya_1527}. 
43 antennas were used with baseline lengths from 15 to 704 m during the first observation, 
while 46 antennas were used with baseline lengths from 21 to 3697 m during the second observation. 
The total observation times were 57 and 122 minutes, 
where the total on-source times were 36 and 41 minutes, 
for the two observations. 
The size of the field of view was 35\arcsec. 
The largest recoverable angular scale was requested to be 4\arcsec. 
The beam size for the CS (\csl) line was 
$\left( 0\farcs459 \times 0\farcs400 \right)$ (P.A. $-20.098$\degr).  
The bandpass calibration and total flux calibration were performed with J0510$+$1800 during the two observations. 
J0438$+$3004 and J0440$+2728$ were observed for the phase calibration 
every 8 and 2 minutes in the first and second observation, respectively. 
In the second observation, 
J0435$+$2532 was also observed for the water vapor radiometer gain calibration. 
During the ALMA Cycle 4 operation, 
the absolute accuracy of the flux calibration is expected to be better than 15\%\ \citep{TechHB_Cyc4}. 

The CS (\csl) line image was obtained with the CLEAN algorithm 
with the Briggs weighting with a robustness parameter of 0.5. 
We prepared the line image by subtracting the continuum component from the visibility data, 
where the continuum component was obtained by averaging line-free channels. 
A primary beam correction was applied to the line image. 
The rms noise level of the CS image is 4 \mJypb\ with the channel width of 0.2 \kmps. 
Since we do not require a higher S/N ratio in this demonstrative study, 
we did not apply self-calibration for this observational data.

\input{part_reference.tex}
\input{part_fig.tex}

\end{document}

%% file: part_preamble.tex
\newcommand{\ifsubmit}{\iffalse}

\newcommand{\iffigure}{\iffalse}
\renewcommand{\iffigure}{\iftrue}

\newcommand{\iffigurechisq}{\iffalse}
\renewcommand{\iffigurechisq}{\iftrue}

\newcommand{\dirnamefig}{./figs/}
\newcommand{\dirnamecm}{/Users/oyayoko/observationData_working/models4paper/ire_cubefits/miriad_Flare30/}
\newcommand{\dirnamePV}{/Users/oyayoko/observationData_working/models4paper/ire_cubefits/models_Flare30/} 
\newcommand{\dirnamechi}{/Volumes/minto/modelpaper/cubechi2/bestFit/chi2result_limI/} 

\renewcommand{\iffigure}{\iftrue}
\renewcommand{\iffigurechisq}{\iftrue}
\renewcommand{\dirnamefig}{}
\renewcommand{\dirnamecm}{}
\renewcommand{\dirnamePV}{}
\renewcommand{\dirnamechi}{}


\ifsubmit
\else
\tableofcontents
\fi


\newcommand{\model}{{\tt FERIA}}
\newcommand{\modellong}{Flat Envelope model with Rotation and Infall under Angular momentum conservation}

\newcommand{\filehead}{{\tt feria.h}}
\newcommand{\fileenv}{{\tt feria\_env.cpp}}
\newcommand{\filefitsio}{{\tt feria\_fitsio.cpp}}
\newcommand{\filemain}{{\tt feria\_main.cpp}}
\newcommand{\filemesh}{{\tt feria\_mesh.cpp}}
\newcommand{\fileparameter}{{\tt feria\_parameter.cpp}}
\newcommand{\filePV}{{\tt feria\_PV.cpp}}
\newcommand{\filesky}{{\tt feria\_sky.cpp}}
\newcommand{\fileinput}{{\tt template.in}}

\newcommand{\githubmodel}{https://github.com/YokoOya/FERIA}
\newcommand{\githubchi}{https://github.com/YokoOya/cubechi2}

\newcommand{\fits}{{\tt FITS}}
\newcommand{\cfitsio}{{\tt CFITS}}

\newcommand{\scikit}{https://scikit-learn.org/stable/model\_selection.html}
\newcommand{\scikitsgd}{https://scikit-learn.org/stable/modules/generated/sklearn.linear\_model.SGDClassifier.html}

\newcommand{\cycCH}{\ce{c-C_3H_2}}	
\newcommand{\CS}{\ce{C^{34}S}}
\newcommand{\cs}{$J=5-4$}
\newcommand{\ocs}{$J=19-18$}
\newcommand{\TFA}{\ce{H_2CS}}	
\newcommand{\FA}{\ce{H_2CO}}	

\newcommand{\csl}{$J=3-2$}

\newcommand{\soelias}{$J_N=6_7-5_6$}

\newcommand{\MN}{\ce{CH_3OH}}	
\newcommand{\MND}{\ce{CH_2DOH}}	
\newcommand{\mnb}{$12_{6,7}-13_{5,8}$; E}

\newcommand{\CO}{\ce{C^{17}O}}	
\newcommand{\ACE}{\ce{(CH_3)_2CO}}	
\newcommand{\ace}{$18_{6,13}-17_{5,12}$}

\newcommand{\hydro}{\ce{H_2}}	

\newcommand{\iras}{IRAS 16293$-$2422}
\newcommand{\irass}{IRAS 15398$-$3359}

\newcommand{\ire}{infalling-rotating envelope}
\newcommand{\irm}{infalling-rotating motion}
\newcommand{\cb}{centrifugal barrier} 
\newcommand{\rcb}{$r_{\rm CB}$}
\newcommand{\rcr}{$r_{\rm CR}$}
\newcommand{\centr}{centrifugal radius}
\newcommand{\cvelcomp}{counter-velocity component}

\newcommand{\chisq}{$\chi^2$}

\newcommand{\logNpix}{lbNpix} 
\newcommand{\logNvel}{lbNvel} 

\newcommand{\rout}{$R_{\rm out}$}
\newcommand{\rin}{$R_{\rm in}$}

\newcommand{\henv}{$H_{\rm env}$}
\newcommand{\fenv}{$F_{\rm env}$}
\newcommand{\hdisk}{$H_{\rm disk}$}
\newcommand{\fdisk}{$F_{\rm disk}$}

\newcommand{\dprofenv}{$\alpha^{\rm dens}_{\rm env}$}
\newcommand{\tprofenv}{$\alpha^{\rm temp}_{\rm env}$}
\newcommand{\dprofdisk}{$\alpha^{\rm dens}_{\rm disk}$}
\newcommand{\tprofdisk}{$\alpha^{\rm temp}_{\rm disk}$}

\newcommand{\nhydro}{$n ({\rm H}_2)$}
\newcommand{\nCB}{$n_{\rm CB} ({\rm H}_2)$}
\newcommand{\nmolCB}{$n_{\rm CB} ({\rm X})$}
\newcommand{\fracd}{$F ({\rm X})$}
\newcommand{\tCB}{$T_{\rm CB}$}

\newcommand{\bmaj}{$\theta_{\rm major}$}
\newcommand{\bmin}{$\theta_{\rm minor}$}
\newcommand{\bpa}{$\theta_{\rm PA}$}

\newcommand{\vshift}{$v_{\rm shift}$}
\newcommand{\vfall}{$v_{\rm fall}$}
\newcommand{\vrot}{$v_{\rm rot}$}
\newcommand{\vkep}{$v_{\rm Kep}$}
\newcommand{\vsound}{$c_{\rm s}$}

\newcommand{\Pstat}{$P_{\rm stat}$}
\newcommand{\Pdyn}{$P_{\rm dyn}$}

\newcommand{\inv}{$^{-1}$}
\newcommand{\kmps}{km s\inv}
\newcommand{\Msun}{$M_\odot$}
\newcommand{\cmsquare}{cm$^{-2}$}
\newcommand{\cmcubic}{cm$^{-3}$}
\newcommand{\mJypb}{mJy beam\inv}

\newcommand{\desys}{disk/envelope system}
\newcommand{\los}{line of sight}
\newcommand{\incRem}{0\degr\ for a face-on configuration}
\newcommand{\am}{angular momentum}
\newcommand{\sam}{specific angular momentum}
\newcommand{\ia}{inclination angle}

\newcommand{\pos}{plane of the sky}

\newcommand{\meshP}{0\farcs05}
\newcommand{\meshV}{0.05 \kmps}
\newcommand{\beamsize}{0\farcs2}
\newcommand{\beamsizeau}{20 au}
\newcommand{\linewidthkmps}{0.2 \kmps}

\newcommand{\Dex}{100 pc}
\newcommand{\Mex}{0.3 \Msun}
\newcommand{\Iex}{90\degr}
\newcommand{\CBex}{100 au}
\newcommand{\Routex}{500 au}

\newcommand{\figchisqexcomma}{\ref{fig:chi2_L1527_example}, \ref{fig:chi2_B335_example}, \ref{fig:chi2_Elias29_example}}
\newcommand{\figchisqvalcomma}{\ref{fig:chi2_L1527_val}, \ref{fig:chi2_B335_val}, \ref{fig:chi2_Elias29_val}}
\newcommand{\figchisqexand}{\ref{fig:chi2_L1527_example}, \ref{fig:chi2_B335_example}, and \ref{fig:chi2_Elias29_example}}
\newcommand{\figchisqvaland}{\ref{fig:chi2_L1527_val}, \ref{fig:chi2_B335_val}, and \ref{fig:chi2_Elias29_val}}

\newcounter{tbnotecount}
\setcounter{tbnotecount}{0}
\newcommand{\tbnote}{\refstepcounter{tbnotecount}\alph{tbnotecount}}
\newcommand{\tbnotemark}{\tablenotemark{\tbnote}}
\newcommand{\tbnotetext}[1]{\tablenotetext{\tbnote}{#1}}
\newcommand{\steptbnote}{\refstepcounter{tbnotecount}}
\newcommand{\resettbnote}[1]{\setcounter{tbnotecount}{0}#1}

%% file: part_chisq.tex
\subsubsection{Chi-Squared Test: L1527 Case} \label{sec:obs_fit_chi2_L1527}

The first case is 
a protostellar source whose \ia\ of the \desys\ is well constrained. 
It is L1527 IRS (IRAS 04365$+$2557), the Class 0/I protostellar source in Taurus. 
{\bff The \desys\ of L1527 has} an almost edge-on configuration 
with the \ia\ of 95\degr\ \citep{Tobin2013, Oya_1527}. 
{\bff This source is already known to have the infalling-rotating envelope based on the previous observational studies. 
The \irm\ of the gas has been detected in various molecular lines: 
C$^{18}$O by \citet{Ohashi2014} and \citet{Aso2017}, 
\cycCH\ and CCH by \citet{Sakai_1527nature, Sakai_1527apjl}, 
and CS by \citet{Oya_1527}. 
Thus, this source is a good test case to confirm that \model\ works for the actual observation data.
}

We perform a \chisq\ test for the CS (\csl) emission observed with ALMA (see Appendix \ref{sec:app_obs}). 
{\bff The CS emission is suitable for this study, 
because it is known to trace the \desys\ of L1527 \citep{Oya_1527} and 
to be free from the contamination by other molecular lines. 
Moreover, our CS (\csl) data observed in ALMA Cycle 4 have a better quality 
than the CS (\cs) data in ALMA Cycle 0 reported by \citet{Sakai_1527apjl} and \citet{Oya_1527}.}
The free parameters for each case are summarized in Table \ref{tb:chi2_L1527_params}. 
The \ia\ ($i$) of 95\degr\ is employed according to the previous reports \citep{Tobin2013, Oya_1527}, 
where the western side of the \desys\ faces us \citep{Oya_1527}. 
\rout\ is fixed to be 500~au according to the distribution of the CS emission. 
The distance ($d$) of 137~pc is employed for the consistency with our previous report \citep{Oya_1527}. 
The emission in the following models is convolved with the Gaussian beam of 0\farcs459~$\times$~0\farcs400 (P.A.~$-20.098$\degr) 
and the intrinsic line width of 0.5~\kmps. 
Since there is a contamination from the outflow component in the CS emission, 
only the data points within the specified region and the velocity range are considered (Figures \ref{fig:chi2_L1527_mom}a, \ref{fig:chi2_L1527_PV}a).

{\bff 
The \chisq\ test is performed for the model of the \ire, 
according to the aforementioned previous reports. 
The free parameters for the \chisq\ test are the protostellar mass ($M$) and the radius of the \cb\ (\rcb). 
Their values for the best-fit model are summarized in Table \ref{tb:chi2_L1527_params}. 

The best-fit model is compared with the observation in Figures \ref{fig:chi2_L1527_mom}$-$\ref{fig:chi2_L1527_PV}. 
The obtained \ire\ model reasonably reproduces the observed flattened distribution in the integrated intensity map. 
As well, 
the velocity gradient along the north-south direction in the observation is reproduced. 
Figure \ref{fig:chi2_L1527_chisqplot} shows the reduced \chisq\ map for the various values for the two free parameters. 
The central mass ($M$) seems well constrained with the \ire\ model. 

It is notable that the values obtained in this analysis ($M$ of 0.15 \Msun, \rcb\ of 100 au) 
are reasonably consistent with the previous reports; 
for instance, \citet{Sakai_1527nature} assumed $M$ of 0.18 \Msun\ and \rcb\ of 100 au 
based on the morphology in one PV diagram prepared along the mid-plane of the envelope. 
Our analysis taking the PPV (cube) data into account indeed reproduces the previous result. 
}

\subsubsection{Chi-Squared Test: B335 Case} \label{sec:obs_fit_chi2_B335}

The {\bff second} case is a source with a very small rotation motion. 
B335 is a Bok globule harboring a low-mass Class 0 protostar IRAS 19347$+$0727 
\citep{Keene_B335}. 
This is an isolated source, 
and thus, its physical and kinematic structures have extensively been studied 
as a good test bed of the protostellar evolution studies 
\citep[e.g.][]{Hirano1988, Hirano1992, Chandler1993, Zhou1993, Wilner2000, Harvey2001, Yen2020, Cabedo2021}. 
\citet{Evans_B335infall} and \citet{Yen_B335nodisk} detected 
the infall motion of the gas associated with the protostar 
with ALMA. 
The small rotation motion on a 10-au scale in its \desys\ was investigated 
with ALMA {\bff at resolutions of 10 au and 3 au} 
by \citet{Imai_B335rotation} and \citet{Bjerkeli2019}, {\bff respectively.} 
{\bff \citet{Imai_B335rotation} reported that the PV diagram observed at a 10 au resolution 
is better fitted by the \irm\ than the Keplerian motion, 
and they derived the protostellar mass of $(0.02-0.06)$ \Msun. 
Meanwhile, \citet{Bjerkeli2019} just assumed the Keplerian motion 
and derived the protostellar mass to be 0.05 \Msun\ by using the PV diagram of their data. 
Hence, the origin of the rotation motion in this source is still controversial.} 
A nearly edge-on geometry is preferable for B335 based on its outflow geometry 
\citep{Hirano1988, Yen2010, Imai_B335HC, Bjerkeli2019}.

We perform a \chisq\ test for {\bff the \ire\ model and the Keplerian disk model on the cube data of} the \MN\ (\mnb) emission reported by \citet{Imai_B335rotation}. 
They reported that this molecular line traces the marginal rotation motion in B335. 
The free parameters for each case are summarized in Table \ref{tb:chi2_B335_params}. 
\rout\ is fixed to 10~au according to the distribution of the \MN\ emission. 
The distance ($d$) to B335 
was reported to be 100~pc by \citet{Olofsson2009}, 
and was later updated to be 164.5~pc by \citet{Watson2020}. 
We employ the former for the consistency with the previous work by \citet{Imai_B335rotation}. 
The emission in the following models is convolved with the Gaussian beam of 0\farcs116~$\times$~0\farcs082 (P.A.~$-56.692$\degr) 
and the intrinsic line width of 1.0~\kmps. 

The best-fit models are compared with the observation in Figures \ref{fig:chi2_B335_mom}$-$\ref{fig:chi2_B335_PV}. 
The observed \MN\ (\mnb) emission has a compact and almost circular distribution, 
which is reproduced by the two models (Figures \ref{fig:chi2_B335_mom}a$-$c). 
\citet{Imai_B335rotation} detected the velocity gradient along {\bff the northwest-southeast direction (Figure \ref{fig:chi2_B335_mom}d),} 
which is tilted from the mid-plane of the \desys\ along the north-south direction, 
and attributed it to the infall motion of the gas. 
It is indeed reproduced by the \ire\ model {\bff (Figure \ref{fig:chi2_B335_mom}e).} 
The channel maps of the models seem to reasonably reproduce the observation 
in Figures \ref{fig:chi2_B335_channelmap-IRE} and \ref{fig:chi2_B335_channelmap-Kep}. 
Figure \ref{fig:chi2_B335_PV} shows the PV diagrams along the mid-plane of the \desys; 
a velocity gradient due to the small rotation motion is fortunately detected in the observation, 
and it is reproduced by either model. 

Figure \ref{fig:chi2_B335_chisqplot} shows the reduced \chisq\ maps for various parameter values. 
The protostellar mass is reasonably constrained by either model. 
\rcb\ and \rin\ are also constrained, 
while the \ia\ is not. 
Although the 
rotation motion of the gas in B335 is marginal in the observation, 
the analysis with the models is found to be useful 
{\bff to constrain the physical parametes.} 

The best-fit results for the \ire\ model and the Keplerian disk model are summarized in Table \ref{tb:chi2_B335_params}. 
The lowest reduced \chisq\ values for these two case are comparable; 
the kinematic structure {\bff observed in the \MN\ (\mnb) emission} can be approximated by either kinematic structure. 
The small rotation motion in B335 is indeed reflected in the parameters of the models: 
the small central mass ($M$) and the small radius of the \cb\ (\rcb). 

{\bff 
The maximum-likelihood value of $M$ for the best-fit models are different 
between the \ire\ model and the Keplerian model (Table \ref{tb:chi2_B335_params}); 
smaller $M$ is obtained for the \ire\ model. 
This is a natural consequence of the different formulation of the models, 
where the mass estimated from the rotation velocity at the \cb\ by assuming the \irm\ is 
the half of that estimated by assuming the Keplerian motion (see Eqs. (\ref{eq:vrot}) and (\ref{eq:kep})).
}

\subsubsection{Chi-Squared Test: Elias 29 Case} \label{sec:obs_fit_chi2_Elias29}

The {\bff third} case is a source with an ordinary \ia. 
Elias 29 is the Class I protostellar source in Ophiuchus. 
In this source, 
the rotation motion of the gas associated with its protostar 
was detected in the SO (\soelias) {\bff emission} by \citet{Oya_Elias29}. 
They reported that 
the observed kinematic structure is reasonably explained by the Keplerian disk model, 
where 
$M$ is 1.0~\Msun, 
$i$ is 65\degr, 
and \rout\ is 100~au. 
In their report, the mid-plane of the \desys\ is assumed to lie along the north-south direction (P.A. 0\degr). 
It was reported that a fully face-on configuration ($i$ = 0\degr) is unlikely for Elias 29 \citep{Lommen2008}, 
and the \ia\ was constrained to be from 65\degr\ to 115\degr\ \citep{Oya_Elias29}. 

{\bff Elias 29 is a relatively evolved Class I source 
judging from its high bolometric temperature \citep{Miotello2014}. 
Thus, it is naturally expected to have a Keplerian disk, as assumed by \citet{Oya_Elias29}. 
However, is this really the case? 
In this study, 
we examine a possibility of an infall motion in this source as well as that of a Keplerian motion with the aid of \model. 
} 

We perform a \chisq\ test for the SO (\soelias) data reported by \citet{Oya_Elias29}. 
The free parameters for the \ire\ model and the Keplerian disk model 
are summarized in Table \ref{tb:chi2_Elias29_params}. 
\rout\ is fixed to be 50~au according to the distribution of the SO emission. 
The distance ($d$) to Elias 29 
was reported to be $(137-147)$~pc by \citet{Ortiz-Leon_Ophiuchus} 
and $141^{+30}_{-21}$~pc by \citet{Dzib2018}. 
We employ $d$ of 137~pc in this study for the consistency with our previous work \citep{Oya_Elias29}.  
We assume that 
the mid-plane of the \desys\ of Elias 29 extends 
along the northeast-southwest direction with the position angle of 30\degr, 
judging from the outflow directions 
\citep{Ceccarelli2002, Ybarra2006, Bussmann2007}. 
The emission in the following models is convolved with the Gaussian beam of 0\farcs832~$\times$~0\farcs488 (P.A.~$-85.833$\degr) 
and the intrinsic line width of 1.0~\kmps. 

The best-fit results for the \ire\ model and the Keplerian disk model are summarized in Table \ref{tb:chi2_Elias29_params}. 
Again, $M$ is obtained to be smaller for the \ire\ model (0.6~\Msun) than the Keplerian disk model (1.4~\Msun). 
This difference can mainly be interpreted as the expected difference by a factor of 2 following Eqs. (\ref{eq:vrot}) and (\ref{eq:kep}), 
as described in {\bff Section \ref{sec:obs_fit_chi2_B335},} 
although the different \ia s between both models also contribute {\bff to} the difference. 

The best-fit models are compared with the observation in Figures \ref{fig:chi2_Elias29_mom}$-$\ref{fig:chi2_Elias29_PV}. 
The observed SO (\soelias) emission has an elliptic distribution tracing the \desys, 
which is reproduced by both models (Figures \ref{fig:chi2_Elias29_mom}a$-$c). 
As well, the velocity gradient along the north-south direction in the observation is reproduced by the models (Figures \ref{fig:chi2_Elias29_mom}e$-$g). 
A skewed velocity feature with respect to the mid-plane direction (the red line in Figure \ref{fig:chi2_Elias29_mom}a; P.A. 30\degr) 
is marginally seen in the observation {\bff (Figure \ref{fig:chi2_Elias29_mom}e).} 
This trend seems a bit overemphasized in the \ire\ model (Figure \ref{fig:chi2_Elias29_mom}f). 
In contrast, the Keplerian model shows a velocity gradient rather close to the mid-plane direction. 
The feature seen in the northwestern {\bff and southeastern parts of the Keplerian model} would not be robust; 
it is interpreted as the effect of the large flat beam, 
because it does not appear with a small circular beam (Figure \ref{fig:result_kepler}). 
The channel maps are shown in Figures \ref{fig:chi2_Elias29_channelmap-IRE} and \ref{fig:chi2_Elias29_channelmap-Kep}, 
and the PV diagrams along the mid-plane of the \desys\ are in Figure \ref{fig:chi2_Elias29_PV}. 
{\bff A small difference of the reduced \chisq\ values (Table \ref{tb:chi2_Elias29_params}) 
suggests that the observed kinematic structure is explained by either model.}

Figure \ref{fig:chi2_Elias29_chisqplot} shows the reduced \chisq\ maps for various parameter values. 
The central mass ($M$) seems well constrained in both the two cases. 
The radius of the \cb\ (\rcb) or the inner radius of the disk (\rin) are 
estimated to be much smaller than the beam size (0\farcs8 $\times$ 0\farcs5 $\sim$ 100~au). 
Meanwhile, 
the \ia\ ($i$) is poorly constrained in this analysis. 
Further discussion for the \desys\ of Elias 29 would become {\bff possible}  
when $i$ is constrained by other approaches, 
for instance the analysis of the outflow structure. 

{\bff 
It should be noted that 
the lowest reduced \chisq\ value is smaller for the \ire\ model than the Keplerian model by 10\%. 
This slight difference may not have a statistical meaning due to the overwhelming contribution from the simple assumptions in the models.
Nevertheless, 
an infalling motion may contribute to the kinematic structure of Elias 29 in addition to a pure Keplerian motion previously assumed by \citet{Oya_Elias29}. 
}

{\bff Based on the above results,} 
one may think that 
the actual situation would be the hybrid case {\bff of the \irm\ and Keplerian motion.} 
Since \model\ can simulate such a hybrid case by the combined model, 
we here present a preliminary attempt to apply it for the observed data to see how it works. 
The results will be useful for the user of \model\ to apply the combined model for more complicated systems. 

{\bff As a demonstration, 
the combined model is} compared with 
{\bff the observation for Elias 29.} 
{\bff In this model,} 
the gas within the radius of the \cb\ (\rcb) is assumed to be the Keplerian disk. 
The inner radius (\rin) is fixed to be 1~au to reduce the parameter space. 
{\bff The result of the reduced \chisq\ test is summarized in Table \ref{tb:chi2_Elias29_params}. 
The obtained \rcb\ of 10 au suggests that 
there are both the contribution from the \ire\ and the Keplerian disk. 
The best-fit model is compared with the observational result 
in Figures \ref{fig:chi2_Elias29_mom}(d, h), \ref{fig:chi2_Elias29_PV}(d), and \ref{fig:chi2_Elias29_channelmap-combined}. 
}

{\bff 
As summarized in Table \ref{tb:chi2_Elias29_params}, 
the fitting is slightly improved with the combined model 
compared with the results for the \ire\ model and the Keplerian model. 
However, 
the improvement of \chisq\ value is too small 
to conclude that the obtained values with the combined model reflect the actual physical conditions. 
Our result indicates that the disk-forming region of this source 
needs to be further characterized 
with a higher angular-resolution observations in the future.  
}

\subsubsection{Chi-Squared Test: Discussion} \label{sec:obs_fit_chi3_disc}

In previous works, 
the observed kinematic structures have typically been analyzed in 2-dimensional manners; 
for instance, PV diagrams or velocity profiles, 
as described in the above sections. 
Such analyses often effectively focus on the characteristic features in the observations, 
and it have helped us to constrain the physical parameters, 
such as the protostellar mass. 
However, these analyses arbitrarily prune much information of the 3-dimensional cube data for simplicity. 
Such dimensional reduction may provide a biased view on the kinematic structure. 
Thus, using all the 3-dimensional information is preferred to characterize the observed kinematic structure. 
{\bff \model\ makes such comparison of the 3-dimensional data easier.} 

In the above sections, 
the \chisq\ test is applied for the 3-dimensional cube data for the three sources. 
We find that it is difficult to discriminate the \irm\ and the Keplerian rotation for {\bff B335 and Elias 29.} 
For these cases, 
the rotation structures are not well resolved, 
and higher spatial resolution observations may work for definitive discrimination. 
{\bff The results for B335 and Elias 29 also imply} 
that we should not too much rely on the \chisq\ test of a single molecular line for discrimination. 
Limited quality {\bff and limited resolutions} of the data {\bff as well as} contamination of the outflow components may affect {\bff the} result. 

Since some molecular emissions selectively trace a particular part of the disk/envelope structure 
just as molecular markers \citep[][]{Sakai_1527apjl, Oya_16293, Oya_483, Oya_unifiedpicture, Oya_16293Cyc4}, 
selection of molecular lines in advance is very important. 
Model analyses would get more successful by applying to molecular lines 
whose kinematic structures are classified into the \irm\ or the Keplerian rotation in preprocess. 
Based on multiple-molecular-line analyses, 
comprehensive consideration should be made for full understanding of the disk/envelope structure. 
With this in mind, 
we present a possible approach to classification of molecular lines in the next section, 
as another application of \model.

%% file: part_fig.tex
\clearpage


\begin{ThreePartTable}
\begin{TableNotes}
\item[a] See Section \ref{sec:howto_inp_head}. 
\item[b] See Section \ref{sec:howto_inp_grid}. 
\item[c] Specified in the header file ({\tt \filehead}). The maximum value is 7 for computers with the memory size (RAM) of 8 GB or larger, and 8 for those with RAM of 64 GB or larger.
\item[d] See Sections {\bff \ref{sec:formulae_vel}, \ref{sec:formulae_int},} \ref{sec:howto_inp_cube}, and \ref{sec:howto_inp_conv}. 
\item[e] See Section \ref{sec:howto_inp_PV}. 
\end{TableNotes}

\begin{longtable}{lll}
\caption{Physical Parameters of the Model
		\label{tb:params}} \\ 
\hline \hline 
Physical Parameter & Type & Remarks \\ \hline 
\endfirsthead
\hline 
Physical Parameter & Type & Remarks \\ \hline 
\endhead
\hline \endfoot
\insertTableNotes \\ 
\endlastfoot
\multicolumn{3}{c}{Parameters for the Output File$^{\rm a}$} \\ \hline 
Name of the output file & string & Arbitrary string with length less than 256 \\ 
& & or 'parameter' for naming after the input parameter values \\ 
Overwrite & char (y/n) & If a file with the same output file name exists, \\ 
& & it is overwritten with `y' and not with `n'. \\ 
\hline \multicolumn{3}{c}{Parameters for the Source and Molecular Line$^{\rm a}$} \\ \hline 
Object & string & Name of the object source \\ 
Coordinate & string & (e.g. ICRS, J2000) \\ 
Field center (RA) & string & Right Ascension of the field center (e.g. 0h0m0.0s) \\ 
Field center (Dec) & string & Declination of the field center (e.g. 0d0m0.0s) \\ 
Systemic velocity & double & Systemic velocity of the object in \kmps \\ 
Line & string & Name of the line and the transition (e.g. CO2-1) \\ 
Rest frequency & double & Rest frequency of the molecular line (GHz) \\ 
\hline \multicolumn{3}{c}{Parameters for the Mesh Size$^{\rm b}$} \\ \hline 
Pixel size & double & Mesh size for the position axes (arcsecond) \\ 
Velocity resolution & double & Mesh size for the velocity axis (\kmps) \\ 
\logNpix$^{\rm c}$
& integer & Number of mesh of the position axes is $2^{\rm \logNpix}$. \\ 
\logNvel$^{\rm c}$ & integer & Number of mesh of the position axes is $2^{\rm \logNvel}$. \\ 
\hline \multicolumn{3}{c}{Parameters for the Envelope/Disk$^{\rm d}$} \\ \hline 
$d$ & double & Distance to the object from the Sun (pc) \\ 
$M$ & double & Central mass (\Msun) \\ 
\rcb & double & Radius of the \cb\ (au) \\ 
$i$ & double & Inclination angle of the envelope (degree)\\ 
& & (\incRem) \\ 
Position angle & double & Position angle of the line \\ 
& & along which the mid-plane of the envelope is extended \\ 
Direction of rotation & int (1/$-1$) & 1 for counterclockwise with the \ia\ of 0\degr \\ 
\rout & double & Outer radius of the envelope (au) \\ 
\rin & double & Inner radius of the envelope (au) \\ 
\henv, \hdisk & double & Thickness of the envelope/disk (au) \\ 
\fenv, \fdisk & double & Flared angle of the scale height of the envelope/disk (au) \\ 
\dprofenv, \dprofdisk & double & Power-law index of the radial density profile \\ 
\tprofenv, \tprofdisk & double & Power-law index of the radial temperature profile \\ 
\nmolCB & double & Molecular density at the \cb\ (\cmcubic) \\ 
\tCB & double & Gas temperature at the \cb\ (K) \\ 
Line width & double & FWHM of the intrinsic Gaussian profile (\kmps) \\ 
\bmaj & double & FWHM of the major axis of the Gaussian beam (arcsecond) \\ 
\bmin & double & FWHM of the minor axis of the Gaussian beam (arcsecond) \\ 
\bpa & double & Position angle of the major axis of the Gaussian beam (degree) \\ 
Normalize & char (y/n) & Calculated intensity is normalized by the maximum value. \\ 
\hline \multicolumn{3}{c}{Parameters for the Position-Velocity Diagram$^{\rm e}$} \\ \hline 
PA & double & Position angle of the position axis of the PV diagram (degree) \\ 
Offset$_{\rm RA}$, Offset$_{\rm Dec}$ & double & Offset of the central position of the PV diagram \\ 
& & to the field center (au) \\ 
\hline 
\end{longtable}
\end{ThreePartTable}

\begin{table}
	\begin{center}
	\begin{threeparttable}
	\caption{Free Parameters and their Best-Fit Values in the \chisq\ Test for L1527 
			\label{tb:chi2_L1527_params}}
	\begin{tabular}{lcccc}
	\hline \hline 
	Model & Parameters & Ranges$^{\rm a}$ & Best & Reduced \chisq\ value \\ 
	\hline 
	Infalling-rotating envelope model 
	& $M$ (\Msun) & $0.05-1.00$ & 0.15 & 2.5 \\ 
	& \rcb\ (au) & $1-400$ & 100 \\ 
	& $i$ (\degr) & $95$ & Fixed \\ 
	& \rin\ (au) & \rcb & Fixed \\ 
	& \rout\ (au) & 500 & Fixed \\ 
	\hline 
	\end{tabular}
	\begin{tablenotes}
	\item[a] Parameter ranges surveyed in the \chisq\ analysis. 
	\end{tablenotes}
	\end{threeparttable}
	\end{center}
\end{table}

\begin{table}
	\begin{center}
	\begin{threeparttable}
	\caption{Free Parameters and their Best-Fit Values in the \chisq\ Test for B335 
			\label{tb:chi2_B335_params}}
	\begin{tabular}{lcccc}
	\hline \hline 
	Model & Parameters & Ranges$^{\rm a}$ & Best & Reduced \chisq\ value \\ 
	\hline 
	Infalling-rotating envelope model 
	& $M$ (\Msun) & $0.005-0.50$ & 0.02 & 0.77 \\ 
	& \rcb\ (au) & $1-7$ & 1 \\ 
	& $i$ (\degr) & $70-90$ & 70 \\ 
	& \rin\ (au) & \rcb & Fixed \\ 
	& \rout\ (au) & 10 & Fixed \\ 
	\hline 
	Keplerian disk model 
	& $M$ (\Msun) & $0.005-0.50$ & 0.04 & 0.78 \\ 
	& \rcb\ (au) & \rout & Fixed \\ 
	& $i$ (\degr) & $70-90$ & 75 \\ 
	& \rin\ (au) & $1-7$ & 1 \\ 
	& \rout\ (au) & 10 & Fixed \\ 
	\hline 
	\end{tabular}
	\begin{tablenotes}
	\item[a] Parameter ranges surveyed in the \chisq\ analysis. 
	\end{tablenotes}
	\end{threeparttable}
	\end{center}
\end{table}

\begin{table}
	\begin{center}
	\begin{threeparttable}
	\caption{Free Parameters and their Best-Fit Values in the \chisq\ Test for Elias 29
			\label{tb:chi2_Elias29_params}}
	\begin{tabular}{lcccc}
	\hline \hline 
	Model & Parameters & Ranges$^{\rm a}$ & Best & Reduced \chisq\ value \\ 
	\hline 
	Infalling-rotating envelope model 
	& $M$ (\Msun) & $0.2-10.0$ & 0.6 & 9.8 \\ 
	& \rcb\ (au) & $1-40$ & 5 \\ 
	& $i$ (\degr) & $60-120$ & 120 \\ 
	& \rin\ (au) & \rcb & Fixed \\ 
	& \rout\ (au) & 50 & Fixed \\ 
	\hline 
	Keplerian disk model 
	& $M$ (\Msun) & $0.2-10.0$ & 1.4 & 10.8 \\ 
	& \rcb\ (au) & \rout & Fixed \\ 
	& $i$ (\degr) & $60-120$ & 80 \\ 
	& \rin\ (au) & $1-40$ & 1 \\ 
	& \rout\ (au) & 50 & Fixed \\ 
	\hline 
	Combined Model 
	& $M$ (\Msun) & $0.2-10.0$ & 0.8 & 9.4 \\ 
	& \rcb\ (au) & $1-40$ & 10 \\ 
	& $i$ (\degr) & $60-120$ & 120 \\ 
	& \rin\ (au) & $1$ & Fixed \\ 
	& \rout\ (au) & 50 & Fixed \\
	\hline  
	\end{tabular}
	\begin{tablenotes}
	\item[a] Parameter ranges surveyed in the \chisq\ analysis. 
	\end{tablenotes}
	\end{threeparttable}
	\end{center}
\end{table}

\begin{table}
	\begin{center}
	\begin{threeparttable}
	\caption{Parameter Ranges for the Mock Data Used in the SVC Analyses for 49 Ceti and \iras\ Source A 
			\label{tb:ML_params}}
	\begin{tabular}{lclc}
	\hline \hline 
	Parameter & Unit & Values & Number of cases \\ \hline 
	\multicolumn{4}{c}{49 Ceti} \\ \hline 
	Protostellar mass ($M$) & \Msun & 0.5, 0.6, ..., 1.9 & 15 \\ 
	Inclination angle ($i$)& \degr & 0, 30, ..., 330 & 12 \\ \vspace*{-5pt}
	Radius of the \cb\ (\rcb) & au & 1, 20, ..., 200 & 11 \\ 
	or inner radius (\rin)$^{\rm a}$ & & & \\
	Outer radius (\rout)$^{\rm b}$ & au & 120, 140, ..., 220 & 6 \\ \hline 
	\multicolumn{4}{c}{\iras\ Source A} \\ \hline 
	Protostellar mass ($M$) & \Msun & 0.1, 0.2, ..., 1.9, & 25 \\ 
	& & 2.0, 2.2, ..., 3.0 & \\
	Inclination angle ($i$)& \degr & 0, 30, ..., 330 & 12 \\ \vspace*{-5pt}
	Radius of the \cb\ (\rcb) & au & 1, 20, ..., 200 & 11 \\ 
	or inner radius (\rin)$^{\rm a}$ & & & \\ 
	Outer radius (\rout)$^{\rm b}$ & au & 50, 100, ..., 300 & 6 \\ \hline 
	\end{tabular}
	\begin{tablenotes}
	\item[a] A model of an \ire\ has \rcb\ as a parameter, while that of a Keplerian disk has \rin. 
			To make the cube data of Keplerian disks with \model, 
			\rcb\ is set to be larger than \rout. 
	\item[b] Models are prepared only for \rout\ larger than \rcb\ or \rin.  
	\end{tablenotes}
	\end{threeparttable}
	\end{center}
\end{table}

\clearpage

\begin{table}
	\begin{center}
	\begin{threeparttable}
	\caption{Confusion Matrix of the Classifier Trained in the SVC Analysis for 49 Ceti
			\label{tb:ML_cfmat-49Ceti}}
	\begin{tabular}{cccc}
	\hline \hline 
	 & & \multicolumn{2}{c}{Predicted Kinematic Structure} \\ 
	 & & IRE$^{\rm a}$ & Keplerian disk \\ \hline 
	 \multirow{2}{*}{True Kinematic Structure} & IRE$^{\rm a}$ & 1834 & 2$^{\rm b}$ \\
	 & Keplerian disk & 2$^{\rm c}$ & 1834 \\ 
	 \hline 
	\end{tabular}
	\begin{tablenotes}
	\item[a] Infalling-rotating envelope.
	\item[b] Both wrongly classified IRE models have the completely face-on configuration ($i$ of 0\degr\ or 180\degr). 
	\item[c] Both wrongly classified Keplerian-disk models have the smallest protostellar mass ($M$ of 0.05 \Msun). 
	\end{tablenotes}
	\end{threeparttable}
	\end{center}
\end{table}

\begin{table}
	\begin{center}
	\begin{threeparttable}
	\caption{Confusion Matrix of the Classifier Trained in the SVC Analysis for \iras\ Source A
			\label{tb:ML_cfmat-16293}}
	\begin{tabular}{cccc}
	\hline \hline 
	 & & \multicolumn{2}{c}{Predicted Kinematic Structure} \\ 
	 & & IRE$^{\rm a}$ & Keplerian disk \\ \hline 
	 \multirow{2}{*}{True Kinematic Structure} & IRE$^{\rm a}$ & 2821 & 59$^{\rm b}$ \\
	 & Keplerian disk & 3$^{\rm c}$ & 2877 \\ 
	 \hline 
	\end{tabular}
	\begin{tablenotes}
	\item[a] Infalling-rotating envelope.
	\item[b] 37 of 59 wrongly classified IRE models have the completely face-on configuration ($i$ of 0\degr\ or 180\degr), 
			and 7 models have the smallest protostellar mass ($M$ of 0.05 \Msun). 
			The other 15 models have small emitting regions; 
			\rcb\ of 140 au and \rout\ of 150 au, or \rcb\ of 180 au and \rout\ of 200 au. 
	\item[c] All the wrongly classified Keplerian-disk models have the smallest protostellar mass ($M$ of 0.05 \Msun). 
	\end{tablenotes}
	\end{threeparttable}
	\end{center}
\end{table}

\begin{table}
	\begin{center}
	\begin{threeparttable}
	\caption{Results of the Classification of the Kinematic Structures Traced by Molecular Emissions in \iras\ Source A
			\label{tb:ML_results-16293}}
	\begin{tabular}{lccc}
	\hline \hline 
	Molecule & Transition & Prediction by the Classifier$^{\rm a}$ & Previous Report$^{\rm a}$  \\ \hline 
	\CO & $J=2-1$ 							& IRE & IRE$^{\rm b}$ \\ 
	\CS & $J=2-1$ 							& IRE & IRE$^{\rm c}$ \\ 
	\TFA & $7_{0,7}-6_{0,6}$ 					& IRE & IRE \& Keplerian disk$^{\rm b}$ \\ 
	\TFA & $7_{2,5}-6_{2,4}$ 					& IRE & IRE \& Keplerian disk$^{\rm b}$ \\ 
	\TFA & $7_{3,5}-6_{3,4}$ 					& IRE & IRE \& Keplerian disk$^{\rm b}$ \\ 
	\TFA & $7_{4,4}-6_{4,3}$, $7_{4,3}-6_{4,2}$ 	& IRE & IRE \& Keplerian disk$^{\rm b}$ \\ 
	OCS & $J=7-6$ 						& IRE & Disk/envelope and outflow$^{\rm c}$ \\ 
	OCS & $J=8-7$ 						& IRE & Disk/envelope and outflow$^{\rm c}$ \\ 
	SO & $J_N = 2_2-1_1$ 					& IRE & Disk/envelope and outflow$^{\rm c}$ \\ 
	\MN & $5_{-1,5}-4_{0,4}$ 					& IRE & (IRE$^{\rm d}$) \\ 
	\ACE & \ace 							& Keplerian disk & - \\ 
	\hline 
	\end{tabular}
	\begin{tablenotes}
	\item[a] `IRE' denotes an infalling-rotating envelope. 
	\item[b] \citet{Oya_16293Cyc4}. 
	\item[c] \citet{Oya_16293Aoutflow}. 
	\item[d] \citet{Oya_16293} reported that the \MN\ ($11_{0,11}-10_{1,10}$; A$^{++}$) emission 
			comes from a ring-like structure around the \cb\ at the innermost part of the \ire. 
	\end{tablenotes}
	\end{threeparttable}
	\end{center}
\end{table}


\begin{landscape}
\begin{figure}
	\iffigure
	\centering \includegraphics[bb = 0 0 1500 660, scale = 0.55]{\dirnamefig 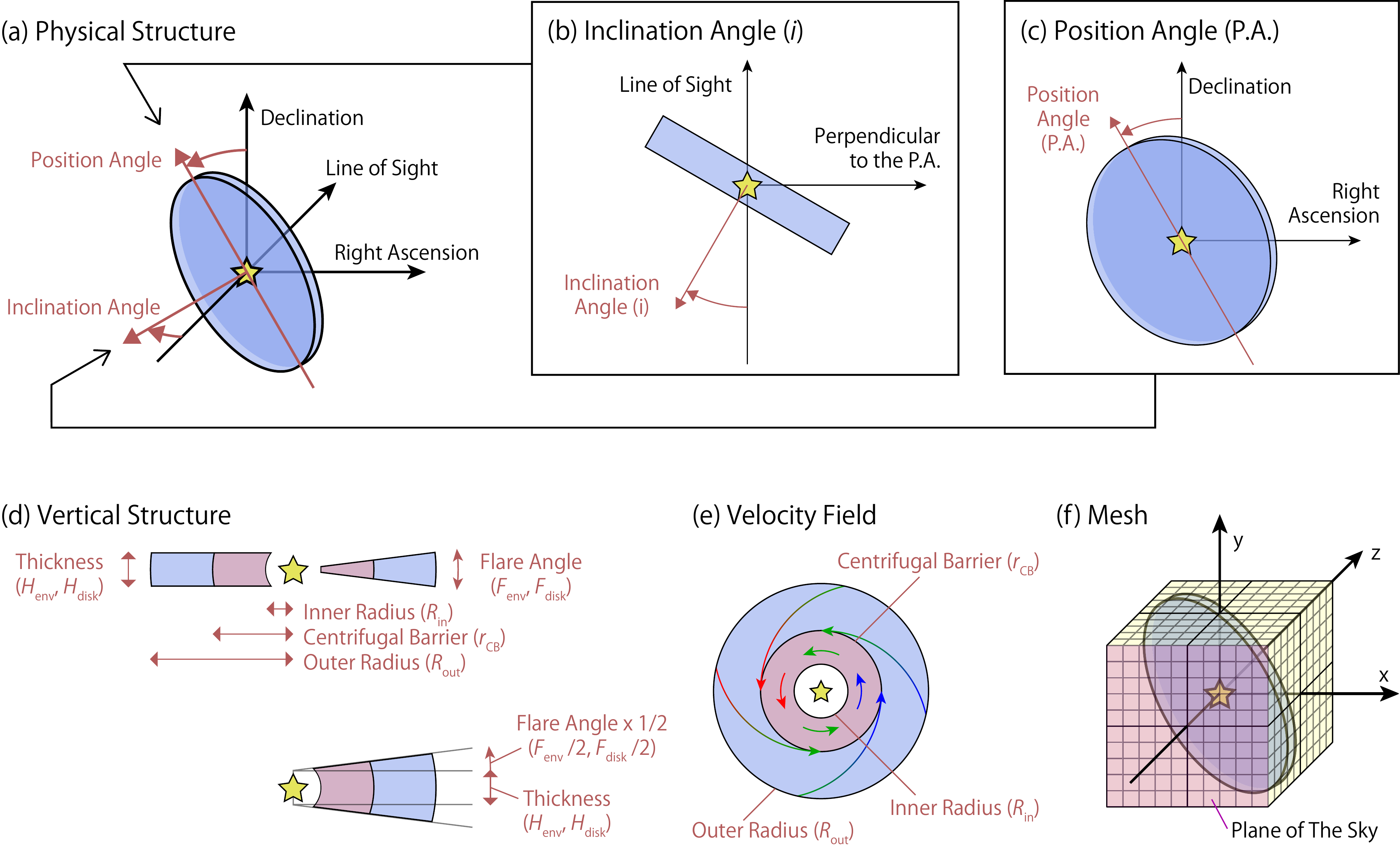} 
	\fi
\end{figure}
\begin{figure}
	\caption{Schematic illustrations of the \ire\ and Keplerian disk models. 
			(a, b, c) The kinematic structure of the model is first calculated in the three-dimensional coordinates. 
			Its directions is determined based on the input values of the \ia\ and the position angle (P.A.). 
			The \ia\ is the angle between the \los\ and the line perpendicular to the mid-plane of the model. 
			An \ia\ of 0\degr\ stands for a face-on configuration, 
			while that of 90\degr\ for an edge-on configuration. 
			With the edge-on configuration, 
			the P.A. is the angle between the vertical line on the plane of sky (i.e., declination) 
			and the line along which the mid-plane of the model extends. 
			{\bff (d)} The vertical structure of the model. 
			Section \ref{sec:howto_inp_cube} describes the details. 
			{\bff (e)} Infall and rotation motion of the gas in the model with the rotation direction of `1' (See Table \ref{tb:params}). 
			At the distance of $r$ from the central protostar, 
			the gas motion is assumed to be 
			the \irm\ for $r$ between the outer radius (\rout) and the radius of the \cb\ (\rcb) 
			and the Keplrian rotation for $r$ between the radius of the \cb\ (\rcb) and the inner radius (\rin). 
			{\bff (f) The three dimensional structure of the model is divided into small elements of gas.} 
			\label{fig:scheme}} 
\end{figure}
\end{landscape}

\begin{figure}
	\iffigure
	\centering \includegraphics[bb = 0 0 800 500, scale = 0.6]{\dirnamefig 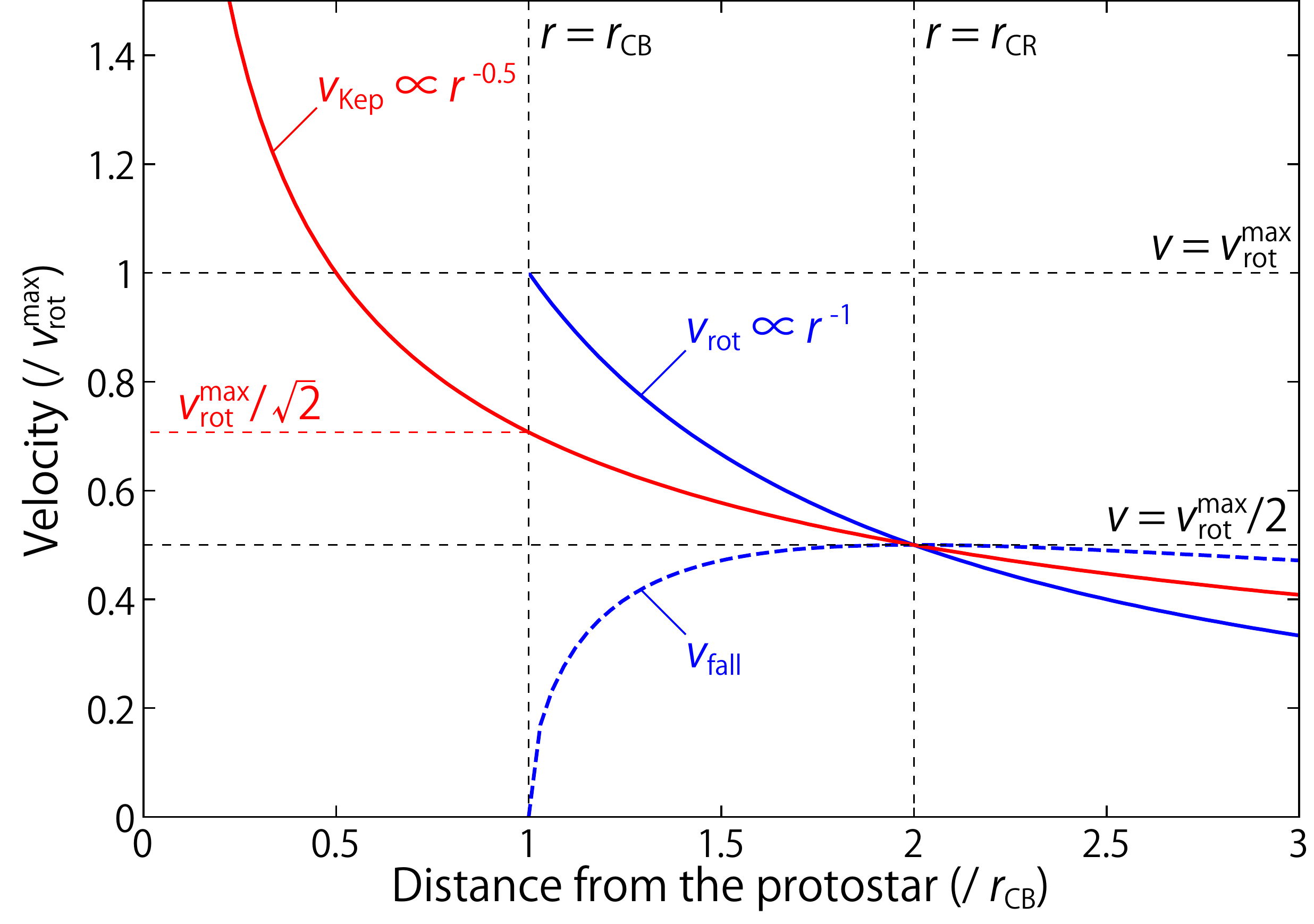}
	\fi
	\caption{Velocity of the gas with the \irm\ and the Kepelrian rotation as a function of the distance from the protostar. 
			The horizontal axis represents the distance from the protostar ($r$) normalized by the radius of the \cb\ (\rcb). 
			The vertical axis represents the velocity ($v$) normalized by the rotation velocity of the \irm\ at the \cb\ ($v^{\rm max}_{\rm rot}$). 
			The blue solid and dashed lines represent the rotation (\vrot) and infall (\vfall) velocities in the \irm, respectively. 
			At the \cb\ ($r =$ \rcb), \vfall\ equals 0 and \vrot\ takes its maximum value. 
			At the \centr\ ($r = 2$\rcb), \vfall\ takes its maximum value. 
			The red solid line represents the Keplerian rotation velocity (\vkep). 
			All of \vrot, \vfall, and \vkep\ take the same value 
			($v = v^{\rm max}_{\rm fall} = v^{\rm max}_{\rm rot} / 2$) 
			at the \centr\ (\rcr). 
			\label{fig:velocityprofile}} 
\end{figure}

\begin{figure}
	\iffigure
	\centering \includegraphics[bb = 0 0 600 200, scale = 0.8]{\dirnamefig 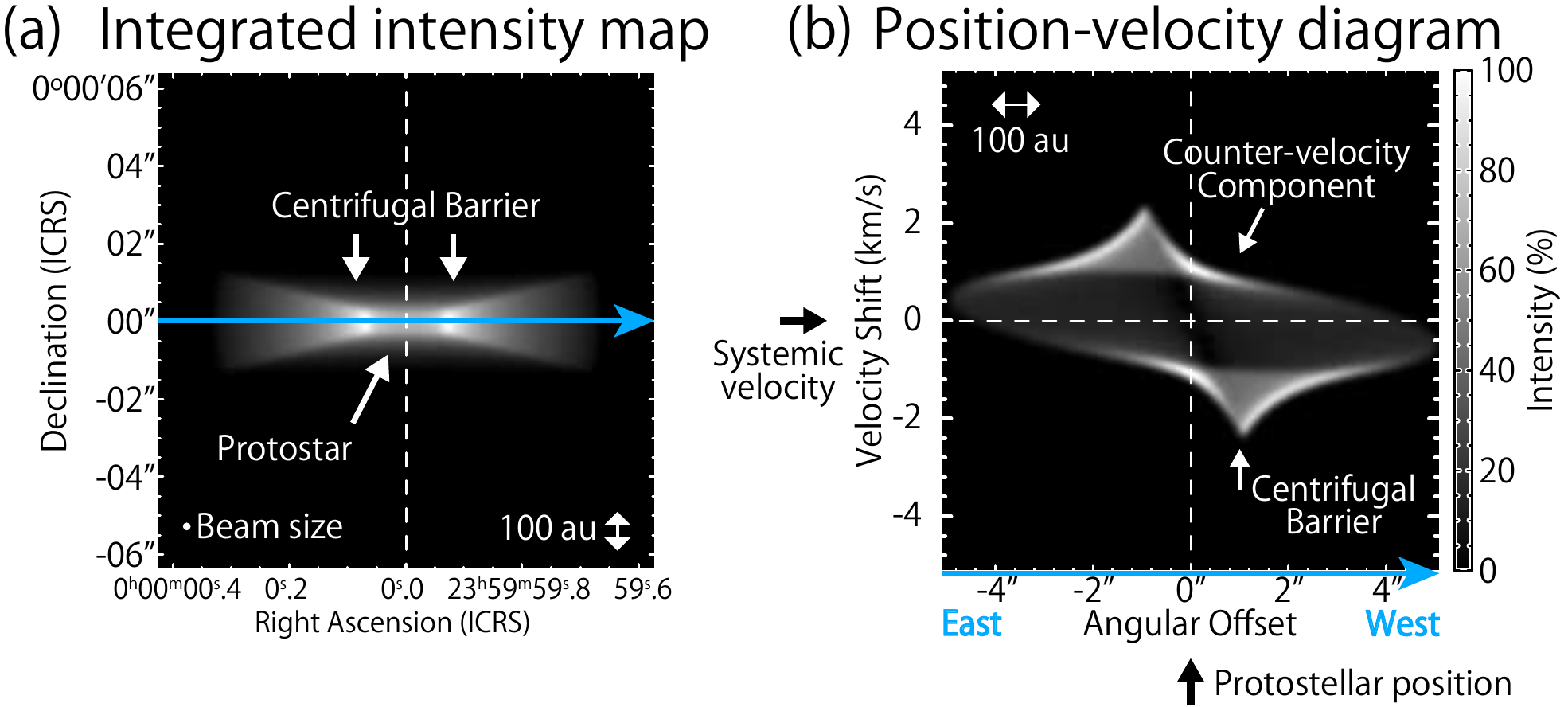}
	\fi
	\caption{Examples of 
			(a) the integrated intensity (moment 0) map
			and (b) the position-velocity diagram 
			of a modeled data cube of an \ire. 
			The position axis for the position-velocity diagram is taken 
			along the mid-plane of the envelope, 
			which is indicated by a cyan arrow in the integrated intensity map (P.A. 270\degr). 
			The parameters for the model are as follows; 
			the distance to the object ($d$) is 100 pc, 
			the central mass ($M$) is 0.3 \Msun, 
			the position angle of the \desys\ is 90\degr, 
			the direction of the rotation is counterclockwise (1 as the input value), 
			the outer radius (\rout) is 500 au, 
			the inner radius (\rin) equals to \rcb, 
			the thickness (\henv) at the protostar is 0 au, 
			the flare angle (\fenv) is 30\degr, 
			the power-law index of the radial density profile (\dprofenv) is $-1.5$, 
			{\bff and that of the gas-temperature distribution (\tprofenv) is $0$.
			It should be noted that the following parameters do not affect the results, 
			and thus their values are just set arbitrarily; 
			the molecular density and the gas temperature at the \cb\ (\nmolCB) 
			are $10^{-2}$~\cmcubic\ and 10~K, respectively.}  
			The emission is convolved with 
			the intrinsic Gaussian profile with the FWHM of 0.2~\kmps\ 
			and the Gaussian beam of (\bmaj $\times$ \bmin) $=$ (0\farcs2 $\times$ 0\farcs2) (P.A. = 0\degr). 
			\label{fig:result_example}} 
\end{figure}

\begin{figure}
	\iffigure
	\centering \includegraphics[bb = 0 0 1250 1380, scale = 0.36]{\dirnamefig 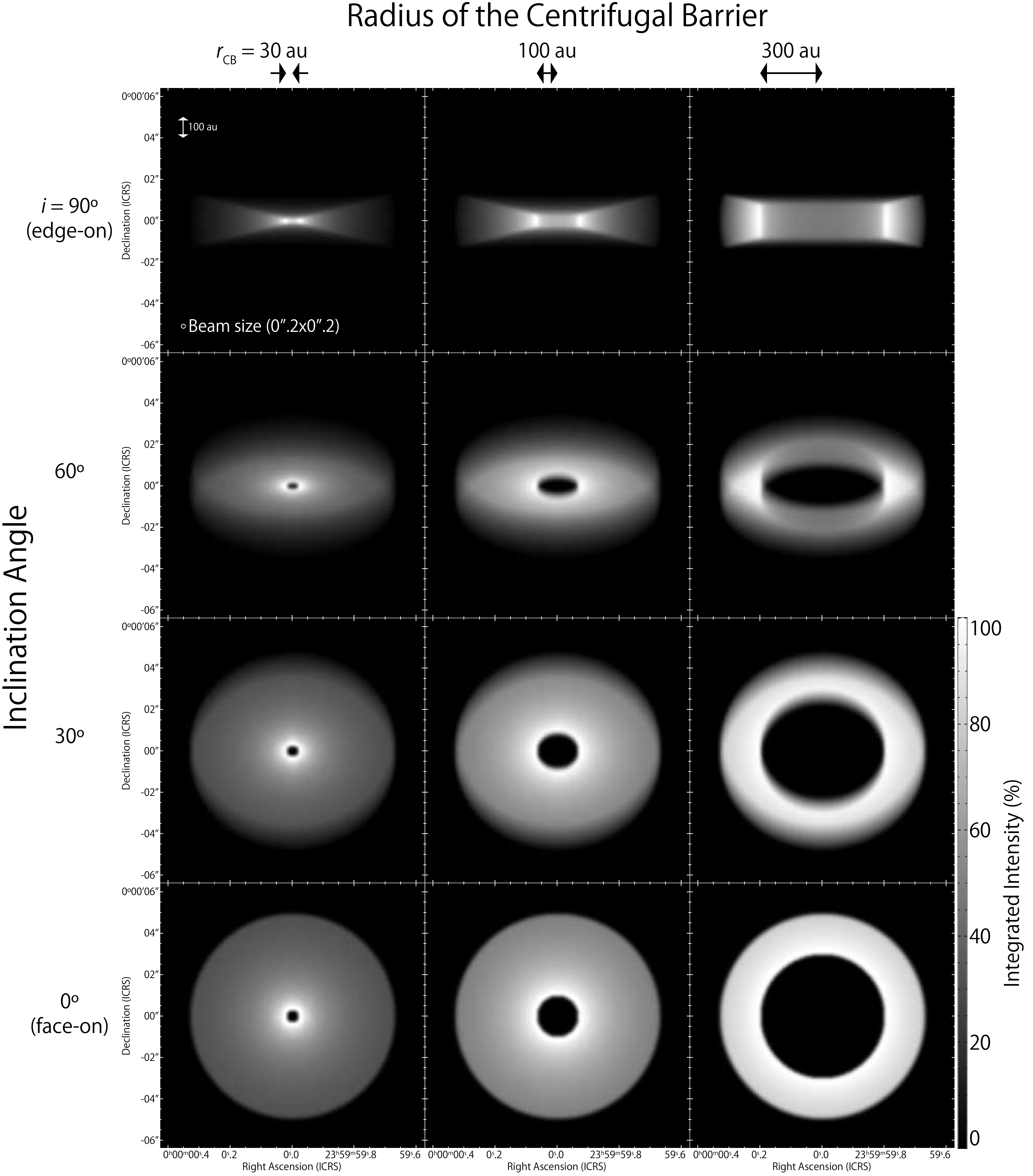}
	\fi
	\caption{Examples of the integrated intensity maps of \ire\ models. 
			{\bff 
			Various values of the radius of the \cb\ (\rcb) and the \ia\ ($i$) are employed, 
			while the other parameters are the same as those summarized 
			in the caption of {\bff Figure} \ref{fig:result_example}.} 
			The intensity is normalized by its peak value in the cube, 
			which does not affect the appearance of the figures. 
			\label{fig:varparams_mom0}} 
\end{figure}

\begin{figure}
	\iffigure
	\centering \includegraphics[bb = 0 0 1250 1380, scale = 0.36]{\dirnamefig 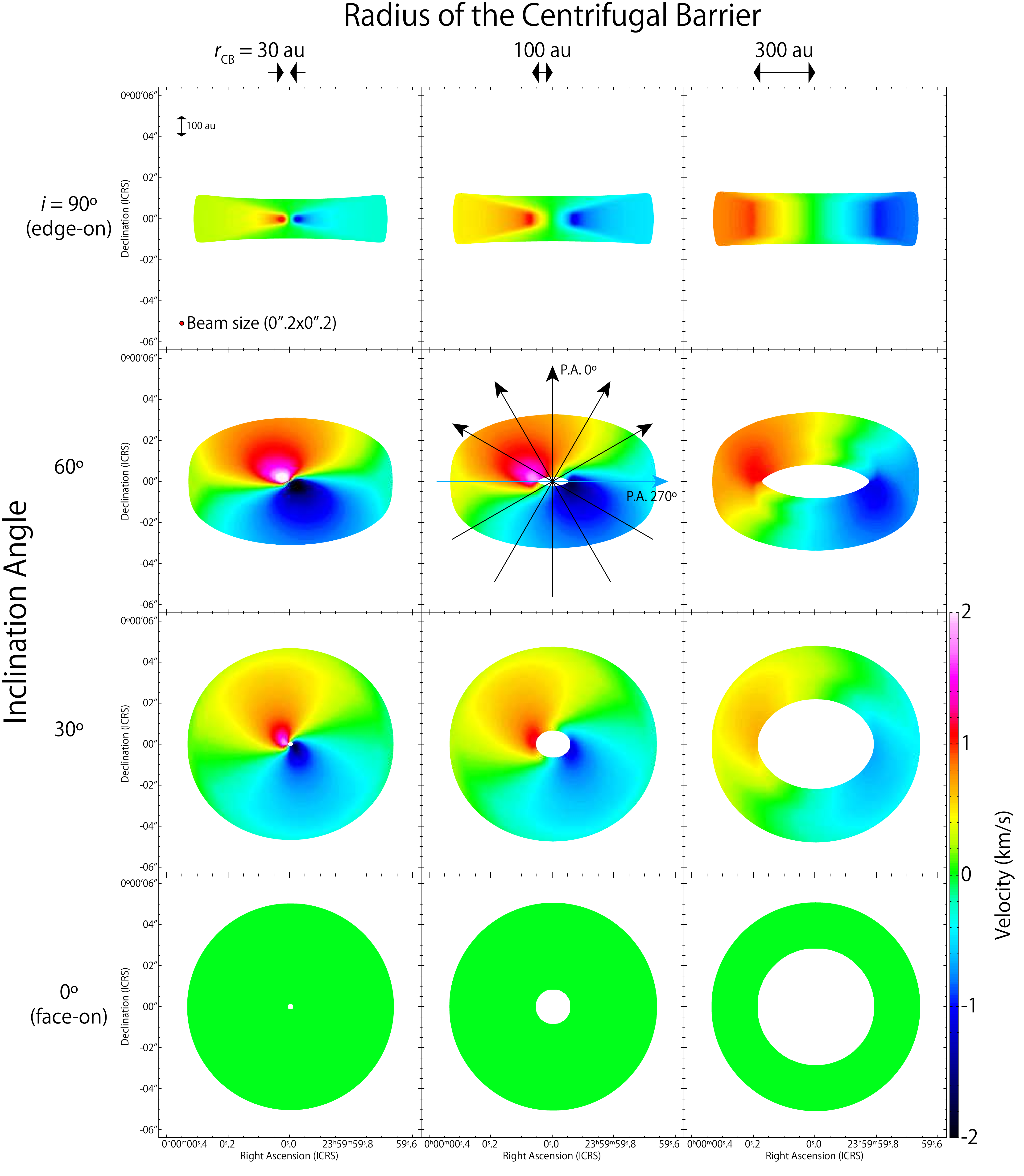}
	\fi
	\caption{Examples of the velocity field (moment 1) maps of \ire\ models.  
			The employed parameter values for the models are summarized 
			in the caption of Figures \ref{fig:result_example} and \ref{fig:varparams_mom0}. 
			The positions with an integrated intensity (Figure \ref{fig:varparams_mom0}) 
			lower than the 5\% relative to the peak integrated intensity are masked. 
			\label{fig:varparams_mom1}} 
\end{figure}

\begin{figure}
	\iffigure
	\centering \includegraphics[bb = 0 0 700 800, scale = 0.62]{\dirnamefig 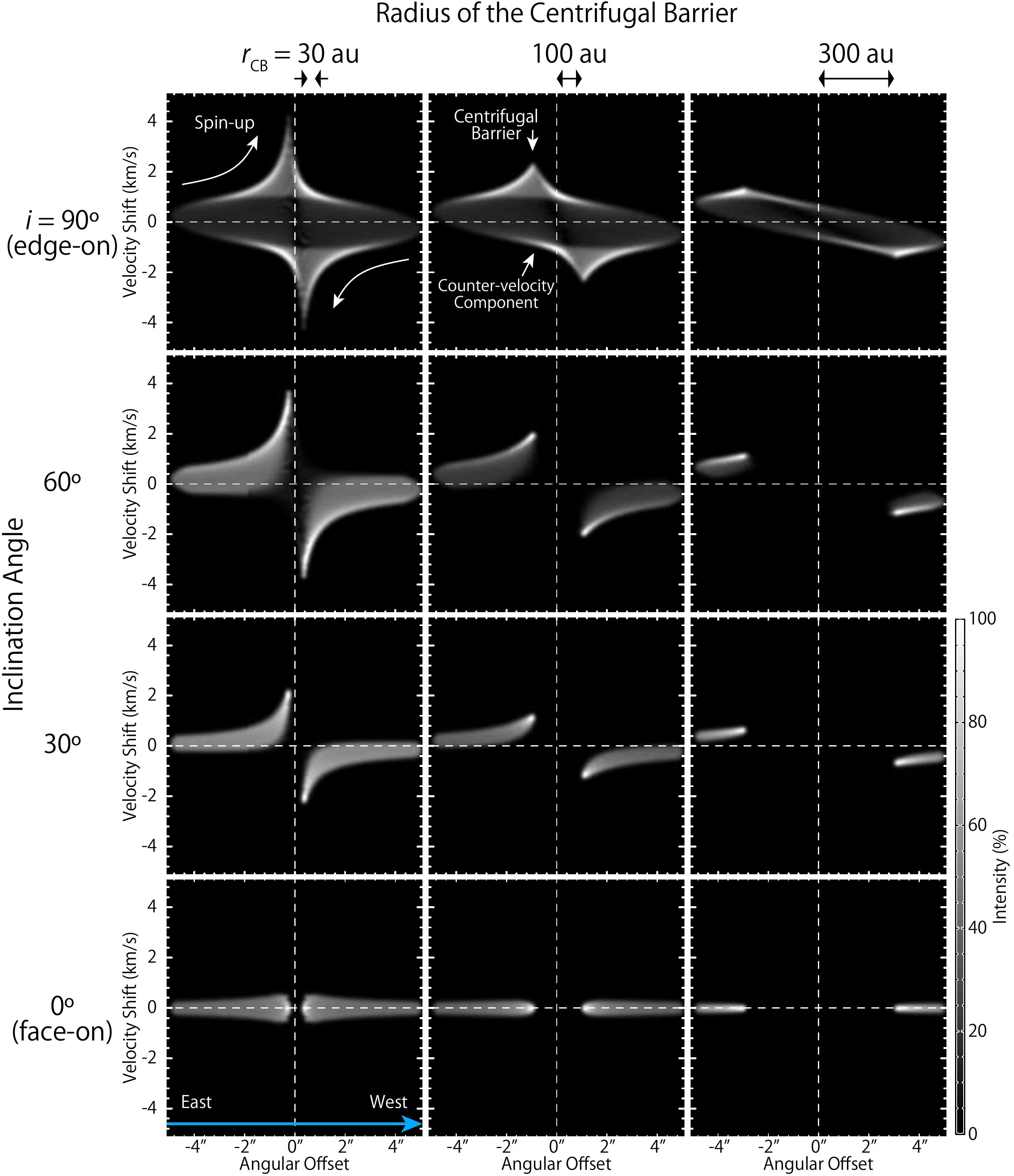} 
	\fi
	\caption{Examples of the position-velocity diagrams of \ire\ models. 
			The position axes are taken along the mid-plane of the envelopes, 
			which is indicated by a cyan arrow in Figure \ref{fig:varparams_mom1} (P.A. 270\degr). 
			The employed parameter values for the models are summarized 
			in the caption of Figures \ref{fig:result_example} and \ref{fig:varparams_mom0}. 
			\label{fig:varparams_PV}} 
\end{figure}

\begin{landscape}
\begin{figure}
	\iffigure
	\centering \includegraphics[bb = 0 0 1350 750, scale = 0.48]{\dirnamefig 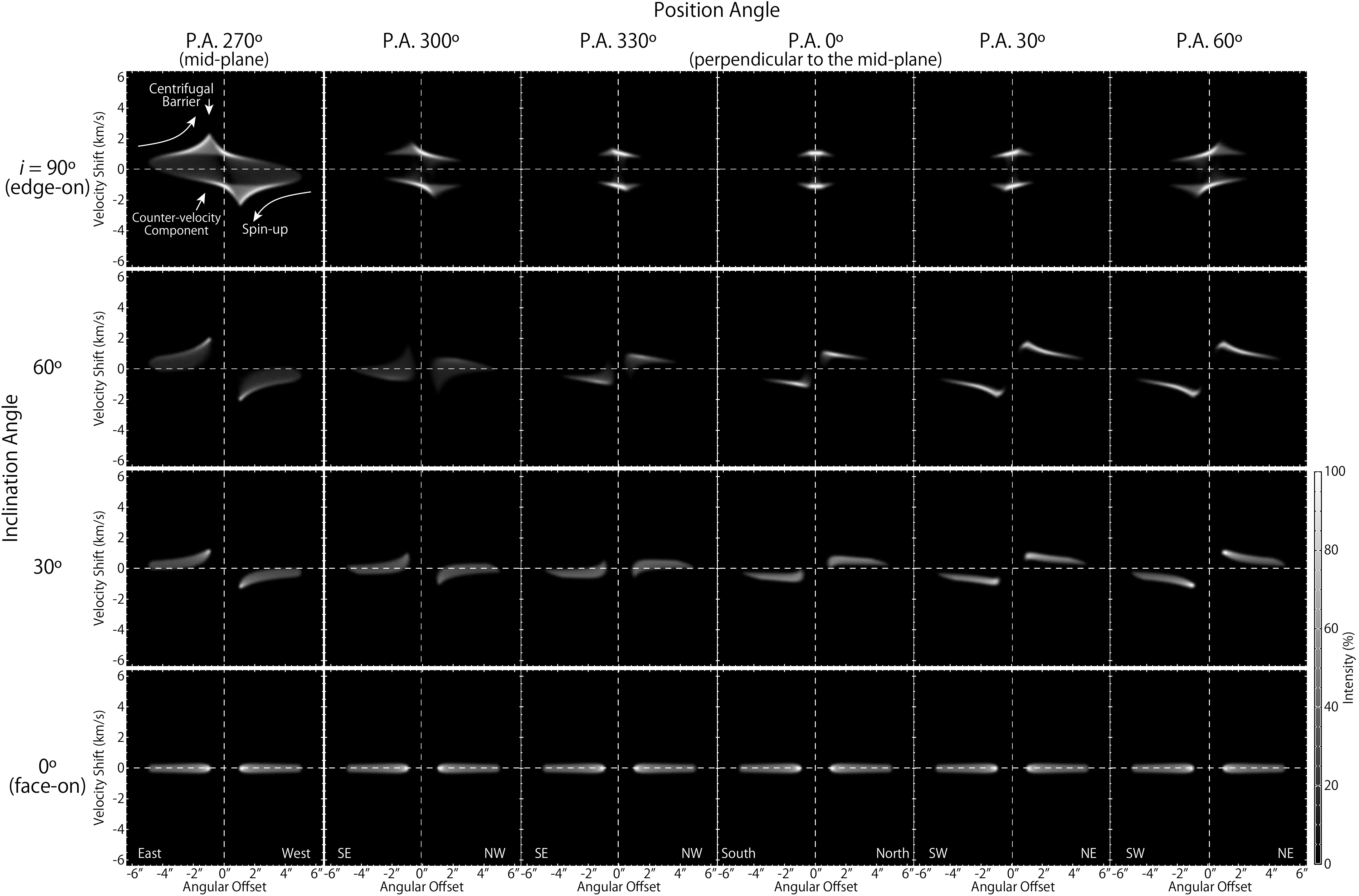} 
	\fi
	\caption{Examples of the position-velocity diagrams of \ire\ models 
			along various directions. 
			The position axes are indicated by arrows in Figure \ref{fig:varparams_mom1}, 
			which are taken for every 30\degr\ in P.A. 
			The employed parameter values for the models are summarized 
			in the caption of Figures \ref{fig:result_example} and \ref{fig:varparams_mom0}. 
			\label{fig:varparams_PV-PA}} 
\end{figure}
\end{landscape}

\begin{figure}
	\iffigure
	\centering \includegraphics[bb = 0 0 1800 1300, scale = 0.27]{\dirnamefig 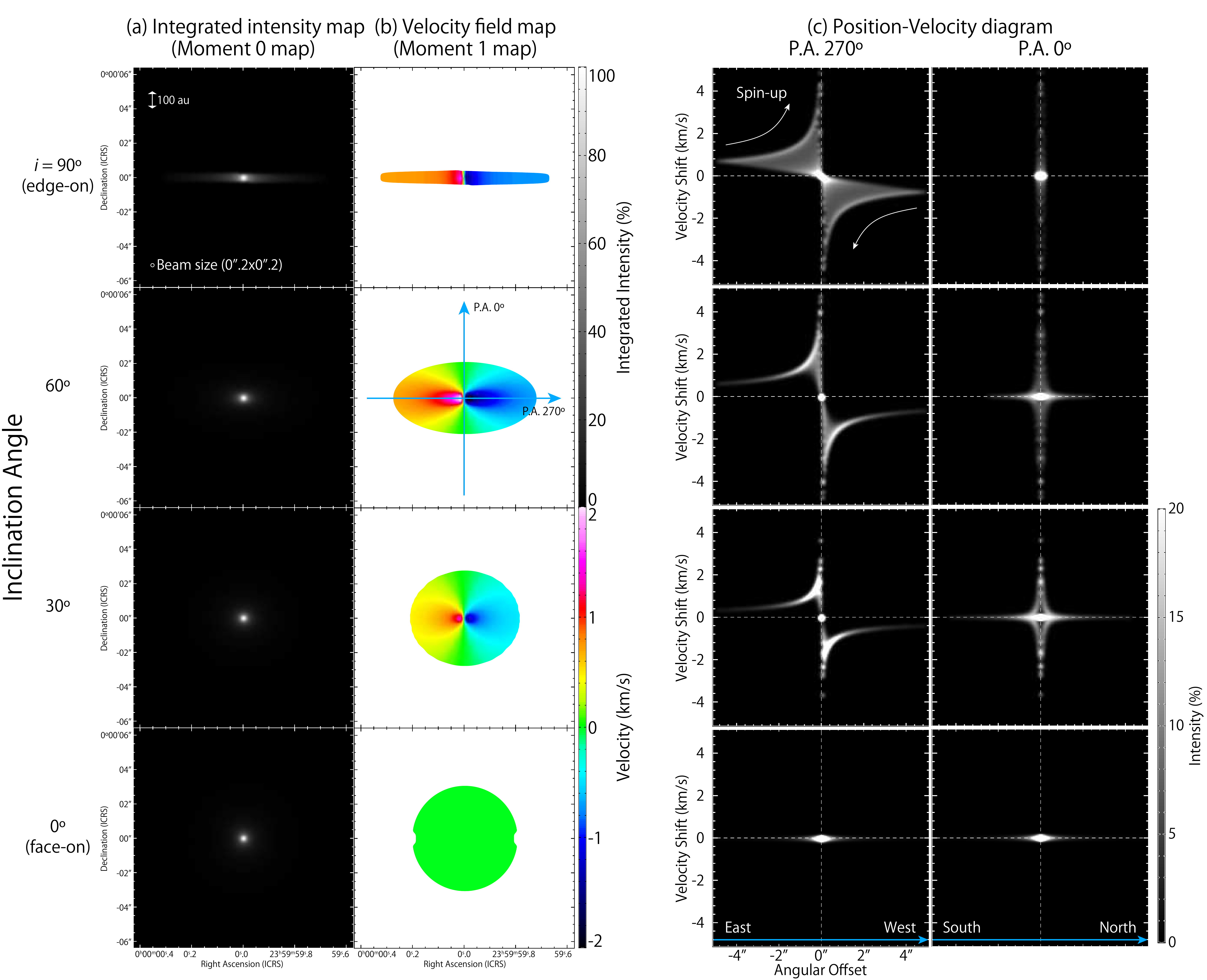} 
	\fi
	\caption{Examples of 
			(a) the integrated intensity (moment 0) maps, 
			(b) the velocity field (moment 1) maps, 
			and (c) the position-velocity diagrams 
			of the Keplerian disk models. 
			The positions with an integrated intensity lower than the 5\%\ relative to the peak value in each model are 
			masked in the velocity field map. 
			The position axes for the {\bff position-velocity diagrams} are taken 
			along (P.A. 270\degr) and perpendicular (P.A. 0\degr) to the mid-plane of the disks, 
			which are indicated by cyan arrows in panel (b). 
			The employed parameter values for the models are summarized 
			in the caption of Figures \ref{fig:result_example} and \ref{fig:varparams_mom0}, 
			except for the \rcb\ and \rin; 
			\rcb\ is set to be larger than \rout, 
			and \rin\ is to be 0 au. 
			\label{fig:result_kepler}} 
\end{figure}

\clearpage
\begin{landscape}
\begin{figure}
	\iffigurechisq
	\centering \includegraphics[bb = 0 0 1000 950, scale = 0.5]{\dirnamechi 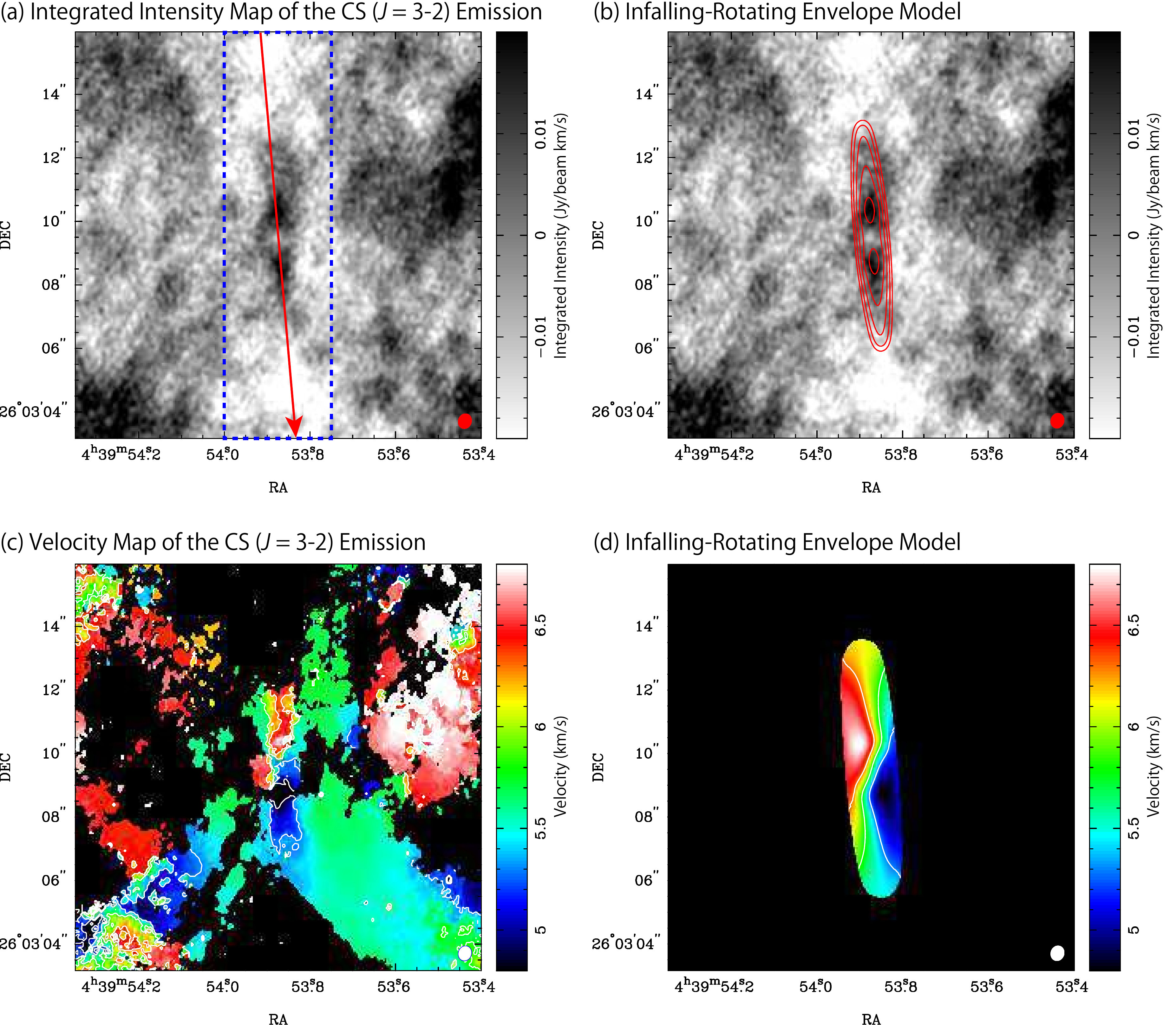} 
	\fi
\end{figure}
\begin{figure}
	\caption{Comparison between the observation and the model results for L1527 (see Section \ref{sec:obs_fit_chi2_L1527}). 
			The CS {\bff (\csl; a, c)} emission data are used for comparison. 
			{\bff The model result is the best-fit one with the \ire\ model (b, d). 
			The} parameter values are summarized in Table \ref{tb:chi2_L1527_params}. 
			{\bff Panels (a, b)} represent the integrated intensity (moment 0) maps, 
			where the model {\bff result} in contours {\bff is} overlaid on the observed {\bff image} in grey scale. 
			Contour levels are 5\%, 10\%, 20\%, 40\%, and 80\%\ of the {\bff peak} intensity  {\bff of the} model. 
			A rectangle enclosed by a dashed blue line in panel (a) represents the region considered in the {\bff \chisq\ test} for the cube data 
			to avoid the contamination from the outflow components. 
			A red arrow in panel (a) represents the position axis 
			along which the PV diagrams in Figure \ref{fig:chi2_L1527_PV} are prepared. 
			{\bff Panels (c, d)} represent the velocity (moment 1) maps. 
			Contour levels are every 0.5~\kmps\ from the systemic velocity of 5.8~\kmps.
			Velocity maps are prepared by using the data points with an intensity equal to or larger than the following thresholds: 
			15~\mJypb\ (3$\sigma$) for the observation, and 0.1\%\ of the maximum value in {\bff the} modeled cube data. 
			Black colored pixels do not have any data points with an intensity larger than the threshold through all the considered velocity range. 
			The beam size is depicted at the bottom right corner of each panel. 
			\label{fig:chi2_L1527_mom}}
\end{figure}
\end{landscape}

\clearpage
\begin{figure}
	\iffigurechisq
	\centering \includegraphics[bb = 0 0 500 660, scale = 0.68]{\dirnamechi 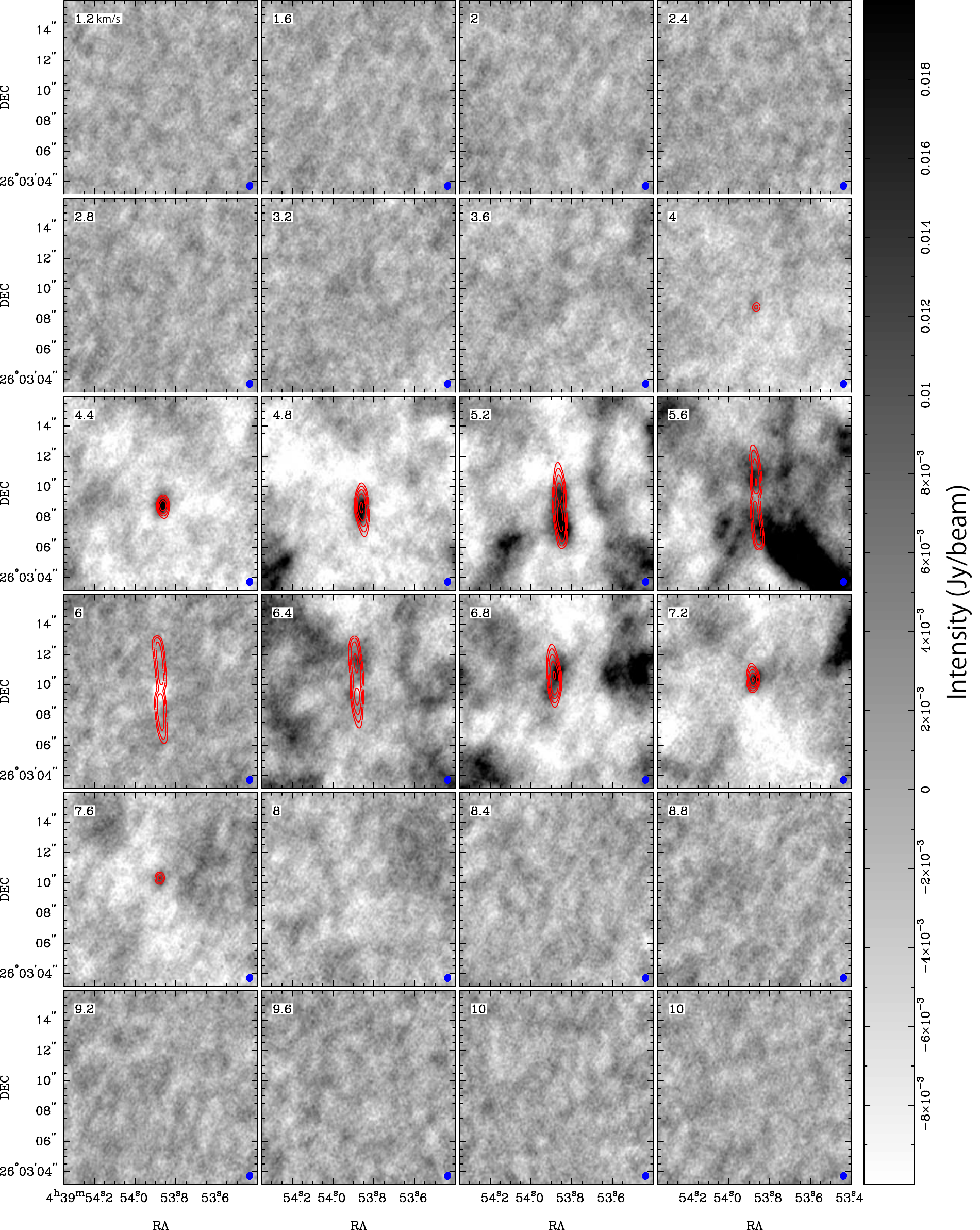}
	\fi
	\caption{Comparison of the velocity channel maps between the observation and the results {\bff of the \ire\ model} for L1527 (see Section \ref{sec:obs_fit_chi2_L1527}). 
			Grey scale maps represent the observed CS (\csl) emission. 
			Contours represent the best-fit model obtained by the \chisq\ test. 
			The parameters for the model are: 
			$M$ of 0.15 \Msun\ and \rcb\ of 100 au, 
			assuming $i$ of 95\degr. 
			Contour levels are 5\%, 10\%, 20\%, 40\%, and 80\%\ of the peak intensity of the model. 
			The original channel width of 0.2~\kmps\ is used in the \chisq\ test, 
			whereas the four successive channels are combined in these velocity channel maps. 
			The central velocity is shown at the upper left corner of each panel. 
			The beam size is depicted at the bottom right corner of each panel. 
			\label{fig:chi2_L1527_channelmap-IRE}}
\end{figure}

\clearpage
\begin{figure}
	\iffigurechisq
	\centering \includegraphics[bb = 0 0 1500 660, scale = 0.4]{\dirnamechi 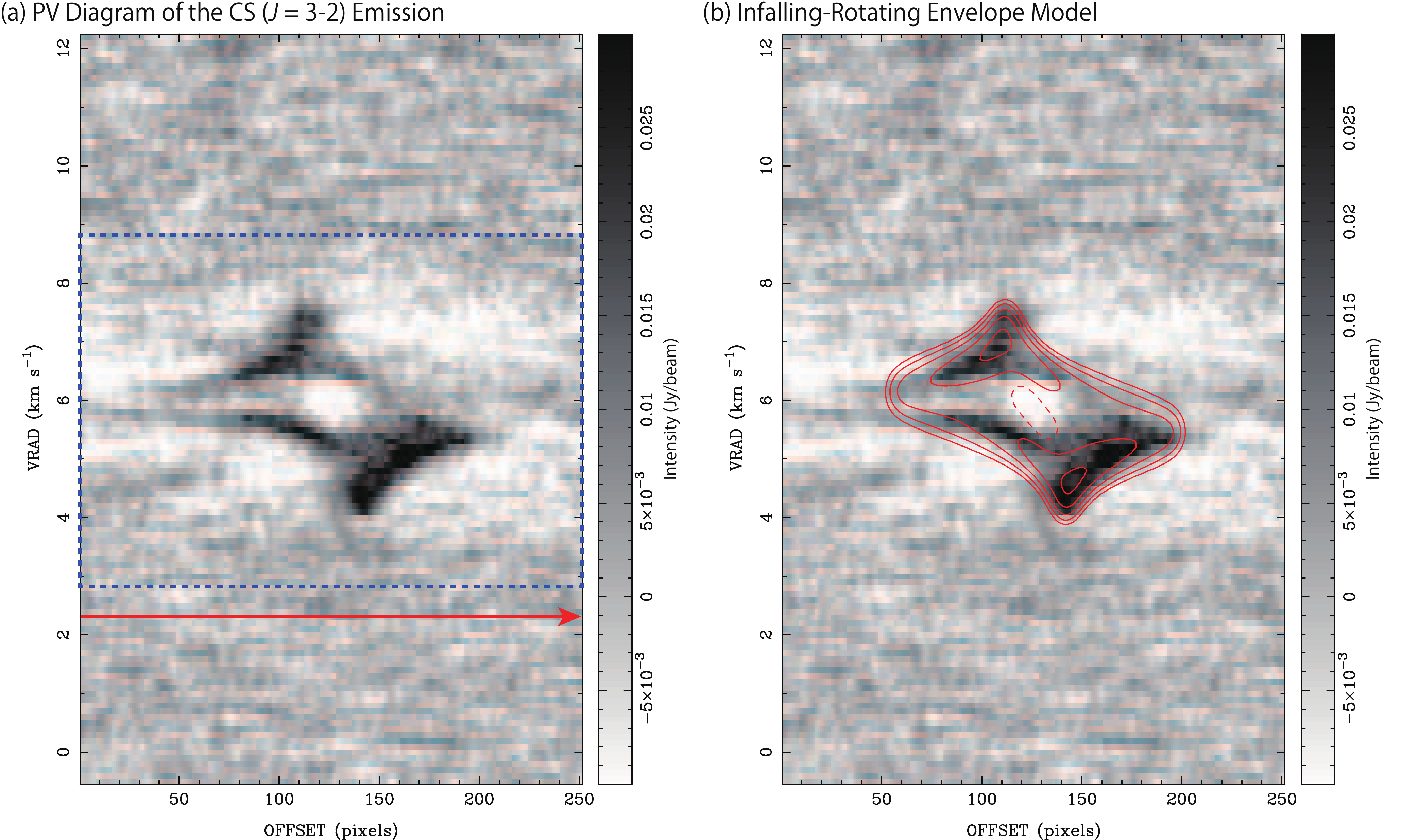} 
	\fi
\end{figure}
\begin{figure}
	\caption{Comparison of the position-velocity diagrams between the observation and the model {\bff result} for L1527 (see Section \ref{sec:obs_fit_chi2_L1527}). 
			Grey scale maps represent the observed CS (\csl) emission. 
			{\bff The model result represented in contours is the best-fit one with the \ire\ model (b). 
			The} parameter values are summarized in Table \ref{tb:chi2_L1527_params}. 
			Contour levels are 5\%, 10\%, 20\%, 40\%, and 80\%\ of the {\bff peak} intensity in {\bff the} model. 
			PV diagrams are prepared along the red arrow in Figure \ref{fig:chi2_L1527_mom}(a); 
			this arrow is taken along the mid-plane of the \desys\ and centered at the protostellar position. 
			A rectangle enclosed by a dashed blue line in panel (a) represents the velocity range considered in the {\bff \chisq\ test} for the cube data, 
			where the velocity shift is within $\pm$3~\kmps\ from the systemic velocity of 5.8~\kmps. 
			Dashed {\bff contour} in {\bff panel (b) represents} the intensity dip. 
			\label{fig:chi2_L1527_PV}}
\end{figure}

\clearpage
\begin{figure}
	\iffigurechisq
	\centering \includegraphics[bb = 0 0 400 200, scale = 1.0]{\dirnamechi 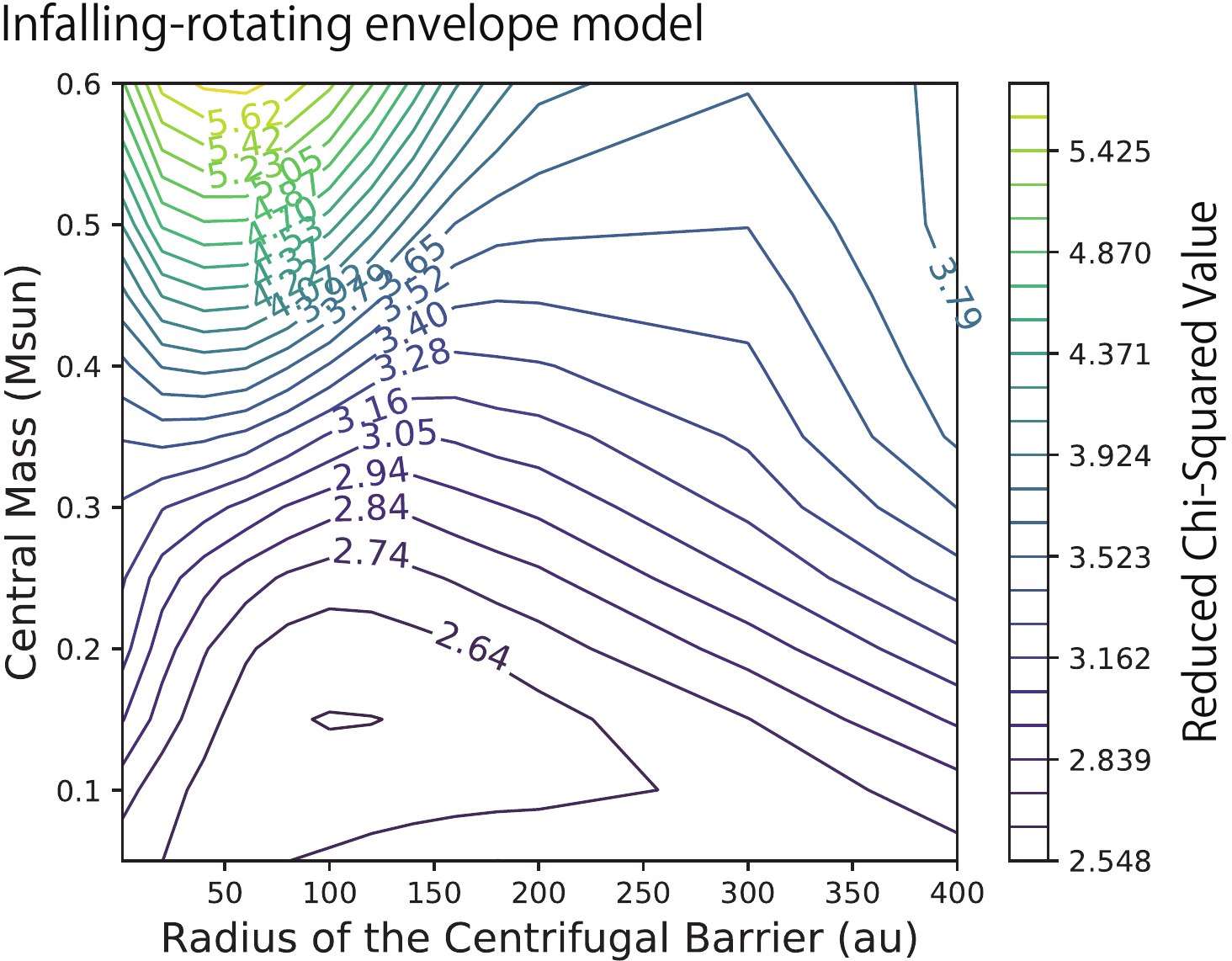} 
	\fi
	\caption{Reduced {\bff \chisq\ map} with various parameter values for L1527. 
			Free parameters and their ranges are summarized in Table \ref{tb:chi2_L1527_params}. 
			\label{fig:chi2_L1527_chisqplot}}
\end{figure}

\clearpage
\begin{landscape}
\begin{figure}
	\iffigurechisq
	\centering \includegraphics[bb = 0 0 1500 950, scale = 0.38]{\dirnamechi 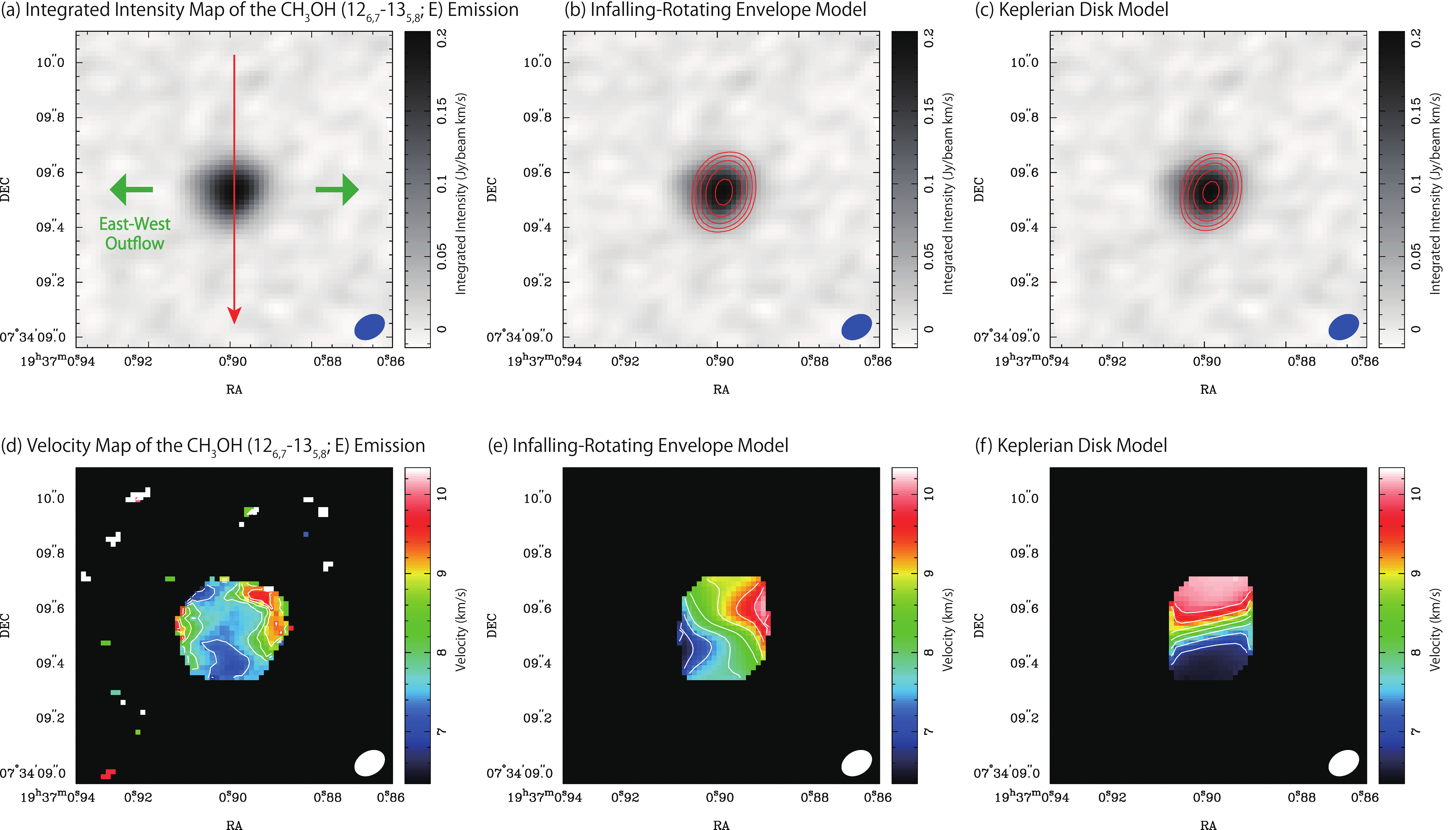} 
	\fi
\end{figure}
\begin{figure}
	\caption{Comparison between the observation and the model results for B335 (see Section \ref{sec:obs_fit_chi2_B335}). 
			The \MN\ {\bff (\mnb; a, d)} emission data are used for comparison. 
			The model results are the best-fit ones with 
			{\bff the \ire\ model (b, e) 
			and the Keplerian disk model (c, f).} 
			Their parameter values are summarized in Table \ref{tb:chi2_B335_params}. 
			{\bff Panels (a)$-$(c)} represent the integrated intensity (moment 0) maps, 
			where the model results in contours are overlaid on the observed {\bff image} in grey scale. 
			Contour levels are 5\%, 10\%, 20\%, 40\%, and 80\%\ of the {\bff peak} intensity in each model. 
			A red arrow in panel (a) represents the position axis 
			along which the PV diagrams in Figure \ref{fig:chi2_B335_PV} are prepared. 
			{\bff This arrow is perpendicular to the East-West outflow of this source 
			\citep{Hirano1988, Yen2010, Imai_B335HC, Bjerkeli2019}.} 
			{\bff Panels (d)$-$(f)} represent the velocity (moment 1) maps. 
			Contour levels are every 0.5~\kmps\ from the systemic velocity of 8.34~\kmps.
			Velocity maps are prepared by using the data points with an intensity equal to {\bff or} larger than the following thresholds: 
			3~\mJypb\ (3$\sigma$) for the observation, and 0.1\%\ of the maximum value in each modeled cube data. 
			Black colored pixels do not have any data points with an intensity larger than the threshold through all the considered velocity range. 
			The beam size is depicted at the bottom right corner of each panel. 
			\label{fig:chi2_B335_mom}}
\end{figure}
\end{landscape}

\clearpage
\begin{figure}
	\iffigurechisq
	\centering \includegraphics[bb = 0 0 500 660, scale = 0.68]{\dirnamechi 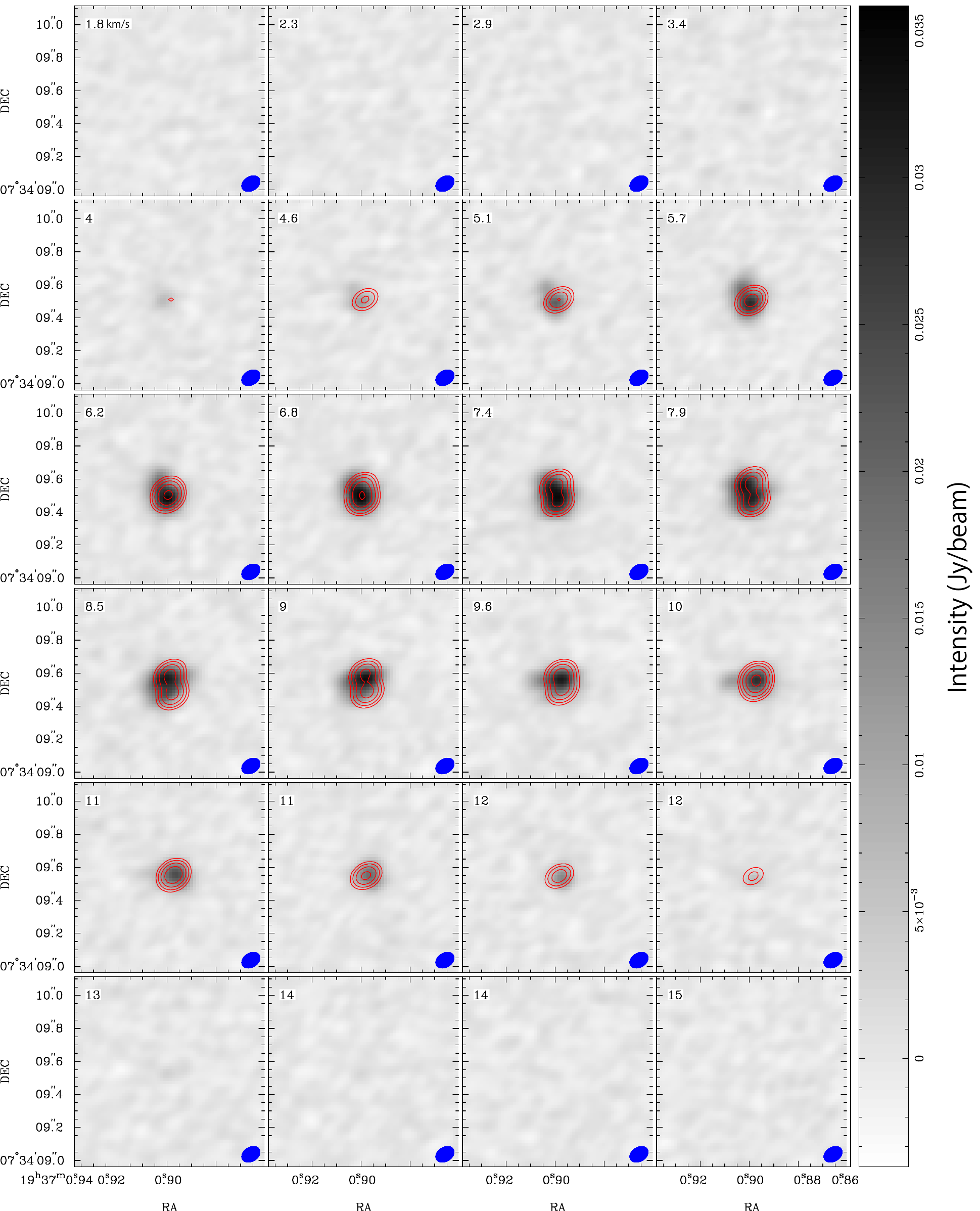}
	\fi
	\caption{Comparison of the velocity channel maps between the observation and the {\bff results of the \ire\ model} for B335 (see Section \ref{sec:obs_fit_chi2_B335}). 
			Grey scale maps represent the observed \MN\ (\mnb) emission. 
			Contours represent the best-fit model obtained by the \chisq\ test for the \ire\ model. 
			The parameters for the model are: 
			$M$ of 0.02 \Msun, \rcb\ of 1 au, and $i$ of 70\degr. 
			Contour levels are 5\%, 10\%, 20\%, 40\%, and 80\%\ of the peak intensity of the model. 
			The original channel width of 0.28~\kmps\ is used in the \chisq\ test, 
			whereas the two successive channels are combined in these velocity channel maps. 
			The central velocity is shown at the upper left corner of each panel. 
			The beam size is depicted at the bottom right corner of each panel. 
			\label{fig:chi2_B335_channelmap-IRE}}
\end{figure}

\clearpage
\begin{figure}
	\iffigurechisq
	\centering \includegraphics[bb = 0 0 500 660, scale = 0.68]{\dirnamechi 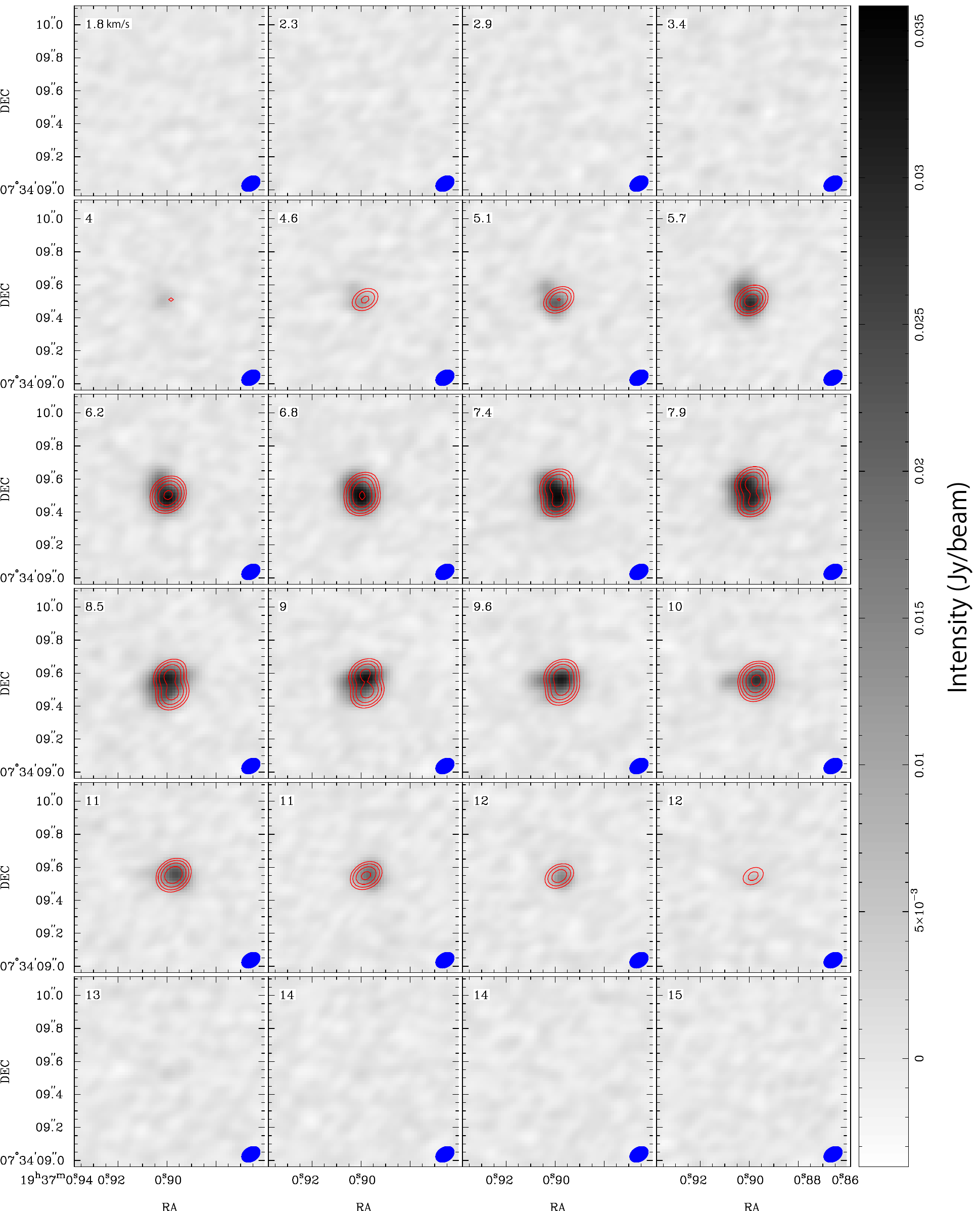}
	\fi
	\caption{Same as Figure \ref{fig:chi2_B335_channelmap-IRE}, 
			but for the Keplerian disk model.  
			Contours represent the best-fit model obtained by the \chisq\ test for the Keplerian disk model. 
			The parameters for the model are: 
			$M$ of 0.04 \Msun, \rin\ of 1 au, and $i$ of 75\degr. 
			\label{fig:chi2_B335_channelmap-Kep}}
\end{figure}

\clearpage
\begin{landscape}
\begin{figure}
	\iffigurechisq
	\centering \includegraphics[bb = 0 0 1750 700, scale = 0.35]{\dirnamechi 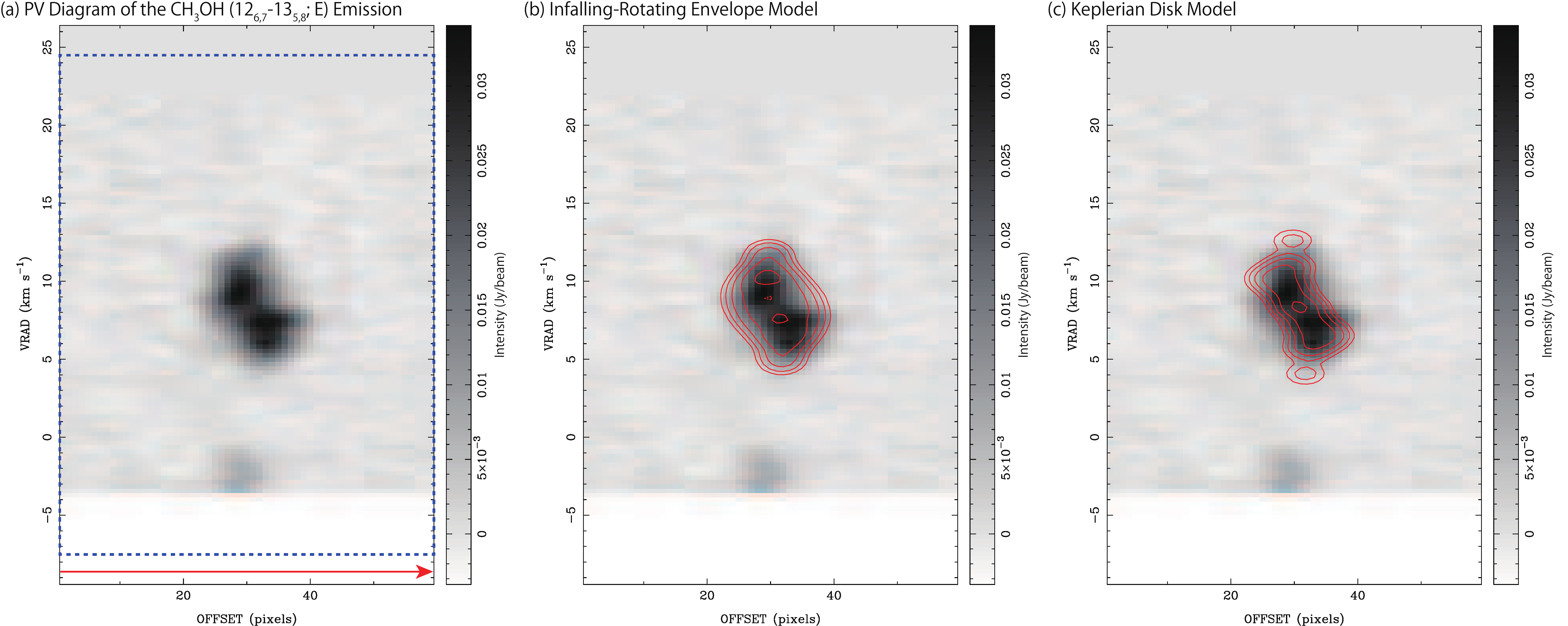} 
	\fi
\end{figure}
\end{landscape}
\begin{landscape}
\begin{figure}
	\caption{Comparison of the position-velocity diagrams between the observation and the model results for B335 (see Section \ref{sec:obs_fit_chi2_B335}). 
			Grey scale maps represent the observed \MN\ (\mnb) emission. 
			{\bff The model results represented in contours are the best-fit ones with 
			the \ire\ model (b) and the Keplerian disk model (c).} 
			Their parameter values are summarized in Table \ref{tb:chi2_B335_params}. 
			Contour levels are 5\%, 10\%, 20\%, 40\%, and 80\%\ of the {\bff peak} intensity in each model. 
			PV diagrams are prepared along the red arrow in Figure \ref{fig:chi2_B335_mom}(a); 
			this arrow is taken along the mid-plane of the \desys\ and centered at the protostellar position. 
			A rectangle enclosed by a dashed blue line in panel (a) represents the velocity range considered in the \chisq\ tests for the cube data, 
			where the velocity shift is within $\pm$16~\kmps\ from the systemic velocity of 8.34~\kmps. 
			Dashed {\bff contour} in {\bff panel (b) represents} the intensity dip. 
			\label{fig:chi2_B335_PV}}
\end{figure}
\end{landscape}

\clearpage
\begin{figure}
	\iffigurechisq
	\centering \includegraphics[bb = 0 0 900 1300, scale = 0.43]{\dirnamechi 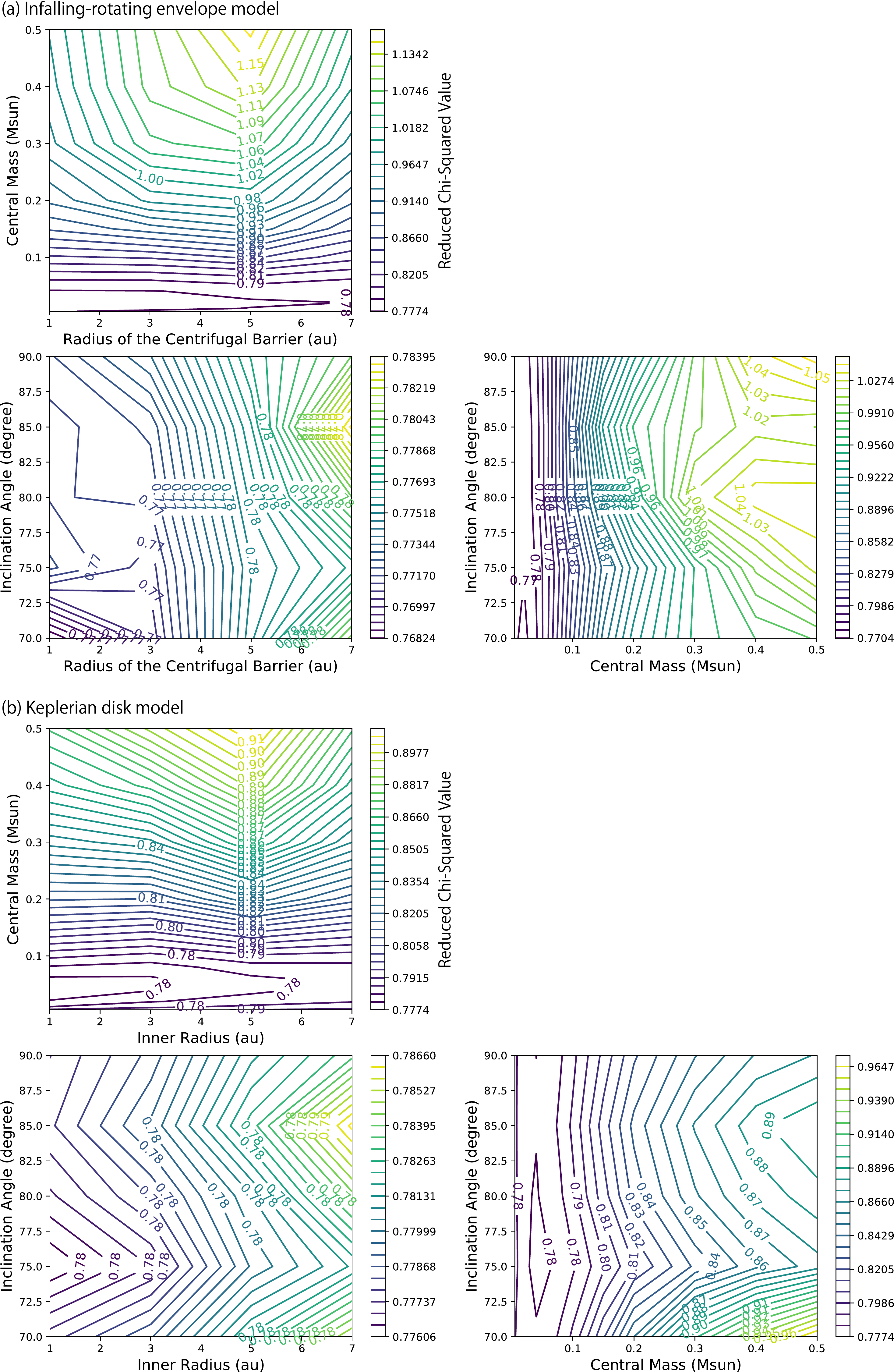} 
	\fi
	\caption{Reduced \chisq\ maps with various parameter values for B335. 
			Free parameters and their ranges are summarized in Table \ref{tb:chi2_B335_params}. 
			\label{fig:chi2_B335_chisqplot}}
\end{figure}

\clearpage
\begin{landscape}
\begin{figure}
	\iffigurechisq
	\centering \includegraphics[bb = 0 0 2500 950, scale = 0.29]{\dirnamechi 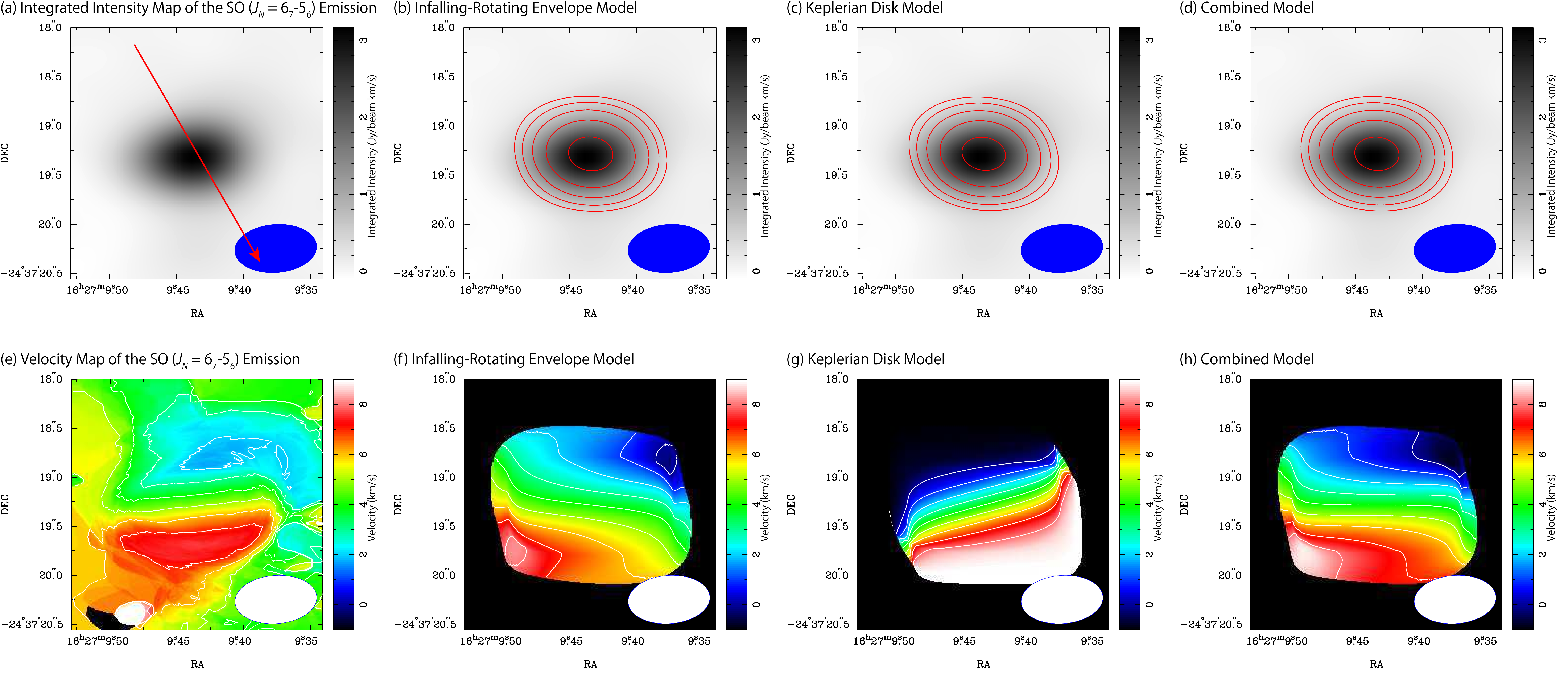} 
	\fi
\end{figure}
\begin{figure}
	\caption{Comparison between the observation and the model results for Elias 29 (see Section \ref{sec:obs_fit_chi2_Elias29}). 
			The SO {\bff (\soelias; a, e)} emission data are used for comparison. 
			The model results are the best-fit ones with 
			the \ire\ model (b, f), 
			the Keplerian disk model (c, g), 
			and the combined model (d, h). 
			Their parameter values are summarized in Table \ref{tb:chi2_Elias29_params}. 
			Panels (a)$-$(d) represent the integrated intensity (moment 0) maps, 
			where the model results in contours are overlaid on the observed {\bff image} in grey scale. 
			Contour levels are 5\%, 10\%, 20\%, 40\%, and 80\%\ of the {\bff peak} intensity in each model. 
			A red arrow in panel (a) represents the position axis 
			along which the PV diagrams of Figure \ref{fig:chi2_Elias29_PV} are prepared. 
			Panels (e)$-$(h) represent the velocity (moment 1) maps. 
			Contour levels are every 1~\kmps\ from the systemic velocity of 4~\kmps.
			Velocity maps are prepared by using the data points with an intensity equal to or larger than the following thresholds: 
			21~\mJypb\ (3$\sigma$) for the observation, and 0.1\%\ of the maximum value in each modeled cube data. 
			Black colored pixels do not have any data points with an intensity larger than the threshold through all the considered velocity range. 
			The beam size is depicted at the bottom right corner of each panel. 
			\label{fig:chi2_Elias29_mom}}
\end{figure}
\end{landscape}

\clearpage
\begin{figure}
	\iffigurechisq
	\centering \includegraphics[bb = 0 0 500 660, scale = 0.68]{\dirnamechi 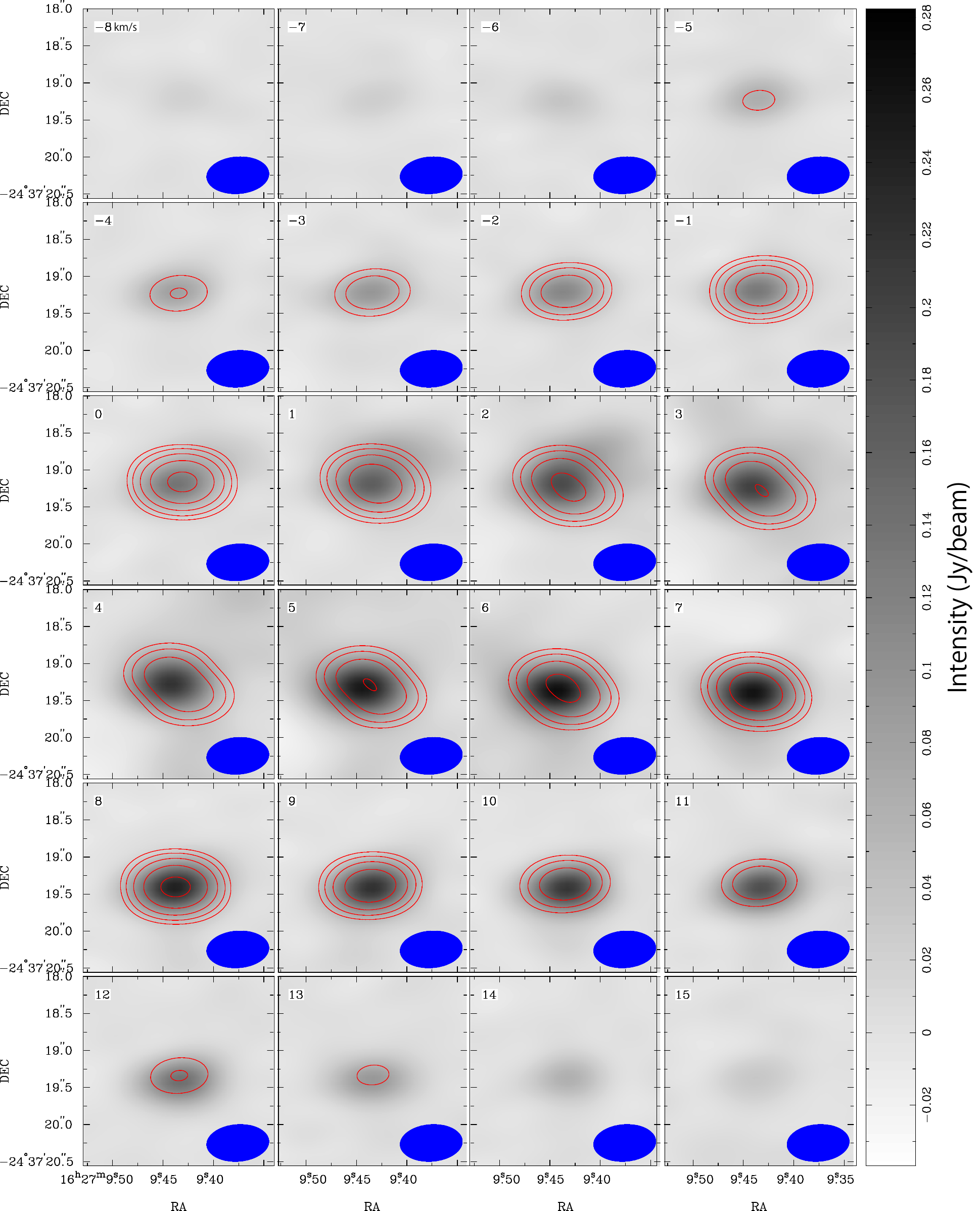}
	\fi
	\caption{Comparison of the velocity channel maps between the observation and 
			the results {\bff of the \ire\ model} for Elias 29 (see Section \ref{sec:obs_fit_chi2_Elias29}). 
			Grey scale maps represent the observed SO (\soelias) emission. 
			Contours represent the best-fit model obtained by the \chisq\ test for the \ire\ model. 
			The parameters for the model are: 
			$M$ of 0.6 \Msun, \rcb\ of 5 au, and $i$ of 120\degr. 
			Contour levels are 5\%, 10\%, 20\%, 40\%, and 80\%\ of the peak intensity of the model. 
			The original channel width of 0.2~\kmps\ is used in the \chisq\ test, 
			whereas the five successive channels are combined in these velocity channel maps. 
			The central velocity is shown at the upper left corner of each panel. 
			The beam size is depicted at the bottom right corner of each panel. 
			\label{fig:chi2_Elias29_channelmap-IRE}}
\end{figure}

\clearpage
\begin{figure}
	\iffigurechisq
	\centering \includegraphics[bb = 0 0 500 660, scale = 0.68]{\dirnamechi 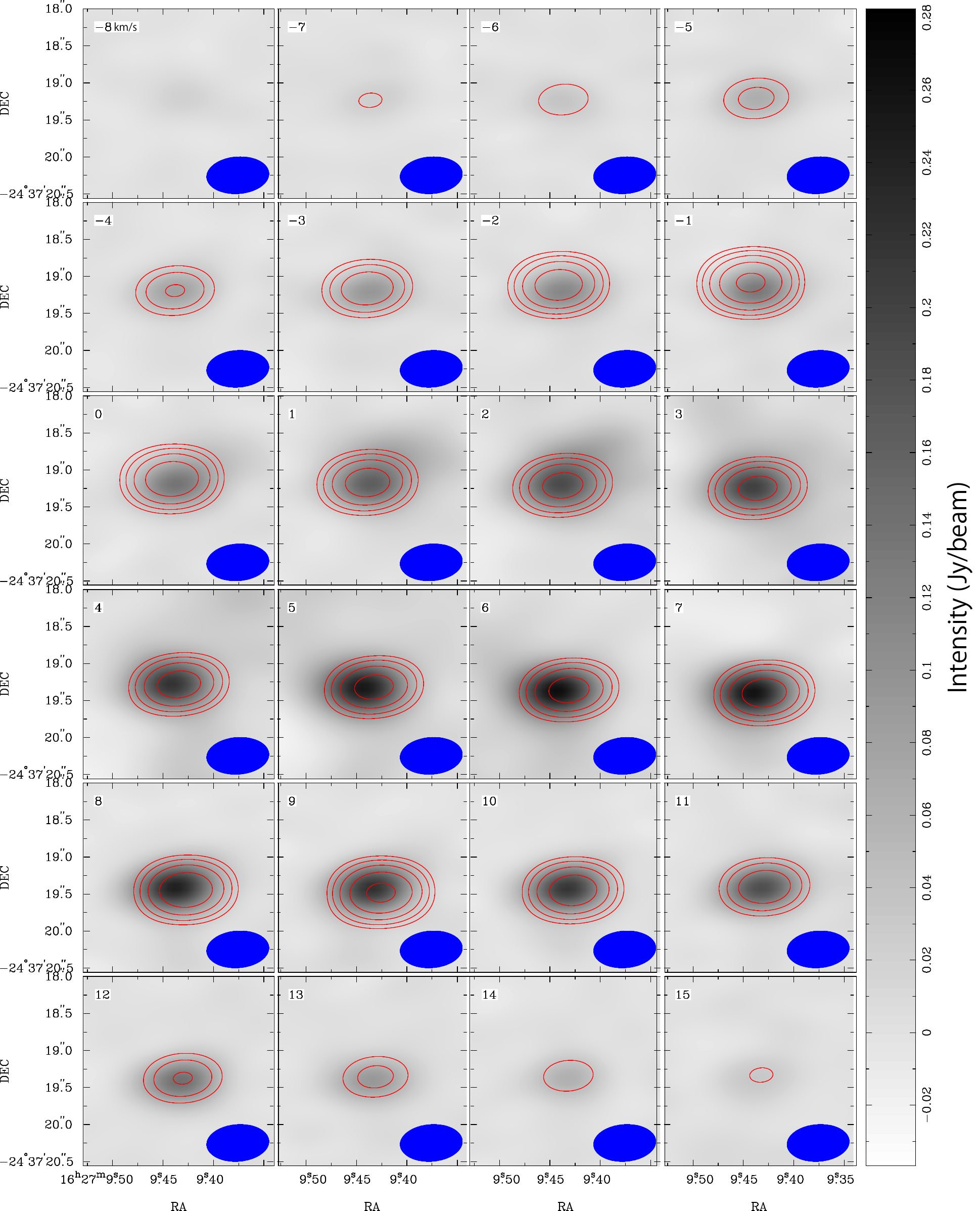}
	\fi
	\caption{Same as Figure \ref{fig:chi2_Elias29_channelmap-IRE}, 
			but for the Keplerian disk model.  
			Contours represent the best-fit model obtained by the \chisq\ test for the Keplerian disk model. 
			The parameters for the model are: 
			$M$ of 1.4 \Msun, \rin\ of 1 au, and $i$ of 80\degr. 
			\label{fig:chi2_Elias29_channelmap-Kep}}
\end{figure}

\clearpage
\begin{figure}
	\iffigurechisq
	\centering \includegraphics[bb = 0 0 1500 1500, scale = 0.4]{\dirnamechi 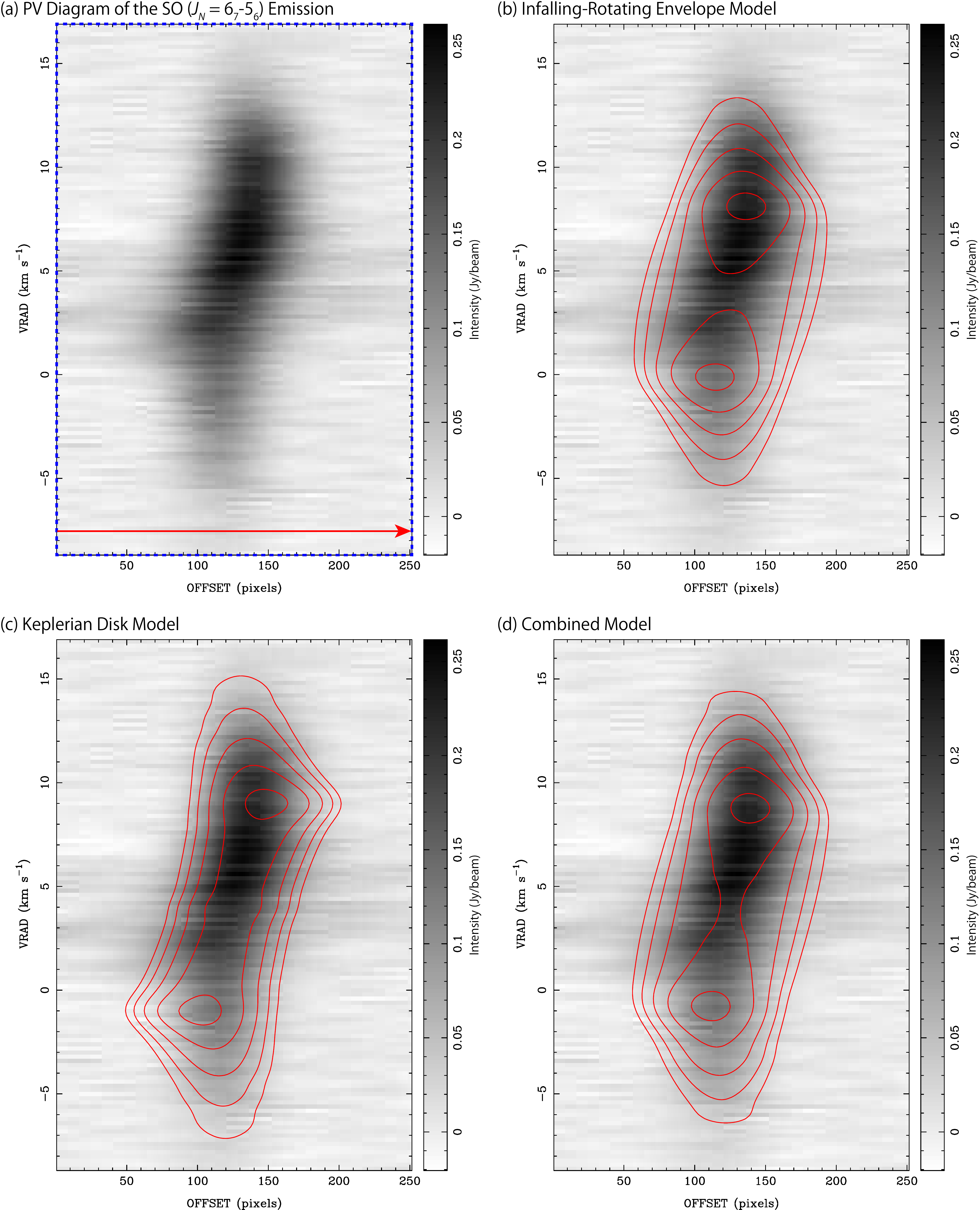} 
	\fi
\end{figure}
\begin{figure}
	\caption{Comparison of the position-velocity diagrams between the observation and the model results for Elias 29 (see Section \ref{sec:obs_fit_chi2_Elias29}). 
			Grey scale maps represent the observed SO (\soelias) emission. 
			The model results {\bff represented in contours} are the best-fit ones with 
			the \ire\ model (b), 
			the Keplerian disk model (c), 
			and the combined model (d). 
			Their parameter values are summarized in Table \ref{tb:chi2_Elias29_params}. 
			Contour levels are 5\%, 10\%, 20\%, 40\%, and 80\%\ of the {\bff peak} intensity in each model. 
			PV diagrams are prepared along the red arrow in Figure \ref{fig:chi2_Elias29_mom}(a); 
			this arrow is taken along the mid-plane of the \desys\ and centered at the protostellar position. 
			A rectangle enclosed by a dashed blue line in panel (a) represents the velocity range considered in the \chisq\ tests for the cube data, 
			where the velocity shift is within $\pm$12.8~\kmps\ from the systemic velocity of 4~\kmps. 
			%
			\label{fig:chi2_Elias29_PV}}
\end{figure}

\clearpage
\begin{figure}
	\iffigurechisq
	\centering \includegraphics[bb = 0 0 900 1300, scale = 0.43]{\dirnamechi 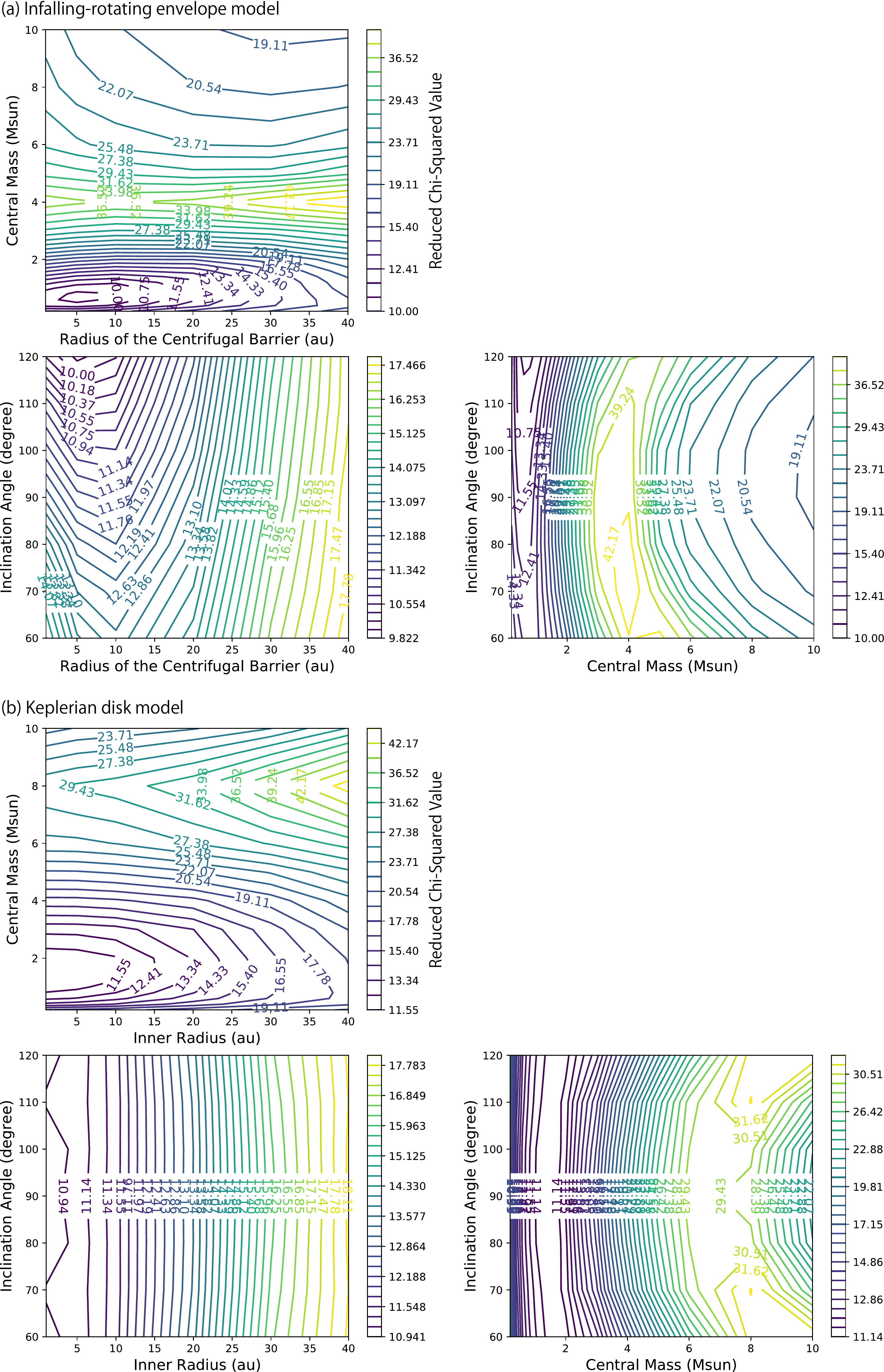} 
	\fi
	\caption{Reduced \chisq\ maps with various parameter values for Elias 29. 
			Free parameters and their ranges are summarized in Table \ref{tb:chi2_Elias29_params}. 
			\label{fig:chi2_Elias29_chisqplot}}
\end{figure}

\clearpage
\begin{figure}
	\iffigurechisq
	\centering \includegraphics[bb = 0 0 500 660, scale = 0.68]{\dirnamechi 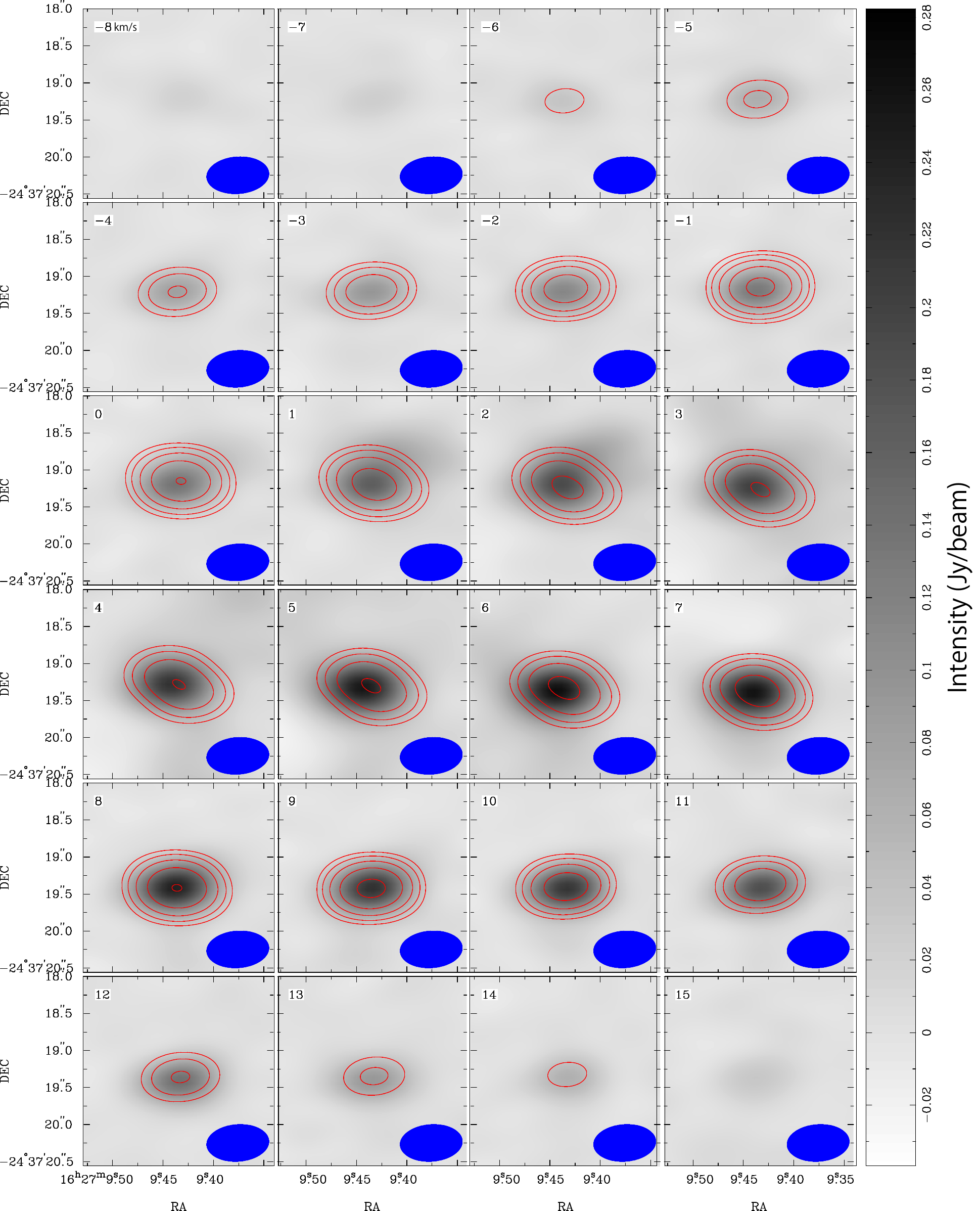}
	\fi
	\caption{Same as Figure \ref{fig:chi2_Elias29_channelmap-IRE}, 
			but for the combined model.   
			Contours represent the best-fit model obtained by the \chisq\ test for the combined model. 
			The parameters for the model are: 
			$M$ of 0.8 \Msun, \rcb\ of 10 au, and $i$ of 120\degr, 
			assuming \rin\ of 1 au. 
			\label{fig:chi2_Elias29_channelmap-combined}}
\end{figure}


\clearpage
\begin{figure}
	\iffigure
	\centering \includegraphics[bb = 0 0 700 500, scale = 0.6]{\dirnamefig 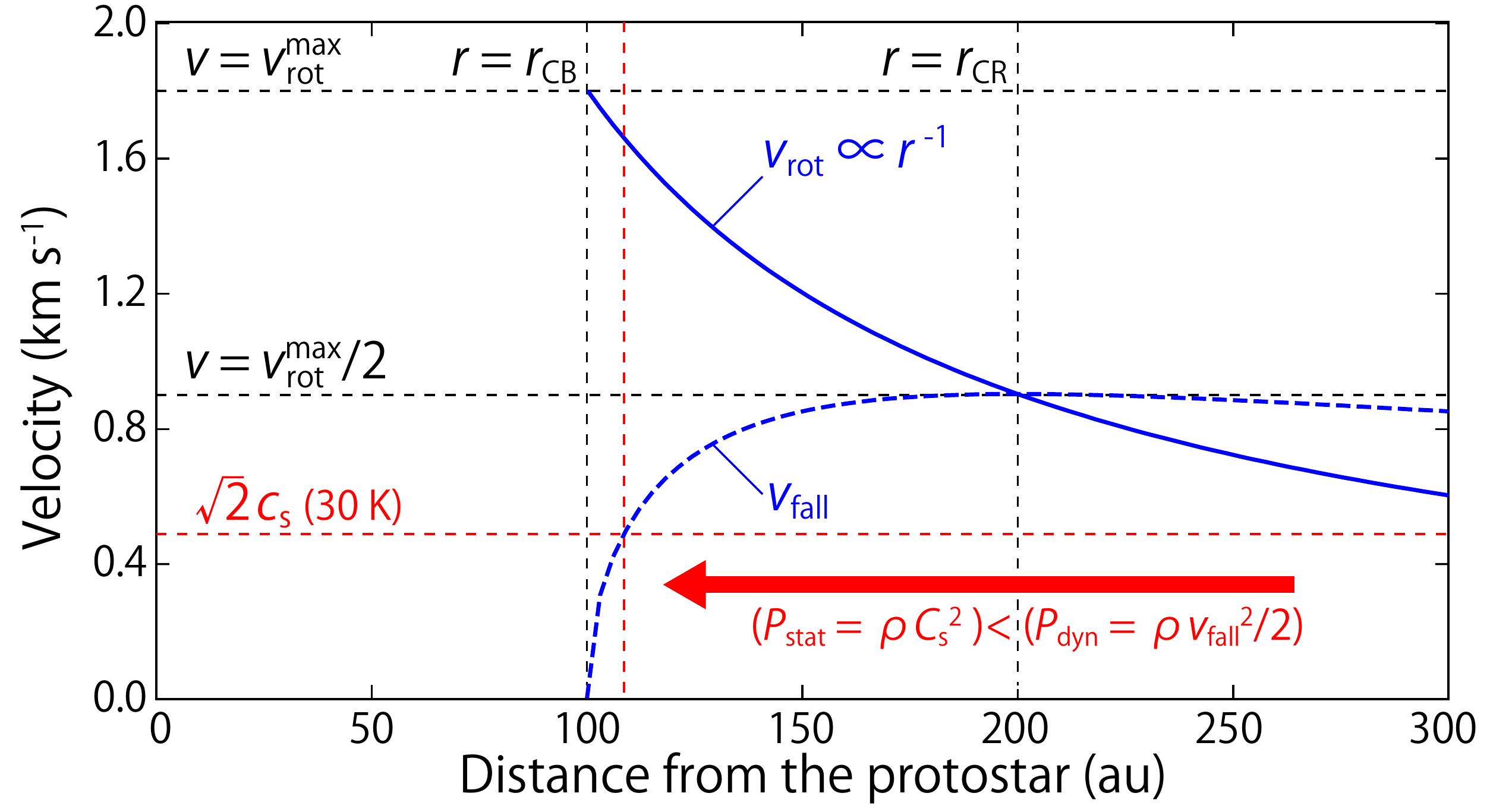} 
	\fi
	\caption{Velocity profile in the \ire\ model for L1527 \citep{Sakai_1527apjl, Oya_1527}. 
			In the \ire\ model, 
			the infall velocity takes its maximum value (\vfall $= 0.9$~\kmps) at the \centr\ (\rcr\ $= 200$ au). 
			The sound speed \vsound\ is assumed to be 0.35~\kmps, 
			which corresponds to the value at the temperature of 30 K. 
			We here employ the protostellar mass of 0.18 \Msun\ and the radius of the \cb\ of 100 au 
			according to the report by \citet{Sakai_1527nature}. 
			\label{fig:pressure}}
\end{figure}

